# Title Page

# Comparative Analysis of Tandem Repeats from Hundreds of Species Reveals Unique Insights into Centromere Evolution

Daniël P. Melters*[1,2], Keith R. Bradnam*[1], Hugh A. Young[3], Natalie Telis[1,2], Michael R. May[4], J. Graham Ruby[5], Robert Sebra[6], Paul Peluso[6], John Eid[6], David Rank[6], José Fernando Garcia[7], Joseph L. DeRisi[5,8], Timothy Smith[10], Christian Tobias[3], Jeffrey Ross-Ibarra#[9], Ian F. Korf#[1] and Simon W.-L. Chan[2,8]

* contributed equally

# corresponding author

1. Department of Molecular and Cell Biology and Genome Center, University of California, Davis, Davis, CA

2. Department of Plant Biology, University of California, Davis, Davis, CA

3. USDA-ARS, Western Regional Research Center, Albany, CA

4. Department of Evolution and Ecology, University of California, Davis, Davis, CA

5. Department of Biochemistry and Biophysics, University of California, San Francisco, San Francisco, CA

6. Pacific Biosciences, Menlo Park, CA

7. Department of Animal Production and Health, Universidade Estadual Paulista, IAEA Collaborating Centre in Animal Genomics and Bioinformatics, Aracatuba, SP 16050-680, Brazil

8. Howard Hughes Medical Institute

9. Department of Plant Sciences and Genome Center, University of California, Davis, Davis, CA

10. USDA-ARS, U.S. Meat Animal Research Center, Clay Center, NE

**Corresponding author contact information:**

Ian F. Korf
Department of Molecular and Cell Biology and The Genome Center
UC Davis
1 Shields Ave.
Davis, CA 95616
USA
ifkorf@ucdavis.edu
(530) 754 4989

**Running title:**

Comparative analysis of tandem repeats and centromere evolution

**Keywords:**

Centromere, tandem repeats, satellite DNA, comparative genomics, eukaryotes, evolution

## Abstract

Centromeres are essential for chromosome segregation, yet their DNA sequences evolve rapidly. In most animals and plants that have been studied, centromeres contain megabase-scale arrays of tandem repeats. Despite their importance, very little is known about the degree to which centromere tandem repeats share common properties between different species across different phyla. We used bioinformatic methods to identify high-copy tandem repeats from 282 species using publicly available genomic sequence and our own data. The assumption that the most abundant tandem repeat is the centromere DNA was true for most species whose centromeres have been previously characterized, suggesting this is a general property of genomes. Our methods are compatible with all current sequencing technologies. Long Pacific Biosciences sequence reads allowed us to find tandem repeat monomers up to 1,419 bp. High-copy centromere tandem repeats were found in almost all animal and plant genomes, but repeat monomers were highly variable in sequence composition and in length. Furthermore, phylogenetic analysis of sequence homology showed little evidence of sequence conservation beyond ~50 million years of divergence. We find that despite an overall lack of sequence conservation, centromere tandem repeats from diverse species showed similar modes of evolution, including the appearance of higher order repeat structures in which several polymorphic monomers make up a larger repeating unit. While centromere position in most eukaryotes is epigenetically determined, our results indicate that tandem repeats are highly prevalent at centromeres of both animals and plants. This suggests a functional role for such repeats, perhaps in promoting concerted evolution of centromere DNA across chromosomes.

## Introduction

Faithful chromosomal segregation in mitosis and meiosis requires that chromosomes attach to spindle microtubules in a regulated manner via the kinetochore protein complex. As the site of kinetochore assembly, the centromere is the genetic locus that facilitates accurate inheritance. Deletion of the centromere or mutation of critical kinetochore proteins results in chromosome loss [1, 2]. Proteins and DNA sequences involved in most essential cellular functions are characterized by their high degree of conservation. Given their conserved function, the observed rapid evolution of kinetochore proteins [3] and lack of homology of centromere repeats thus poses somewhat of a paradox [4].

Centromeres differ greatly in their sequence organization among species. In the budding yeast *Saccharomyces cerevisiae* a 125-bp sequence is sufficient to confer centromere function, and essential kinetochore proteins bind to this "point centromere" in a sequence-dependent manner [5]. Point centromeres are a derived evolutionary characteristic, as ascomycete fungi more distantly related to *S. cerevisiae* have much longer centromere DNAs and do not rely on specific sequences to recruit kinetochore proteins [5, 6]. In the limited set of plant and animal species that have been previously analyzed, centromere DNAs consist of megabase-sized arrays of simple tandem repeats (or satellite DNA), sometimes interspersed with long terminal repeat transposons [7-9]. Some taxa exhibit higher order repeat structures (HORs), in which multiple polymorphic monomers make up a larger repeating unit [10, 11]. When centromeric tandem repeat sequences of different species are compared, sequence similarity appears limited to short evolutionary distances [4, 5]. In fact, specific DNA sequences are probably dispensable for centromere function in most eukaryotes, as kinetochore proteins in diverse organisms can assemble on non-centromeric sequences [2, 12-16]. In humans, these "neocentromeres" have been found through karyotype analysis and can arise at many different loci [17]. In some animals and plants, individual chromosomes — or even the entire chromosome complement — may lack high-copy tandem repeat arrays [2, 13, 15, 16] and in rare cases centromere repeat sequences differ between chromosomes [18, 19] The epigenetic nature of centromere location may be explained by the fact that kinetochores assemble on nucleosomes containing a centromere-specific histone H3 variant, CENH3 (CENP-A in human). Extreme cases of kinetochore protein assembly on diverse sequences are seen in polycentric [18] and holocentric chromosomes [20]. The former has a single very large primary constriction that contains three-to-five CENH3 foci [18], whereas the latter has CENH3-bound sequences and microtubule attachment sites along the entire length of mitotic chromosomes [21]. Despite their dispensable nature, the presence of tandem repeats at the centromere locus of most animals and plants suggests that they serve a function.

Many questions about centromere repeat evolution remain unanswered. How prevalent are high-copy tandem repeat arrays at the centromeres of different animal and plant species? Studies of centromere DNA in animals and plants have so far focused on single organisms or on small clades [5, 22] and few review articles have been dedicated to a broad survey of tandem repeats [23]. No conserved motif has been found for centromere DNA except in small clades (for example, the CENP-B box found in mammalian centromeres [24]). Are there shared properties among centromeric tandem repeats from diverse animals and plants? In *Saccharomyces cerevisiae* and closely related yeast species, short centromere DNA sequences evolve three times faster than other intergenic regions of its genome [25, 26]. How rapidly do centromere tandem repeats evolve and which molecular processes govern their evolution? We performed a survey of tandem repeats in a large and phylogenetically diverse set of animal and plant species in order to address these questions.

Conventional methods used to identify centromeric tandem repeats, particularly CENH3 chromatin immunoprecipitation, are labor intensive and thus difficult to do on a large scale. In this paper, we identified and quantified the most abundant tandem repeats from 282 animal and plant species using a newly developed bioinformatic pipeline. Our method can utilize shotgun whole genome sequence (WGS) data from various sequencing platforms with varying read lengths, including Sanger, Illumina, 454, and Pacific Biosciences. Candidate centromere repeat sequences were characterized by a seemingly unbiased nature. Repeat monomers varied widely in length, GC composition and genomic abundance. Despite great differences in sequence composition, centromere DNAs appeared to evolve by expansion and shrinkage of arrays of related repeat variants (the "library" hypothesis [27]). Using Pacific Biosciences (PacBio) single molecule real-time sequencing to span many contiguous monomers, we characterized the mixing of repeat variants within a single array and the presence of higher-order repeating units. Our data greatly broaden the phylogenetic sampling of centromere DNA, putting evolutionary conclusions about this fast evolving chromosome region on a firmer footing.

## Results

### A bioinformatic pipeline to identify candidate centromere tandem repeats.

Centromere DNAs in most animals and plants share two distinctive properties: the presence of tandem repeats, and their extremely high repeat abundance (often >10,000 copies per chromosome). Therefore, we hypothesized that the most abundant tandem repeat in a given genome would be the prime candidate for the centromere repeat (our method is not designed to find centromere-specific retrotransposons or chromosome-specific repeat sequences). To find such sequences *de novo* from whole genome shotgun (WGS) sequence data, we developed a bioinformatic pipeline that identifies tandem repeats from a variety of sequencing technologies with different read lengths (see Methods) (Fig. 1A). For example, the 171 bp human centromere repeats [11] were identified from Sanger reads, the ~1,400 bp Bovidae repeats [28-30] were identified from PacBio reads (Supplementary Fig. S5), and the 728 bp monkeyflower (*Mimulus guttatus*) repeats [31] were identified from assemblies of Illumina reads. In each case,the most abundant tandem repeat unit was considered to be the candidate centromere DNA.

### Validating the bioinformatic pipeline by identifying known centromere tandem repeats

To validate candidate centromere tandem repeats, we compared our results to sequences described in the literature (Supplementary Table S2). Centromere DNAs have been characterized by restriction enzyme-based methods (e.g. laddering on ethidium bromide-stained gels) combined with fluorescence in situ hybridization (FISH), and by chromatin immunoprecipitation (ChIP) with antibodies raised against a kinetochore protein (typically the centromere-specific histone CENH3). Overall, centromere DNA sequences have been described from 43 of the 282 species in this study. In 38 out of 43 cases, we identified a similar repeat to that reported in the literature (Supplementary Table S2). In the case of opossum (*Monodelphis domestica*) and elephant (*Loxodonta africana*), centromere repeat monomers are believed to be very long (528 and 936 bp respectively) [32] and therefore cannot be found using Sanger reads. We lacked suitable Illumina or 454 data to allow assembly of long tandem repeats from these species, and did not have PacBio data to find long repeats directly. Potato and pea are unusual in that centromere repeats differ across chromosomes [18, 19], with some potato chromosomes lacking tandem centromere repeats entirely [19]. These repeats are too diverse and too long to be identified by our pipeline (upper limit of 2 kbp). Other discrepancies between our candidate centromere repeats and published sequences may be explained by the fact that many previous studies used experimental methods that did not quantify all tandem repeats in the genome (see Supplementary Table S2 for a per species explanation).

In limited cases, an assembled reference genome can assist in identifying a bona fide centromere tandem repeat. As expected for a true centromere DNA sequence, the 1,419 bp repeat from cattle is generally clustered into one large array on all 30 chromosomes in the UMD3.0 genome assembly [33]. These putative centromere arrays contain hundreds of repeat copies (notably, secondary arrays elsewhere in this genome assembly contain only 5–10 copies of the monomer).

CENH3 ChIP followed by sequencing is the most definitive method to confirm that a given sequence underlies the functional kinetochore. Only 13 species out of the 43 had CENH3 ChIP-seq data, and our method correctly identified the published centromere tandem repeat in 10 out of 13 of these cases. The three exceptions were opossum, elephant, and potato where we lacked appropriate sequencing reads to find long tandem repeats (opossum and elephant) or the tandem repeats were too diverse (potato). In summary, our bioinformatic pipeline identified the correct centromere tandem repeat in the large majority of cases where experimental data was available.

In two cases, the most abundant tandem repeat was not the known centromere DNA sequence. In the sequenced maize strain B73 (*Zea mays*) [34], heterochromatic “knobs” contain highly abundant tandem repeats that outnumber the centromere tandem repeat CentC [35]. Knob number, size, repeat abundance and distribution can differ depending on the particular maize variety analyzed, as repeat abundance is variable between isolates [36, 37]. A 178 bp tandem repeat is present at the centromere of the Tammar wallaby (*Macropus eugenii*), but this sequence was only the third most abundant tandem repeat in our analysis [38]. By mammalian standards, Tammar wallaby centromeres are unusually small (~450 kbp per chromosome), and tandem repeats make up a minority of this chromosome region because it is also populated by a centromere-specific retroelement [38].

### Candidate centromere tandem repeats from many uncharacterized animal and plant species

To detect candidate centromere repeats**,** we analyzed a total of 282 species, comprising 78 plants and 204 animals spanning 16 phyla (Fig. 2, Supplementary Table S1). Sanger, Illumina, and 454 sequences were obtained from public databases, and we also performed our own PacBio sequencing. The WGS data included 171 species from Sanger sequencing, 132 from Illumina, 13 from 454, and 9 from PacBio (Fig. 1C). For the 37 species which had both Sanger and assembled Illumina data, both data types yielded the same candidate centromere repeat in the majority of cases (28 out of 37) (Fig. 1C). In most cases where analysis of unassembled Sanger reads revealed a different repeat to Illumina data, individual Sanger reads were too short to find the long repeat monomers (see Supplementary Table S4 for a per species explanation).

Many species whose centromere DNAs had not been previously characterized showed a single tandem repeat whose abundance was much greater than all other tandem repeats in the genome. For example, the American pika (*Ochotona princeps*), *Hydra* (*Hydra magnipapillata*), and Colorado Blue Columbine (*Aquilegia caerulea*) had candidate centromere DNAs of 341 bp, 183 bp, and 329 bp respectively (Fig. 1C).

The most accurate measurements of centromere tandem repeat array size in animals and plants are generally in the range of ~500 kbp to several Mbp [10, 38-40]. Although estimated repeat abundance is subject to several experimental biases, we calculated the average amount of repeat per chromosome, and most organisms in our survey were estimated to contain hundreds of kbp (Supplementary Table S1). Since our analysis was based on WGS data, it is not possible to detect chromosome-to-chromosome variation [2, 15, 16].

### How rapidly do centromere DNA sequences evolve?

An all-vs-all BLAST search of our consensus repeats revealed that sequence conservation was limited to only very closely related species. We found 26 groups of species that showed sequence similarity between centromere tandem repeats (Fig. 2, Supplement Fig. S1). Notable groupings of species with substantial sequence similarity included the primates (Supplement Fig. S2), cichlids (Fig. 6A-B) and grasses (Fig. 6C-D).

The well-studied nature of human centromeres, and the availability of many closely related species, make primates an excellent clade to illustrate the evolution of centromere DNAs [11, 41, 42]. Candidate centromeric tandem repeats in primates showed similarity between monkeys and apes (Fig. 3A, Fig. 4A), but these candidate centromere DNAs were unrelated to those in more basal primates (tarsiers and prosimians). We inspected lower abundance tandem repeat sequences from the Tandem Repeats Finder (TRF) output, and no tandem repeat in tarsiers or prosimians was found to have sequence similarity to the primate candidate centromeric tandem repeat. These results reinforce recent findings showing that the aye-aye (*Daubentonia madagascariensis*) has centromere repeats with no similarity to monkeys and apes [42].

Cichlid fish are another clade in which we identified both conservation and rapid divergence of centromere repeats. Lake Malawi cichlids and the Nile tilapia (*Oreochromis niloticus*) had candidate centromere DNAs that shared 78% sequence similarity, although tilapia diverged from other cichlids 45 MYA. The Princess cichlid *Neolamprologus brichardi* (from Lake Tanganyika) had a candidate centromere repeat with no sequence similarity to either the Lake Malawi cichlids or Nile tilapia, though *Neolamprologus* diverged from Lake Malawi cichlids only 30 MYA. Similar patterns of both conservation and rapid change can be seen in the grasses (Fig. 6C-D). A maize-like centromere repeat can be found in *Panicum*, *Setaria*, and even in a species as distant as rice (*Oryza*), which diverged from maize ~41 MYA. In contrast, sorghum-maize (9 MYA) and *Hordeum-Aegilops* (14 MYA) comparisons show little to no sequence similarity.

To evaluate the rate of sequence evolution across the entirety of our sampled taxa, we assessed the conservation of sequence identity across the phylogeny using a node-averaged comparative analysis (Fig. 3A). We fit a model of exponential decay with divergence, finding that on average sequence identity falls rapidly to background levels (i.e. random 25% identity) after ~50 MYA.

**Candidate centromere tandem repeats from 282 animals and plants display no readily apparent conserved characteristics**

If centromere DNAs are fast evolving, do their repeat monomers at least possess other conserved properties? As our survey is the broadest phylogenetic analysis of tandem repeats to date, we asked if candidate centromere DNAs from 282 species shared common characteristics. Our analyses showed that this was not the case.

First, centromere tandem repeat monomer length is not conserved. As CENH3 is essential for kinetochore nucleation, it has been hypothesized that centromere repeat monomers may tend to be about the size of one nucleosomal DNA [9, 43], as is seen with human (171 bp), *Arabidopsis thaliana* (178 bp), and maize (156 bp) centromere DNAs. This is clearly not a universal rule, as some centromere tandem repeat monomers are much shorter and longer than nucleosomal sizes (e.g. soybean at 92 bp [44, 45] and cattle at 1,419 bp [30]) (Fig. 4A). Plant species tended to have repeat sequences with lengths of approximately 180 bp, whereas we found a broader length distribution in animals. Modest trends in our data, however, may reflect sampling bias in the species for which WGS data was available in public archives rather than biologically meaningful preferences in centromere tandem repeat length.

Second, GC content of centromere tandem repeats is not conserved. Based on limited analysis of animal centromere repeats, it was suggested that centromeric DNA is AT-rich [4]. Our analysis of 282 species revealed that centromeric DNA can be very GC-rich (Fig. 4B), although a slight preference for AT-rich tandem repeats was observed in animals. Plant species do not appear to have a preference for AT- or GC-rich centromere tandem repeats.

Third, the abundance of centromere tandem repeats varies widely (Fig. 4C). We calculated repeat abundance by finding the proportion of reads that matched the repeat monomer (using a set of randomly sampled reads, see Methods and Supplemental Methods). Tandem repeat abundance can be compared between species, but is subject to variability introduced by different library construction protocols at particular sequencing centers, and by biases in the way different sequencing technologies capture high-copy repeats. We compared repeat abundance of 40 species for which there was sequence data from multiple sequencing technologies. On average, sequences derived from Illumina sequencing had higher estimated repeat abundances compared to Sanger, 454 or PacBio data. For most species we estimated that at least 0.5% of the genome was comprised of the candidate centromeric tandem repeat, but the overall percentage was highly variable (Fig. 2, Fig. 4C).

Simple nonphylogenetic correlations found no relation between repeat length, GC content, and genomic fraction of candidate centromere tandem repeats (Supplementary Fig. S3). Similarly, we did not find a correlation between these factors and genome size, genome-wide GC content or chromosome number.

To explicitly test for conservation of sequence characteristics at a finer phylogenetic resolution, we searched for signals in phylogenetic trees that represented the grass and primate clades (Fig. 3B). Both clades are of a similar age (40–45 MYA for the most divergent species) and show substantial sequence similarity among taxa. We calculated Blomberg's *K* statistic [46], a measure of phylogenetic conservation, for various tandem repeat characteristics. The *K* statistic indicates the amount of phylogenetic signal in the data. Values of *K* greater than one suggest that related taxa resemble each other more than would be expected given a null model in which the trait evolves along the tree according to Brownian motion. Values of *K* less than one are observed when related taxa are less similar than expected under the null model. Although repeat monomer length, GC content, and genome fraction all had values of $K<1$ in the grasses, none were statistically significant. In contrast, values for all three characteristics were significantly different from the null model in primates, with GC content and repeat monomer length showing $K>1$ and repeat abundance $K<1$. These data suggest that individual clades likely differ in terms of their tendency for closely related species to have centromere repeats that share conserved sequence characteristics.

**Which species lack candidate high-copy tandem repeats at their centromeres?**

Which animal and plant genomes lack high-copy centromere tandem repeats? The nematode *Caenorhabditis elegans* is a useful negative control for measuring tandem repeat abundance (see red dashed line in the genomic fraction column of Fig. 2), because it has holocentric chromosomes and has been reported to lack centromere tandem repeat arrays in its genome [21]. In total 41 species had a lower abundance of tandem repeats than in *C. elegans*, and these could be assumed to lack high-copy centromere tandem repeats. Nine of these species are known to be holocentric [47] and are not expected to have large tandem arrays. Fungi such as *Saccharomyces* *cerevisiae*, *Candida albicans*, and *Schizosaccharomyces pombe* have small genomes and do not contain high-copy tandem repeat arrays at their centromeres [5]. Many of the other genomes that exhibited low tandem repeat abundance also had small genomes including 7 species of basal plants (green algae, moss, and liverwort) and 11 animals. A few species exhibited low tandem repeat abundance despite possessing large genomes (hedgehog (*Erinaceus europaeus*), tenrec (*Echinops telfairi*), seal (*Leptonychotes weddellii*) and dolphin (*Tursiops truncatus*). This may be due to these species having large repeat units that could not be identified in the available Sanger reads. While a definitive answer is not possible yet, it appears that species lacking large tandem arrays tend to have holocentric centromeres or small genomes.

**Higher order repeat structure and evolution of novel repeat monomers.**

Primate centromeres contain higher order repeat (HOR) structures [11, 48], in which multiple repeat monomers with specific polymorphisms form a unit that itself is repeated (Fig. 5A). HOR structure was easiest to observe in Sanger data, which combines relatively long reads with high sequence accuracy. We used the output from Tandem Repeats Finder (TRF) [49] to identify higher order repeat structures among Sanger sequences from the NCBI trace archive. TRF reports both the repeat monomer, as well as repeating units carrying multimers of the monomer that may represent HOR structure. TRF-defined repeats that occupied approximately the same coordinates within a single read were compared to identify whether longer repeats were dimers of the basic monomer. In true HOR structures, the percentage identity between adjacent multimers should be much higher than between individual monomers (TRF should also report higher scores for the repeats with the longer monomer). Therefore, we filtered TRF output to detect these multimers that had both a higher percentage identity and a higher TRF score compared tothe monomeric repeat that spanned the same coordinates.

Clear cases of HOR structure were identified in 76 of the 171 species with Sanger data. Phylogenetic trees constructed with individual monomers extracted from a single read showed that the “A” monomers and “B” monomers from a dimeric “AB” structure that clustered separately (Fig. 5B-C), confirming that the AB structure indeed represented a HOR unit. HOR structure has been previously described in primates, but our analysis show that it is widespread across both plant and animal kingdoms. The capability to detect HOR units is limited by Sanger read length, so shorter repeat monomers were more likely to display HOR structures. We rarely identified HOR structures that had three or more copies of a repeat monomer, because such structures require at least six monomers to be found in a single Sanger read.

Can HOR structure result in evolution of a new centromere tandem repeat? The centromere repeat monomer has only been reported for one New World Monkey and its length (343 bp) is essentially double the size of human alpha satellite [50]. We extended this analysis to three New World Monkeys and fifteen Old World Monkeys and apes (Fig. 5D, Supplementary Fig. 2). All Old World Monkeys and apes had a 171 bp candidate centromeric tandem repeat, whereas New World Monkeys had a 343 bp candidate centromeric tandem repeat. If the 343 bp repeat is split into two equal halves and aligned to the 171 bp repeat, both halves align, but each has specific polymorphisms and indels (Fig. 5D, Supplementary Fig. S1). These data suggest that in the New World Monkey clade, a doubled version or dimer of the ancestral 171 bp repeat became the dominant centromeric tandem repeat. Such patterns of evolution are likely to be general, as they depend only on acquisition of polymorphism and a particular pattern of recombination within a repeat array [51].

Where HOR structure is present, it means that our calculated values for the abundance of candidate centromere repeats are most likely underestimates. A notable example of this occurs in the gorilla genome (*Gorilla gorilla gorilla*). We correctly identify the 171 bp centromere repeat as the most abundant repeat and this accounts for approximately 1.3% of the genome. However, we also identify a separate, but related, 340 bp repeat which represents a doubled version of the 171 bp repeat. This second repeat accounts for a further 1.2% of the genome, showing that dimeric HOR structure may be especially common among gorilla centromere repeats.

**Coexistence of related repeats support the "library" hypothesis**

The "library" hypothesis aims to explain how centromere DNA evolves so rapidly [27]. This hypothesis assumes that variants of centromere tandem repeats co-exist within the same tandem arrays. Over time, the abundance of particular variants stochastically changes through both expansion and shrinkage [52, 53], resulting in replacement of the most abundant variant with a different variant. Centromere repeat variants could arise by point mutation, deletion, insertion or by mixing of different parental sequences during allopolyploid formation (in all cases, a process such as gene conversion would be required to transfer variants between chromosomes) [54]. Are there cases in our data set that support the "library" hypothesis? Specifically, do repeat variants differ and are there cases where such a repeat was able to colonize a genome and replace the original monomer?

Several Lake Malawi cichlids contained a 237 bp candidate centromeric tandem repeat, whereas the closely related Nile tilapia contained a shorter repeat of 206 bp (Fig. 6A-B). However, the Nile tilapia did contain a less abundant, 237 bp repeat that was similar to the Lake Malawi cichlid repeat (Supplementary Fig. S4). This suggests that the centromere tandem repeat in the common ancestor of Lake Malawi cichlids and Nile tilapia was replaced by a related sequence (having either an insertion or deletion of 29 bp) in one of the two modern clades.

More support for the "library" hypothesis was seen in the grasses (family *Poaceae*); this was the largest plant clade in our dataset that exhibited sequence similarity among most of its members. The modal length of repeat monomers in grasses was 156 bp, but deletions and insertions were found in several species (an 80 bp conserved motif between rice and maize was previously noted within this sequence [40]). Eight of the fifteen grass species had candidate centromere repeats that displayed no similarity to the common 156 bp sequence (Fig. 6C-D). We then searched our data for less abundant tandem repeats related to the dominant repeat monomer. Sanger sequence data for four grass species revealed distinct centromere tandem repeat variants. Maize and foxtail millet (*Setaria italica*) only contained one variant each (variants A and D respectively), witchgrass (*Panicum capillare*) had two variants (B and C) and the switchgrass genome contained three variants (A, B, and C). Variant B itself consists of two distinct repeats, one of 175 bp (variant B1) and another of 166 bp (variant B2). B2 differs from B1 by the deletion of 9 bp, but these two subvariants are otherwise very similar in sequence, so we consider them as one variant (variant B). The existence of related repeat variants in switchgrass and witchgrass is similar to our observations in Lake Malawi cichlids and Nile tilapia, and both these cases further support the "library" hypothesis [27].

Next we asked if switchgrass repeat variants occupied the same tandem repeat arrays by using computationally derived repeat monomers as probes in fluorescent in situ hybridization (FISH) experiments. FISH analysis confirmed that these repeat variants were found at centromeres (Fig. 7). Variants not found in a given genome did not stain chromosomes from that species, showing that our hybridization conditions were specific. The variant A probe only hybridized strongly to one switchgrass chromosome. Variant B in switchgrass was comprised of two repeats (B1 = 175 bp and B2 = 166 bp) and FISH experiments revealed that all switchgrass chromosomes showed hybridization to variants B1, B2 and C, but with differing hybridization intensities (Fig. 8A-B). These data indicate that specific chromosomes harbor different amounts of particular repeat variants, again suggesting that repeat arrays can grow and shrink over evolutionary time.

**Pacific Biosciences sequencing reveals that switchgrass repeat arrays are homogeneous and contain long higher order repeat structures.**

Centromere repeat variants in switchgrass were found on the same chromosomes using FISH (Fig. 8A-B), but the resolution of these experiments could not distinguish large homogeneous arrays of two variants (in close proximity) from arrays that showed more significant mixing of repeat variants. Theoretical simulations predict that an array of polymorphic repeats can become rapidly homogenized by unequal crossing over [51]. Conversely, gene conversion can introduce novel variants into the middle of a repeat array. To determine the degree to which variants were mixed in a given array, we used the PacBio sequencing platform, which yields much longer reads (up to 16.5 kbp) than other sequencing technologies (Fig. 8C) [55]. As PacBio sequencing has a very high indel rate, we focused on repeat variants that differ by indels of at least 9 bp. Switchgrass genomic DNA was sequenced on four runs of the PacBio RS system using the C2 chemistry and a ~10 kbp insert library (see methods for details). All switchgrass chromosomes stained positive for both variant B1 and B2 FISH probes and both repeat variants were present in the PacBio sequence data. However, individual PacBio sequencing reads never contained a mixture of the two variants..This shows that centromere repeat arrays in switchgrass are composed of long homogeneous arrays variants, but that these arrays are mixed together on the same chromosome.

Another benefit of PacBio sequence reads is their ability to detect HOR structure that extends beyond the dimer and trimer structures typically visible in shorter Sanger reads (Fig. 5). We found a novel pattern of HOR structure in switchgrass centromeres using PacBio sequencing: large repeating units that contain deleted versions of a canonical centromere repeat (Fig. 8D). A 2,491 bp read contained a higher order repeat comprised of 4 B1-type monomers followed by a truncated variant approximately half the size of the B1 repeat. The B1 repeat is 175 bp long, and the HOR repeat is 792 bp, too long to be detected by Sanger sequencing. Similarly, a 7,032 bp PacBio read contained a 1,131 bp HOR repeat made of 6 B1-type monomers and a truncated B1 repeat of 53 bp. In this case, the HOR repeat itself is longer than almost all Sanger sequence reads. This application shows that long reads have the benefit of directly revealing long repeat structures that could previously only be seen through painstaking and indirect assembly strategies or by chromosome-specific cytogenetic methods [11, 42, 48, 56].

## Discussion

The ready availability of whole genome shotgun sequence from a wide variety of eukaryote genomes makes comparative genomics an appealing way to study rapidly-evolving tandem repeat sequences, such as those commonly associated with centromeres. Animals and plants are evolutionarily distant, so previous studies showing the presence of high-copy centromere tandem repeats in these organisms raised the question of whether this was indeed a general property. Recently, bioinformatic methods for identifying centromere tandem repeats have been described, and applied to several previously uncharacterized mammals [32, 57, 58] and plants [59]. We have performed the largest survey of animal and plant tandem repeats to date, encompassing every species with sufficient whole genome shotgun sequence in the NCBI trace archive and DDBJ sequence read archive. The bioinformatic methods we used are amenable to every available DNA sequencing technology, making our study expandable as future DNA sequences are generated. In species with previously reported centromere repeats, the most abundant tandem repeat identified in our analysis matched the published sequence in almost every case. The presence of highly abundant tandem repeats in the large majority of species that we analyzed suggests that tandem repeats likely underlie the functional centromere in most animals and plants. Candidate centromere tandem repeats did not share conserved properties such as monomer length, GC content, or common sequence motifs. We found that higher-order tandem repeat structures were prevalent across a broad phylogenetic distribution, as was the evolution of repeats by mutation and indel acquisition. This confirms theoretical predictions that the tandem repeat nature of centromere DNA in animals and in plants can facilitate the rapid evolution of these sequences [51].

As centromeres can form on non-centromeric DNA sequences in both animals and plants, the function of tandem repeats at centromeres is enigmatic [12, 13, 17, 60]. Our finding that centromere tandem repeats are common reinforces the argument that they have a functional, albeit subtle role, although careful experiments may be required to detect this *in vivo*. Further evidence for this comes from both evolutionary and functional experiments. Neocentromeres formed during evolution eventually acquire tandem repeats [61], and neocentromeres lacking tandem repeats are subtly defective in one human cell culture assay [62]. It is possible that centromere specification will be a balance between epigenetic and genetic factors in most plants and animals, although it is clear that epigenetic memory provided by the centromere-specific histone CENH3 is the most important factor.

High-copy tandem repeats have a propensity to form heterochromatin [63], but it is unlikely that this property alone explains their presence at centromeres. Transposons in pericentromeric regions are also highly heterochromatic, and there is little in the chromatin landscape of large repeat-rich genomes such as maize that distinguishes centromeres from similarly gene-poor regions. Transposons inserted into the tandem repeat arrays of cereals and other plant genomes have not been shown to have a function in centromere biology, although they are bound by CENH3 [45, 64, 65] and centromere-specific transposons localize exclusively to the centromeres of close relatives [66]. Most interestingly, the tandem repeats within the CENH3-binding domain of the centromere have significantly different chromatin modifications from typical heterochromatin [67]. In *A. thaliana* and maize, tandem repeats at the functional centromere have been observed to have lower DNA methylation that those at the edge of the repeat array [68]. Extended chromatin fiber microscopy has shown that centromeres in *Drosophila melanogaster* and humans contain some modifications typical of euchromatin (e.g. lack of H3K9 di- or trimethylation), in addition to those associated with gene silencing (hypoacetylation of H3 and H4) [67]. Tethering a transcriptional silencer to a human artificial chromosome or altering its acetylation/methylation balance can lead to centromere inactivation [69, 70]. Lastly, it is possible that non-coding RNAs may have a role in centromere function, and transcription of such molecules may not be compatible with heterochromatic marks [38, 71-73].

If specific DNA sequences play a role at centromeres, and heterochromatin is not needed for kinetochore function, why do so many animal and plant centromeres contain high-copy tandem repeats? The lack of conserved properties among these sequences suggests that it is the tandem nature of the repeats that in itself is useful. Nucleosome phasing may be beneficial for centromeres, and the sequence preferences of histones should lead to phasing on any tandem repeat even if this is a subtle property. Although one study failed to detect nucleosome phasing (translational positioning) at the maize centromere tandem repeats, periodicity based on AA/TT dimers (rotational positioning) within CentC repeats, which suggests that CentC repeats could contribute to a highly stable nucleosome arrangment in centromeres [74]. Nucleosome phasing over the entire centromere should be dominated by nucleosomes containing conventional histone H3, as CENH3 nucleosomes bind to only a small fraction of the tandem repeat array. In a phasing model, the aquisition and accumulation of tandem repeat arrays would be fostered by the chromatin arrangement of centromeres. The phenomenon of centromere reactivation, in which a centromere first loses kinetochore-nucleating activity, then regains it, could suggest that tandem repeats encourage centromeric chromatin states. Notably, centromere reactivation has been observed in both maize [75, 76] and possibly in humans [77].

Rapid evolution itself may explain the fact that centromere DNA in so many animals and plants is comprised of tandem repeats. A prevailing model to explain fast evolution of centromere DNA sequences and CENH3 is that asymmetric meiosis during oogenesis encourages centromeric drive [4, 78]. In this model, competition of centromeres for preferential segregation into the single meiotic cell that survives to become the egg can drive rapid sequence evolution. Eventually, centromere DNA and CENH3 differences could introduce reproductive barriers, causing speciation. CENH3 binding domains in animal and plant chromosomes cover many kbps of DNA. How is it possible that these large stretches of DNA could co-evolve with a histone H3 variant? Similarly, how do centromere DNA sequences on different chromosomes co-evolve? In a tandem repeat array, CENH3 is necessarily binding to the same sequences throughout the centromere, and all chromosomes in the cell typically share versions of the same repeat monomer [79]. In addition, tandem repeats foster rapid evolution, and this property may be favored by meiotic drive [4, 51]. A mutation that arises in any copy of a tandem repeat can be amplified and spread throughout the array by unequal crossing over [51] or by replication fork collapse [80]. Repeat variants can move between different chromosomes in the cell via gene conversion, or possibly through the mobilization of retrotransposons inserted into tandem repeat arrays [81, 82]. As we have shown, the centromere tandem repeat array can be a "library" of sequence variants that show expansion and shrinkage [52, 53], creating opportunities for new variants to colonize a chromosome, likely via concerted evolution or molecular drive [83]. Centromeres with sequence differences would be immediately exposed to selection in organisms with asymmetric female meiosis. Thus, the ability of tandem repeats to facilitate concerted evolution may explain their prevalence at animal and plant centromeres. Yeast species with symmetrical meiosis lack high copy tandem repeats at centromeres [5]. Similarly, the centromere-specific histone does not show positive selection in *Tetrahymena* species with symmetrical meiosis [84]. In the future, it will be interesting to test whether tandem repeats are found at centromeres of diverse eukaryotes that lack asymmetric meiosis.

## Methods

### Obtaining sequence data from online archives

Only whole genome shotgun (WGS) or whole chromosome shotgun (WCS) data were used in our analysis. Sanger DNA sequences (FASTA and corresponding ancillary files) were downloaded from the NCBI Trace Archive (http://www.ncbi.nlm.nih.gov/Traces/home/). For each of the 170 species with WGS or WCS Sanger data, we downloaded up to 5 randomly selected FASTA files (up to 500,000 sequences/file). Illumina and 454 data were downloaded from the DDBJ Sequence Read Archive (http://trace.ddbj.nig.ac.jp/DRASearch/). As of April 1, 2012, 146 species had WGS Illumina or 454 data. For these species, two random FASTQ files were downloaded (one per direction, on average 2Gb/file). For 37 species both Sanger and Illumina data were obtained. A complete list of species, and associated sequence data, that were used in our study can be found in Supplementary Table S1.

### Bioinformatics pipeline for Sanger and Pacific Biosciences data

WGS or WCS data were processed using a Perl-based bioinformatics pipeline. First, Sanger sequences were clipped for quality and/or vector contamination. Subsequently, sequences that had >5% Ns were removed, as were any sequences shorter than 100 bp (Sanger) or 1,000 bp (PacBio). Low complexity sequences were then masked using the DUST filter. The remaining sequences were analyzed by Tandem Repeats Finder (TRF) (http://tandem.bu.edu/trf/trf.html) [49] to identify tandem repeats. We assumed that candidate centromeric tandem repeat arrays should be continuous and occupy the majority — if not all — of any individual read. We therefore excluded repeats that accounted for <80% of the entire read. TRF sometimes predicted multiple tandem repeats occupying the same span within a read (with different repeat monomer lengths). In these situations we only retained the shortest repeat for further analysis. Very short repeats, with monomer lengths <50 bp, were also excluded from further analysis.

After producing a set of tandem repeats for each species of interest (using the consensus repeat sequence from TRF), we then used WU-BLASTN [[NO STYLE for: Gish]] with parameters M=1 N=-1 Q=3 R=3 W=10 (with post-processing from various Perl scripts) to produce a set of 'global' and 'local' clusters of repeats in each species (see Supplemental Methods for full details). Global clusters contained repeats with very similar sequences that also had near-identical lengths. This clustering step used just a sample of the total number of tandem repeats produced by TRF and we identifed the source reads of all of the sample repeats. This allowed us to identify what fraction of the input sample reads were represented by each global or local cluster. Repeats in the top clusters are presumed to be the candidate centromeric repeat.

**Bioinformatics pipeline for Illumina and 454 data**

Illumina and 454 reads are often too short to contain at least two copies of a tandem repeat. Therefore, these shorter reads have to be assembled to create contigs that contain at least two copies of a tandem repeat (even if such contigs are not biologically real). To assemble contigs containing tandem repeats, repeat monomers must be polymorphic (a property shared by all centromere tandem repeats described so far [4]). Some short read assemblers do not work well with sequences containing polymorphisms. To assemble polymorphic centromere tandem repeats, we used the short read assembler PRICE (Paired-Read Iterative Contig Extension) (http://derisilab.ucsf.edu/software/price/index.html). For most of the Illumina and 454 data we used PRICE beta version 0.6. This version could only handle paired-end Illumina and 454 data. The later PRICE beta version 0.13 and subsequent versions also allowed for use of single end Illumina and 454 data. For each species, we used 200,000 randomly selected reads which were assembled on 20,000 seed sequences (see PRICE manual) with at least 85% sequence similarity. The contigs were analyzed for the presence of tandem repeats by TRF, allowing for a tandem repeat monomer of 2,000 bp (upper limit of TRF). To determine genomic fraction, 1,000,000 short reads were aligned to the obtained tandem repeat monomers (see Supplementary Methods for more details).

**Data analysis of centromere tandem repeats**

To compare candidate centromeres from all species to each other, we performed a BLASTN [86] search. We used WU-BLAST version 2.0 with parameters M=1 N=-1 Q=2 R=2 W=8. Since tandem repeat boundaries are arbitrary, it is possible for related repeats to align in a staggered fashion and align over only a fraction of their true length. We therefore aligned a file of repeats to a file of duplicated repeats. Since BLAST produces local alignments and we were interested in overall similarity, we calculated a global percent identity by adding additional alignment length assuming a 25% match rate in unaligned regions.

To assess the rate at which sequence similarity decays on phylogenetic timescales, we performed node-averaged phylogenetically independent contrasts [87, 88]. In order to account for shared history in comparisons of sequence similarity, this method calculates the average sequence similarity between each pair of taxa spanning a node to generate a single value for each node in the tree. Since the taxa of interest span a wide range of eukaryotes and our analyses are relatively insensitive to branch length estimates, we used a tree based on the NCBI taxonomy [89] and repeated our analyses on ten random resolutions of the tree in order to accommodate unresolved relationships. As most unresolved nodes were shallow, these random resolutions had little effect on the quantitative results of the analyses performed (data not shown). All phylogenetic analyses were conducted using the R package APE [90]. We then performed regression analysis in order to determine the relationship between node age (as determined with TimeTree, [91]) and node-averaged sequence similarity. We used the R package bbmle2 to fit the simple exponential model $H \sim \alpha t^{\lambda}$, where H is the node-averaged homology and t is node age, and α is the intercept.

To determine the conservation of several repeat characteristics on a finer scale, we performed phylogenetic comparative analysis using the R packages GEIGER (http://cran.r-project.org/web/packages/geiger/index.html) [92] and picante [93, 94]. We estimated Blomberg's *K* measure of phylogenetic conservation for repeat length, GC content, and repeat abundance using chronograms estimated for primates (http://10ktrees.fas.harvard.edu/index.html) and grasses [95].

**Pacific Biosciences single molecule real time sequencing**

Switchgrass (tetraploid *Panicum virgatum* AP13) DNA was isolated using a modified protocol for Chen and Ronald [96] (Supplementary methods). Library preparation and sequencing was performed according to manufacturer's instructions (Pacific Biosciences, Menlo Park, CA). In short, 3-10 µg of genomic DNA was isolated and fragmented to 7-10 kbp fragments using HydroShear for 15 minutes (switchgrass), or Covaris G-tube (cattle, yak, water buffalo). The first of five Ampure XP bead purifications was performed (0.45X Ampure beads added to DNA dissolved in 200 µL EB, vortexed for 10 minutes at 2,000 rpm, followed by two washes with 70% alcohol and finally diluted in EB) After each Ampure XP purification step a quality control was performed comprised of DNA concentration determination by nanodrop and fragment size distribution by bioanalyzer. Next, the DNA fragments were repaired using DNA Damage Repair solution (1X DNA Damage Repeat Buffer, 1X NAD+, 1 mM ATP high, 0.1 mM dNTP, and 1X DNA Damage Repeat Mix with a final volume of 85.3 µL) with a volume of 21.1 µL and incubated at 37ºC for 20 minutes. DNA ends were repaired next by adding 1X End Repeat Mix to the solution, which was incubated at 25ºC for 5 minutes, followed by the second Ampure XP purification step. Next, 0.75 µM of Blunt Adapter was added to the DNA, followed by 1X template Prep Buffer, 0.05 mM ATP low and 0.75 U/µL T4 ligase to ligate (final volume of 47.5 μL) the bell adapters to the DNA fragments. This solution was incubated at 25ºC for 30 minutes, followed by a 65ºC 10 minute heat-shock. The exonuclease treatment to remove unligated DNA fragments consists of 1.81 U/μL Exo III and 0.18 U/μL Exo IV (final volume of 3.8 μL) which is incubated at 37ºC for 1 hour. Next, three Ampure XP purifications steps were performed. Finally, the bell primer is annealed to the PacBio bell with inserted DNA fragment (80ºC for 2 minute 30 followed by decreasing the temperature by 0.1º/s to 25Cº). This complex was loaded into PacBio RS SMRT cells, which were loaded onto the machine for either 2 x 30, 2 x 45, 1 x 75, or 1 x 90 minute runs. Four cells each were used for *Zea mays*, *Zea luxurians, Panicum virgatum, Bos taurus taurus, Bos taurus indicus, Bos grunniens, Bison bison* and *Bubalus bubalis*, while two cells were sufficient for *Panicum capillare*.

**Fluorescence in situ hybridization (FISH)**

Mitotic chromosome spreads were generated following a protocol by Zhang and Friebe [97] with a few modifications (Supplementary methods). Plasmid vectors containing a single copy of each repeat variant (A, B1, B2, C, or D) were synthesized by Bio Basic Inc. (Ontario, Canada) and used as probes for FISH analyses. Probe hybridization signals were detected using anti-digoxigenin (dig) conjugated FITC (green), anti-dig conjugated Rhodamine (red), or Streptavidin conjugated Rhodamine (red) antibodies (Roche Applied Sciences). Chromosomes were counter-stained with 4',6-diamidino-2-phenylindole (DAPI). Digital images were recorded using an Olympus BX51 epifluorescence microscope (Olympus Corporation, Center Valley, PA) (Suplementary methods for more details).

## Data Access

Pacific Biosciences sequences for *Panicum capillare* and *Panicum virgatum* were deposited in the NCBI SRA under accession number SRA052051. A list of GenBank and Sequence Read Archive accession numbers for all sequences used in this study are provided in the supplementary material. A spreadsheet containing all of the tandem repeat information for each species in this study, along with copies of all Perl scripts used are available to download online (http://korflab.ucdavis.edu/Datasets/).

## Acknowledgments

This work would not have been possible without the guidence, vision, and boundless enthusiasm of Simon Chan who sadly passed away on August 28th, 2012 at the age of 38.

We thank Reneé Godtel for technical assistance with bovid DNA sequencing, the International Nellore Genome Sequencing Consortium for prepublication access to sequence data, Michael Heaton for access to DNA samples of bison, yak, and water buffalo, Lauren Sagara and Henriette

O'Geen for technical assistance with Pacific Biosciences sequencing, and Felicia Tsang, Alan Raetz and Alex Han for technical assistance with bioinformatic analysis.

The U.S. Department of Agriculture, Agricultural Research Service, is an equal opportunity/ affirmative action employer and all agency services are available without discrimination. Mention of commercial products and organizations in this manuscript is solely to provide specific information. It does not constitute endorsement by USDA-ARS over other products and organizations not mentioned.

This work was supported by a Joint USDA/DOE Office of Science Feedstock genomics grant DE-AI02-09ER64829 (to C.T.), and by National Science Foundation grants IOS-0922703 to J.R.-I. and IOS-1026094 to S.C. D.M. was supported by training grant T32-GM008799 from NIH-NIGMS. Its contents are solely the responsibility of the authors and do not necessarily represent the official views of the NIGMS or NIH.

J.D. is an investigator of the Howard Hughes Medical Institute. S.C. is a Howard Hughes Medical Institute and Gordon and Betty Moore Foundation investigator.

## Author Contributions

D.M., K.B., N.T., J.R-I., I.K. and S.C. designed experiments and interpreted bioinformatic results.

D.M., K.B., N.T., M.M. and I.K. performed bioinformatic analyses.

H.Y. performed FISH experiments.

G.R. and J.D. provided the PRICE short read assembler.

D.M., R.S., P.P., J.E. and D.R. performed PacBio sequencing.

J.G. contributed genome sequence data for Nellore.

T.S. contributed genome sequence data for cattle, bison, yak and water buffalo.

D.M., K.B., J.R.-I., I.K. and S.C. wrote the paper, with substantial assistance from H.Y., T.S. and C.T.

**Figure Legends.**

**Figure 1. A bioinformatic pipeline to identify candidate centromere DNAs based on their tandem repeat nature and abundance.**
**A.** Random shotgun sequences from a variety of platforms can be used to identify the most common tandem repeat monomer. Sanger and PacBio reads are usually long enough to contain multiple copies of a tandem repeat. Illumina and 454 reads are generally too short, and must be assembled to create longer sequences. Tandem repeat monomers were identified by Tandem Repeats Finder (TRF).
**B.** Identification of known centromere tandem repeats from three species. The human centromere repeat is 171 bp in length. The 728 bp monkeyflower centromere repeat is too long to be found in Sanger reads, but a PRICE assembly of Illumina reads reveals the known repeat . The 1,419 bp cattle centromere repeat and a less abundant 680 bp tandem repeat were directly identified from PacBio reads. Note that the graph for monkeyflower has no background of low abundance tandem repeats because these were not assembled by PRICE.
**C.** Three examples of *de novo* identification of centromere tandem repeats. Sanger WGS reads from the American pika, Hydra, and Colorado Blue Columbine revealed 253 bp, 183 bp, and 329 bp repeat monomers, respectively.

**Figure 2. Centromere tandem repeat details from diverse animal and plant genomes.**
The phylogenetic relationships between 282 species (204 Animalia and 78 Plantae) are shown. For each species, the figure shows tandem repeat length, GC content, and genomic fraction (log 2 scale) for the (candidate) centromere repeat monomer. Taxonomic relationships were derived from the NCBI taxonomy website. Approximately one third of the species (84 out of 282) could be clustered into 26 groups (light red bars) that exhibited sequence similarity of the tandem repeat monomer within each group. No sequence similarity was found outside these groups, or between them. The most distantly related species within a group diverged about 50 million years ago.

**Figure 3. Centromere tandem repeat monomers are conserved only between closely related species.**
**A.** Percent identity between candidate centromere repeat sequences plotted against estimated divergence time. We averaged percentage identity between comparisons to generate a single value for each node in the phylogenetic tree (Figure 2). To accommodate unresolved relationships, we repeated the analysis on random resolutions of the tree. One such analysis is shown (quantitative results were very similar between analyses).
**B.** For primates and grasses, the phylogenetic signal was tested using Blomberg's *K* analysis for three different parameters: repeat monomer length, repeat monomer GC content and genomic abundance. In primates both repeat length and GC content were more conserved than expected ($K>1$), whereas genomic abundance was less conserved than expected by a model of Brownian evolution ($K<1$). Though $K<1$ for all three traits in the grasses, none were significantly different from 1. P values are shown in brackets.

**Figure 4. Centromere tandem repeats lack conserved sequence properties.**
No strong bias was observed in distribution of centromere repeat monomer length (**A**), GC content (**B**), or genomic fraction (**C**).

**Figure 5. Higher order repeat structures are prevalent in diverse animals and plants.**
**A.** Graphical represention of higher order repeat structure compared to simple monomer repeats. In the higher order repeat, two variants A and B form a single dimer repeat that is repeated in tandem. When plotting repeat monomer length by GC content by genomic fraction, two distinct peaks are seen for *Sorghum bicolor*. The second peak (2) is exactly double the length of the first peak (1).
**B.** Sequence alignment of repeat units from a single *Sorghum bicolor* Sanger read that exhibits a higher order repeat structure consisting of an AB dimer. The arrows point to SNPs unique for either the A or B repeat of the dimer.
**C.** Neighbor joining analysis showing grouping of A and B repeats from sequence alignment in B. Bootstrap numbers are shown.
**D.** Higher order repeat structures can lead to novel centromere repeats. In New World Monkeys, the two halves of the 343 bp monomer are weakly related to each other and to the 171 bp repeat in Old World Monkeys and Apes.

**Figure 6. Evolution by indel acquisition and coexistence of repeat variants support the "library" hypothesis.**
**A.** Candidate centromere repeat sequences of eight cichlids were analyzed for interspecies sequence similarity. The Princess cichlid *Neolamprologus brichardi* lacked centromere repeat similarity with its sister clade of Lake Malawi cichlids (shown in orange, and also including Nile tilapia).
**B.** Sequence alignment of candidate centromere repeats shows that Nile tilapia (*Oreochromis niloticus*) has a deletion relative to other cichlid species.
**C.** Candidate centromere repeat sequences of fifteen grass species were analyzed for interspecies sequence similarity. We found two groups of species with centromere repeat sequences that were similar. The closely related *Sorghum* and *Miscanthus* species have similar 137 bp repeats (blue bars). The clade shown by red bars contains *Oryza sativa* (rice), which is relatively distant from the other species that have similar centromere tandem repeats (red bars). Although the centromere repeats of *Oryza brachyantha* and *Brachypodium distachyon* have repeat monomer length similar to the orange-highlighted group, no sequence similarity was found between them. Interestingly, no sequence similarity was found between the closely related *Zea* species and *Sorghum* species or between *Oryza* species and *Brachypodium*, *Aegilops*, or *Hordeum*.
**D.** Sequence alignment of candidate centromere repeats from 8 grass species. Switchgrass (*Panicum virgatum*) is distinguished by the presence of a short insertion relative to the other species.

**Figure 7. Chromosomal localization of repeat variants in grasses is consistent with repeat abundance measured by our bioinformatic pipeline.**
Chromosomal localization of the different grass repeat variants (maize variant A, switchgrass variants B1 and B2, witchgrass variant C, and foxtail millet variant D) was determined by FISH on metaphase chromosomes of maize (*Zea mays*), switchgrass (*Panicum virgatum*), witchgrass (*Panicum capillare*), and foxtail millet (*Setaria italica*). Switchgrass variants B1 and B2 differ by a 9 bp deletion, whereas both variants differ from maize, witchgrass and foxtail millet by a 20 bp insertion. Maize and foxtail millet chromosomes hybridized only to variants A and D respectively. Only one switchgrass chromosome hybridized to variant A (arrow), but variants B1, B2 and C labeled most chromosomes (arrowheads indicate chromosomes that showed weaker hybridization to variant C). Witchgrass chromosomes were most consistently labeled by variant C, but showed chromosome-specific hybridization to variants B1 and B2, consistent with their lower abundance in the genome.

**Figure 8. Pacific Biosciences sequencing shows homogeneity of repeat arrays and detects long higher order repeat structures.**
**A.** Switchgrass variant B1 hybridized to all switchgrass chromosomes, whereas witchgrass variant C hybridized to all but three switchgrass chromosomes. The three chromosomes that only showed hybridization of variant B1 (arrows) were stained green (see merged).
**B.** Although both switchgrass variants B1 and B2 co-hybridize to all switchgrass chromosomes, the hybridization signal showed a chromosome specific pattern. The arrows highlight chromosomes with stronger hybridization signal for one sub-variant over the other.
**C.** The strength of PacBio sequencing is the extreme length of a small fraction of the reads. In the AP13 switchgrass PacBio sequencing run, the longest inserted sequence was almost 12 kbp in length, although the mean of all the PacBio reads was about 2 kbp. Sanger reads are shorter, but have a more consistant length, whereas both Illumina and 454 reads are very short and very homogeneous in length (longest reads in our study only shown).
**D.** Although no repeat variant mixing was detected in the PacBio reads, several high order repeat structures were found in longer PacBio reads. These HOR structures consisted of a mixture of complete and truntcated repeats. Two switchgrass variant B1 centromere reads with higher order structure and one switchgrass variant B2 centromere repeat are shown. The 1131 bp HOR structure consisted of six repeat monomers and a truncated repeat (about 1/3 the size of 175 bp repeat). In total five-and-half copies of the 1131 bp repeat were found within the 7 kbp read. One variant B2-containing read is shown, containing three copies of a 886 bp HOR structure (comprised of six 166 bp repeats).

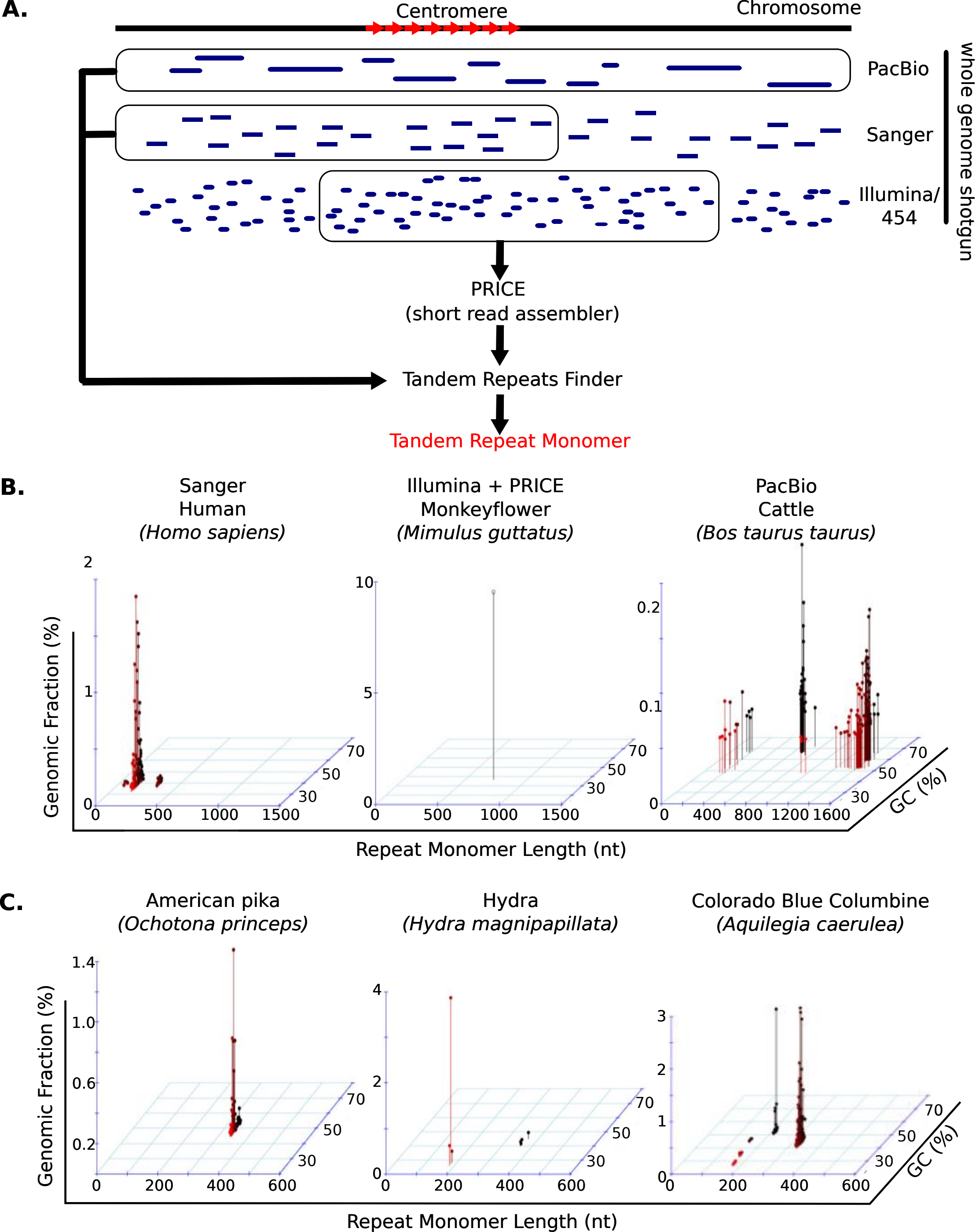

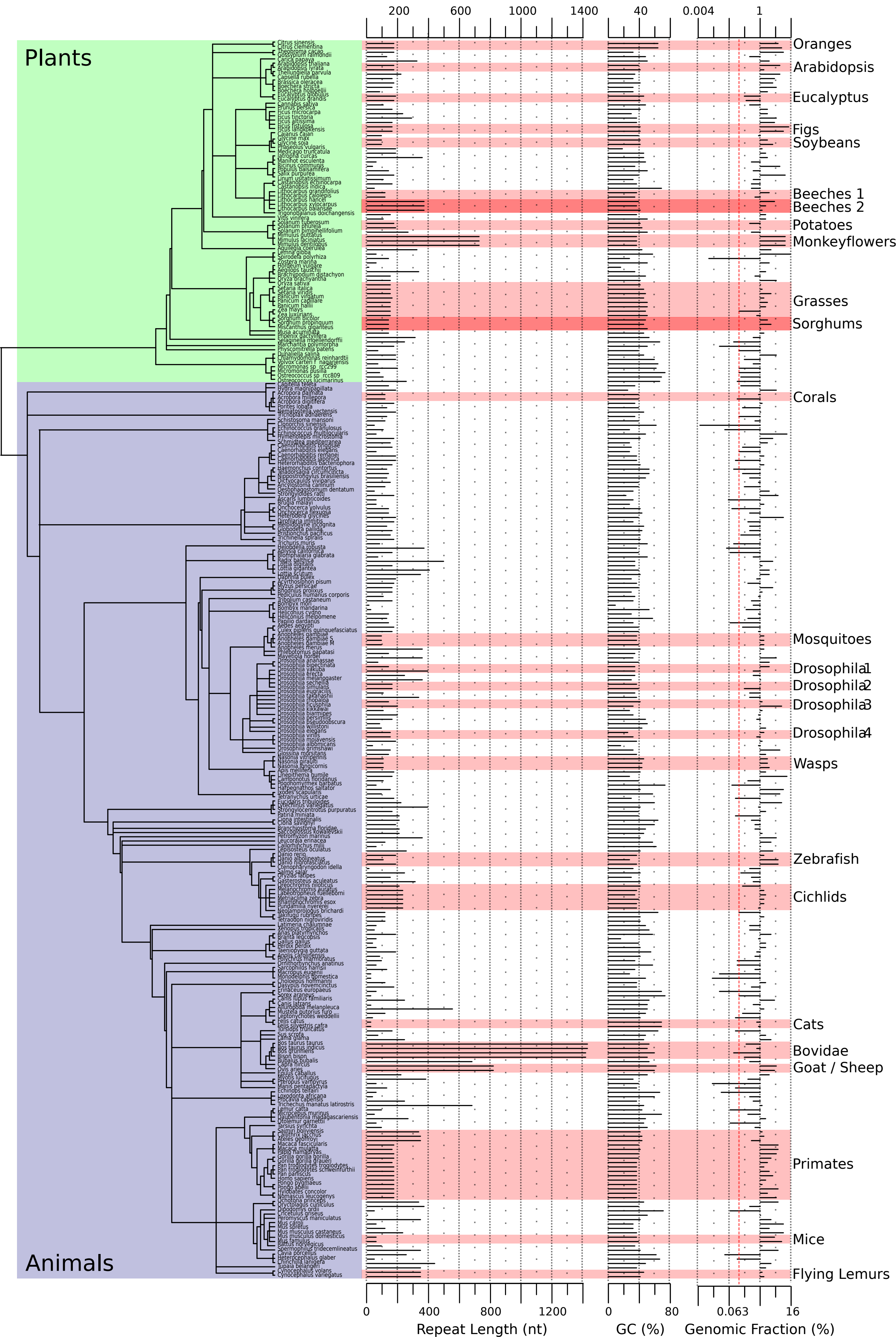

A

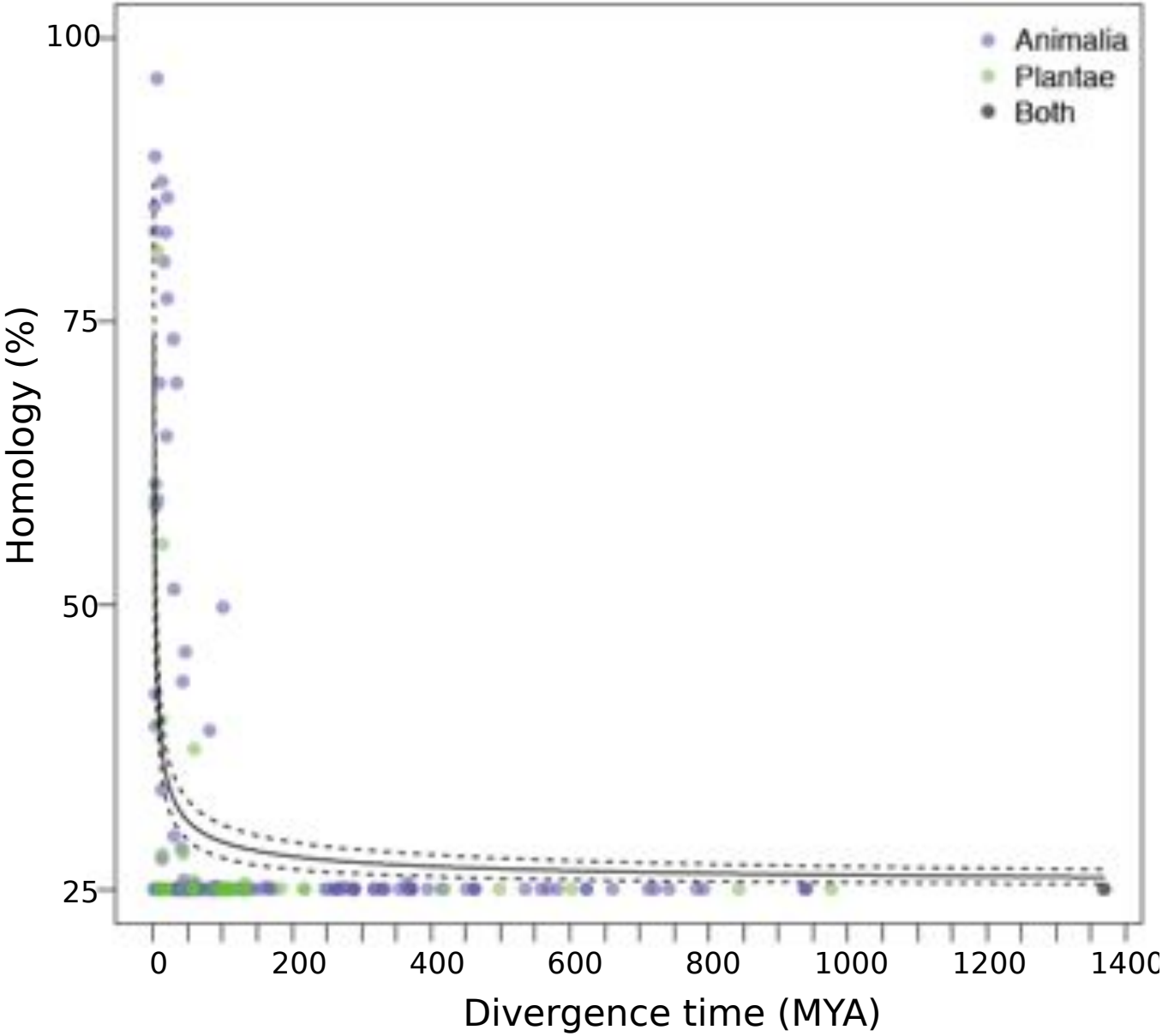

B

| Blomberg's K | Monomer length | GC content | Genomic fraction |
|---|---|---|---|
| Primates | 1.852265 (p = 0.001) | 1.615503 (p = 0.001) | 0.7551643 (p = 0.004) |
| Grasses | 0.414999 (p = 0.427) | 0.6775028 (p = 0.158) | 0.361093 (p = 0.645) |

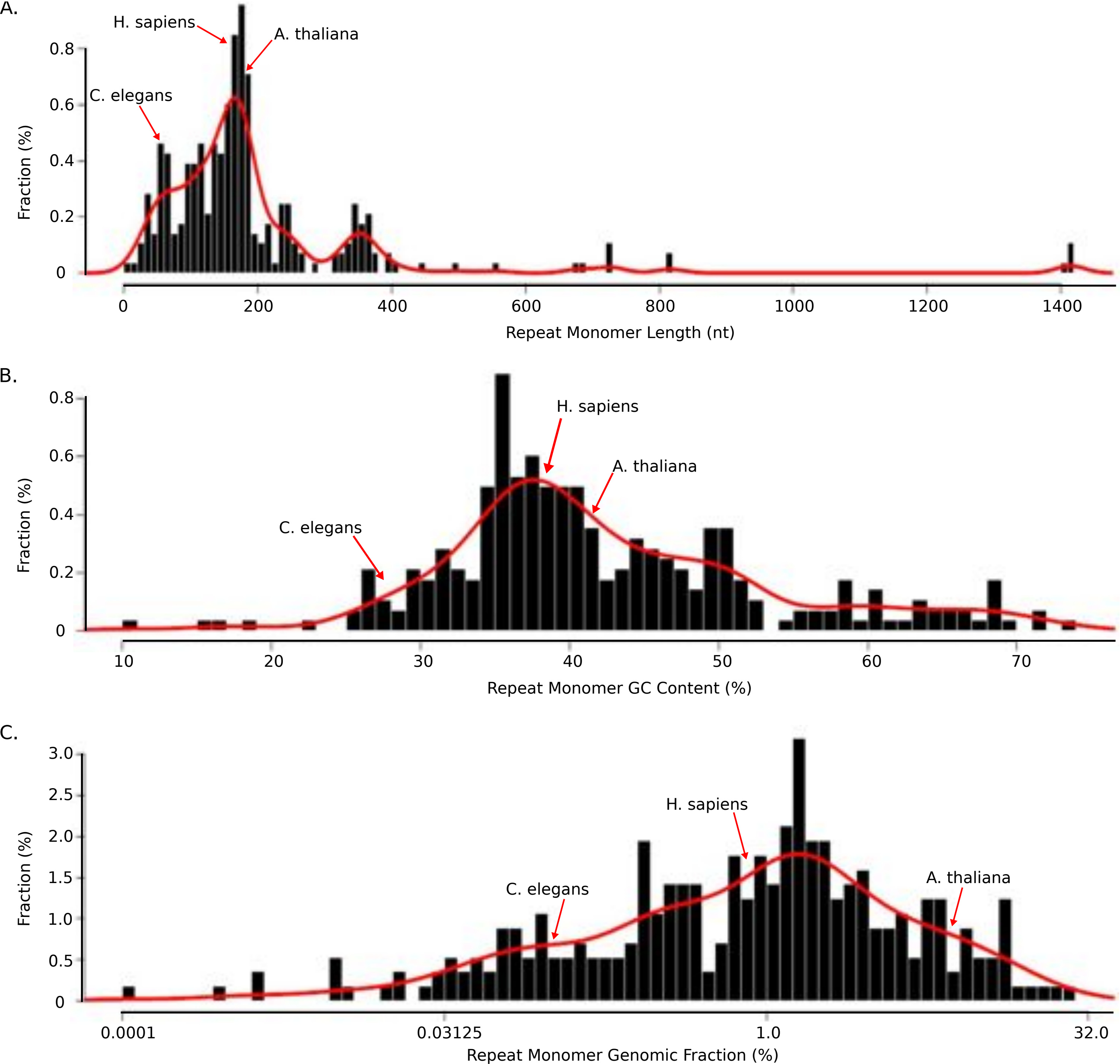

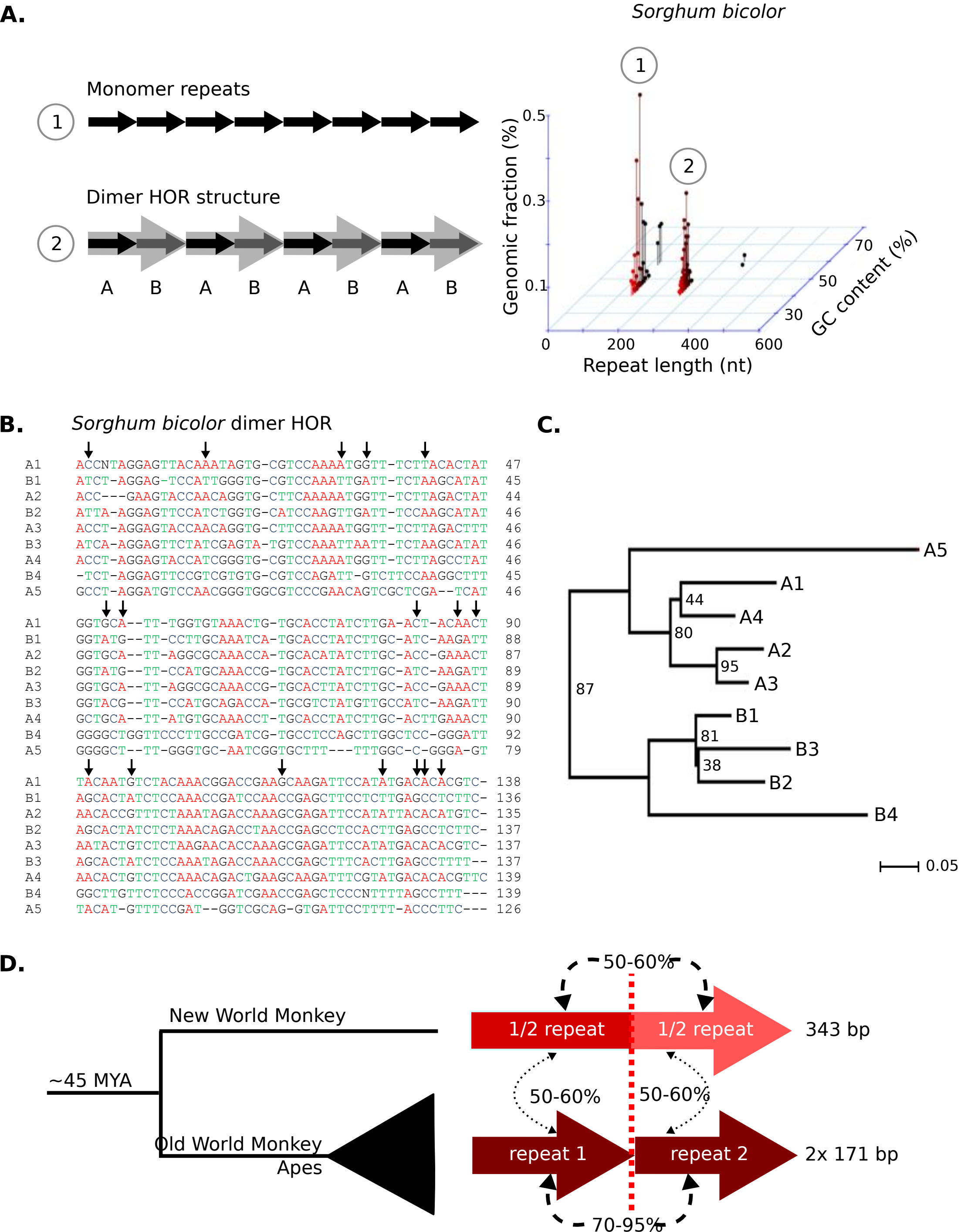
A.
Sorghum bicolor
Monomer repeats
1
Dimer HOR structure
2
A B A B A B A B
Genomic fraction (%)
0.5
0.3
0.1
0
200
400
600
Repeat length (nt)
30
50
70
GC content (%)
B.
Sorghum bicolor dimer HOR
A1 ACCNTAGGAGTTACAAATAGTG-CGTCCAAAATGGTT-TCTTACACTAT 47
B1 ATCT-AGGAG-TCCATTGGGTG-CGTCCAAATTGATT-TCTAAGCATAT 45
A2 ACC---GAAGTACCAACAGGTG-CTTCAAAAATGGTT-TCTTAGACTAT 44
B2 ATTA-AGGAGTTCCATCTGGTG-CATCCAAGTTGATT-TCCAAGCATAT 46
A3 ACCT-AGGAGTACCAACAGGTG-CTTCCAAAATGGTT-TCTTAGACTTT 46
B3 ATCA-AGGAGTTCTATCGAGTA-TGTCCAAATTAATT-TCTAAGCATAT 46
A4 ACCT-AGGAGTACCATCGGGTG-CGTCCAAAATGGTT-TCTTAGCCTAT 46
B4 -TCT-AGGAGTTCCGTCGTGTG-CGTCCAGATT-GTCTTCCAAGGCTTT 45
A5 GCCT-AGGATGTCCAACGGGTGGCGTCCCGAACAGTCGCTCGA--TCAT 46
A1 GGTGCA--TT-TGGTGTAAACTG-TGCACCTATCTTGA-ACT-ACAACT 90
B1 GGTATG--TT-CCTTGCAAATCA-TGCACCTATCTTGC-ATC-AAGATT 88
A2 GGTGCA--TT-AGGCGCAAACCA-TGCACATATCTTGC-ACC-GAAACT 87
B2 GGTATG--TT-CCATGCAAACCG-TGCACCTATCTTGC-ATC-AAGATT 89
A3 GGTGCA--TT-AGGCGCAAACCG-TGCACTTATCTTGC-ACC-GAAACT 89
B3 GGTACG--TT-CCATGCAGACCA-TGCGTCTATGTTGCCATC-AAGATT 90
A4 GCTGCA--TT-ATGTGCAAACCT-TGCACCTATCTTGC-ACTTGAAACT 90
B4 GGGGCTGGTTCCCTTGCCGATCG-TGCCTCCAGCTTGGCTCC-GGGATT 92
A5 GGGGCT--TT-GGGTGC-AATCGGTGCTTT---TTTGGC-C-GGGA-GT 79
A1 TACAATGTCTACAAACGGACCGAAGCAAGATTCCATATGACACACGTC- 138
B1 AGCACTATCTCCAAACCGATCCAACCGAGCTTCCTCTTGAGCCTCTTC- 136
A2 AACACCGTTTCTAAATAGACCAAAGCGAGATTCCATATTACACATGTC- 135
B2 AGCACTATCTCTAAACAGACCTAACCGAGCCTCCACTTGAGCCTCTTC- 137
A3 AATACTGTCTCTAAGAACACCAAAGCGAGATTCCATATGACACACGTC- 137
B3 AGCACTATCTCCAAATAGACCAAACCGAGCTTTCACTTGAGCCTTTT-- 137
A4 AACACTGTCTCCAAACAGACTGAAGCAAGATTTCGTATGACACACGTTC 139
B4 GGCTTGTTCTCCCACCGGATCGAACCGAGCTCCCNTTTTAGCCTTT--- 139
A5 TACAT-GTTTCCGAT--GGTCGCAG-GTGATTCCTTTT-ACCCTTC--- 126
C.
A5
A1
44
A4
80
A2
95
A3
87
B1
81
B3
38
B2
B4
0.05
D.
New World Monkey
~45 MYA
Old World Monkey
Apes
50-60%
1/2 repeat
1/2 repeat
343 bp
50-60%
50-60%
repeat 1
repeat 2
2x 171 bp
70-95%

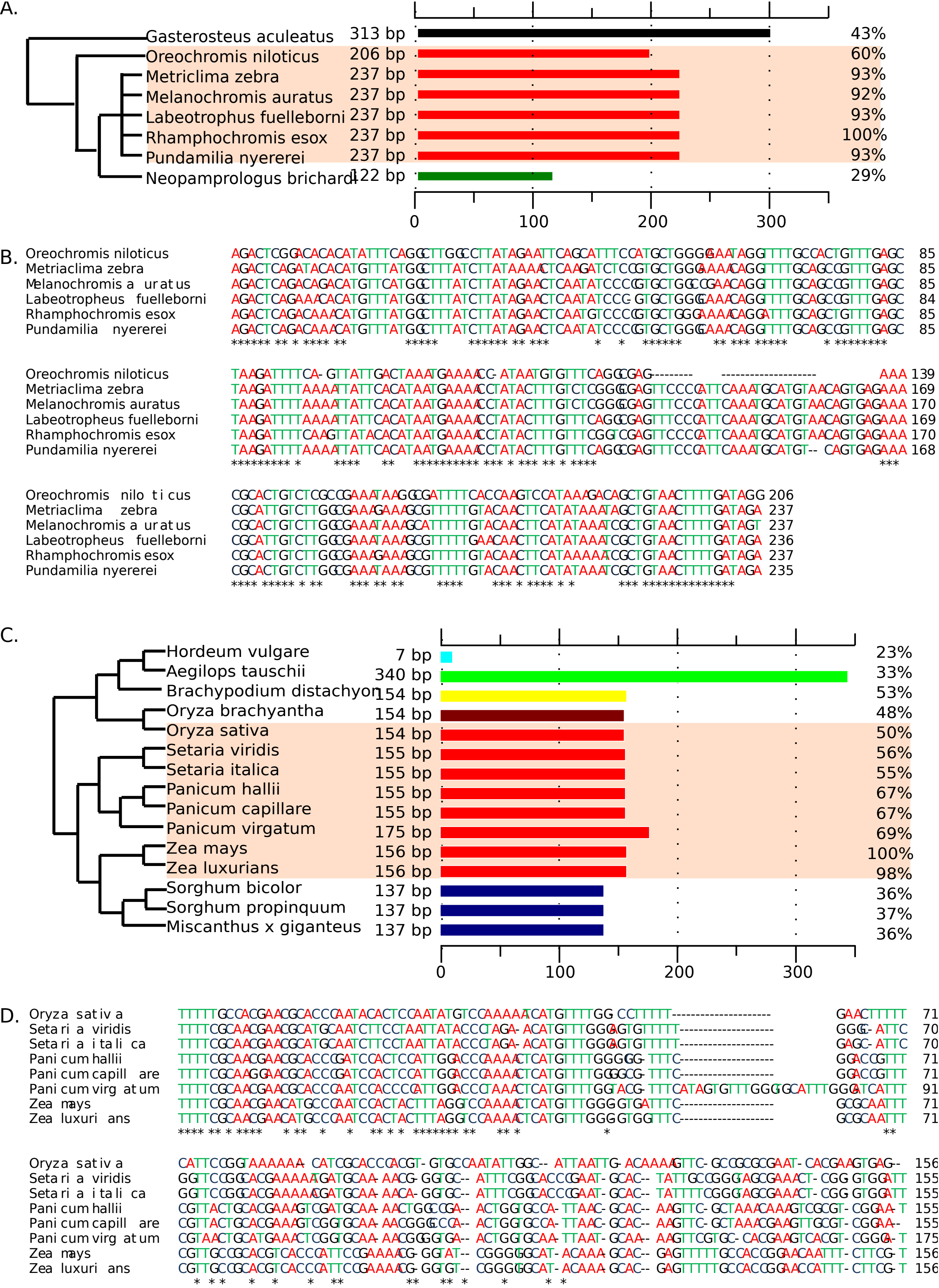

A.
Gasterosteus aculeatus 313 bp 43%
Oreochromis niloticus 206 bp 60%
Metriclima zebra 237 bp 93%
Melanochromis auratus 237 bp 92%
Labeotrophus fuelleborni 237 bp 93%
Rhamphochromis esox 237 bp 100%
Pundamilia nyererei 237 bp 93%
Neopamprologus brichardi 122 bp 29%
0 100 200 300
B.
Oreochromis niloticus AGACTCGGACACACATATTTCAGGCTTGGCCTTATAGAATTCAGCATTTCCATGCTGGGGAAATAGGTTTTGCCACTGTTTGAGC 85
Metriaclima zebra AGACTCAGATACACATGTTTATGGCTTTATCTTATAAAACTCAAGATCTCCGTGCTGGGAAAACAGGTTTTGCAGCCGTTTGAGC 85
Melanochromis auratus AGACTCAGACAGACATGTTCATGGCTTTATCTTATAGAACTCAATATCCCGTGCTGGCCGAACAGGTTTTGCAGCCGTTTGAGC 85
Labeotropheus fuelleborni AGACTCAGAAACACATGTTTATGGCTTTATCTTATAGAACTCAATATCCC-GTGCTGGGGAAACAGGTTTTGCAGCCGTTTGAGC 84
Rhamphochromis esox AGACTCAGACAAACATGTTTATGGCTTTATCTTATAGAACTCAATGTCCCCGTGCTGGGAAAACAGGATTTGCAGCTGTTTGAGC 85
Pundamilia nyererei AGACTCAGACAAACATGTTTATGGCTTTATCTTATAGAACTCAATATCCCCGTGCTGGGGAAACAGGTTTTGCAGCCGTTTGAGC 85
Oreochromis niloticus TAAGATTTTCA-GTTATTGACTAAATGAAAACC-ATAATGTGTTTCAGGCGAG--------- -------------------- AAA 139
Metriaclima zebra TAAGATTTTAAAATATTCACATAATGAAAACCTATACTTTGTCTCGGGGGAGTTCCCCATTCAAATGCATGTAACAGTGAGAAA 169
Melanochromis auratus TAAGATTTTAAAATATTCACATAATGAAAACCTATACTTTGTCTCGGGGGAGTTCCCATTCAAATGCATGTAACAGTGAGAAA 170
Labeotropheus fuelleborni TAAGATTTTAAAATATTCACATAATGAAAACCTATACTTTGTTTCAGGCGAGTTCCCATTCAAATGCATGTAACAGTGAGAAA 169
Rhamphochromis esox TAAGATTTTCAAGTTATACACATAATGAAAACCTATACTTTGTTTCGGTCGAGTTCCCCATTCAAATGCATGTAACAGTGAGAAA 170
Pundamilia nyererei TAAGATTTTAAAATATTCACATAATGAAAACCTATACTTTGTTTCAGGCGAGTTCCCATTCAAATGCATGT-- CAGTGAGAAA 168
Oreochromis niloticus CGCACTGTCTCGCGGAAATAAGGCGATTTTCACCAAGTCCATAAAGACAGCTGTAACTTTTGATAGG 206
Metriaclima zebra CGCATTGTCTTGGCGAAAGAAAGCGTTTTTGTACAACTTCATATAAATAGCTGTAACTTTTGATAGA 237
Melanochromis auratus CGCACTGTCTTGGCGAAATAAAGCATTTTTGTACAACTTCATATAAATCGCTGTAACTTTTGATAGT 237
Labeotropheus fuelleborni CGCATTGTCTTGGCGAAATAAAGCGTTTTTGAACAACTTCATATAAATCGCTGTAACTTTTGATAGA 236
Rhamphochromis esox CGCACTGTCTTGGCGAAAGAAAGCGTTTTTGTACAACTTCATAAAAATCGCTGTAACTTTTGATAGA 237
Pundamilia nyererei CGCACTGTCTTGGCGAAATAAAGCGTTTTTGTACAACTTCATATAAATCGCTGTAACTTTTGATAGA 235
C.
Hordeum vulgare 7 bp 23%
Aegilops tauschii 340 bp 33%
Brachypodium distachyon 154 bp 53%
Oryza brachyantha 154 bp 48%
Oryza sativa 154 bp 50%
Setaria viridis 155 bp 56%
Setaria italica 155 bp 55%
Panicum hallii 155 bp 67%
Panicum capillare 155 bp 67%
Panicum virgatum 175 bp 69%
Zea mays 156 bp 100%
Zea luxurians 156 bp 98%
Sorghum bicolor 137 bp 36%
Sorghum propinquum 137 bp 37%
Miscanthus x giganteus 137 bp 36%
0 100 200 300
D.
Oryza sativa TTTTTGCCACGAACGCACCCAATACACTCCAATATGTCCAAAAATCATGTTTTGGCCTTTTT-------------------- GAACTTTTT 71
Setaria viridis TTTTCGCAACGAACGCATGCAATCTTCCTAATTATACCCTAGA-ACATGTTTGGGAGTGTTTTT------------------- GGGC-ATTC 70
Setaria italica TTTTCGCAACGAACGCATGCAATCTTCCTAATTATACCCTAGA-ACATGTTTGGGAGTGTTTTT------------------- GAGC-ATTC 70
Panicum hallii TTTTCGCAACGAACGCACCCGATCCACTCCATTGGACCCAAAACTCATGTTTGGGGGG-TTTC------------------- GGACCGTTT 71
Panicum capillare TTTTCGCAAGGAACGCACCCGATCCACTCCATTGGACCCAAAACTCATGTTTGGGGCG-TTTC------------------- GGACCGTTT 71
Panicum virgatum TTTTCGCAACGAACGCACCCAATCCACCCCATTGGACCCTAAACTCATGTTTGGTACG-TTTCATAGTGTTTGGGTGCATTTGGGATCATTT 91
Zea mays TTTTCGCAACGAACATGCCCAATCCACTACTTTAGGTCCAAAACTCATGTTTGGGGTGATTTC------------------- GCGCAATTT 71
Zea luxurians TTTTCGCAACGAACATGCCCAATCCACTACTTTAGGTCCAAAACTCATGTTTGGGGTGGTTTC------------------- GCGCAATTT 71
Oryza sativa CATTCCGGTAAAAAA-CATCGCACCCACGT-GTGCCAATATTGGC--ATTAATTG-ACAAAAGTTC-GCCGCGCGAAT-CACGAAGTGAG-- 156
Setaria viridis GGTTCCGGCACGAAAAATGATGCAA-AACG-GGTGC--ATTTCGGCACCCGAAT-GCAC--TATTGCCGGGTAGCGAAACT-CGGGTGGATT 155
Setaria italica GGTTCCGGCACGAAAAACGATGCAA-AACA-GGTGC--ATTTCGGCACCCGAAT-GCAC--TATTTTCGGGTAGCGAAACT-CGGGTGGATT 155
Panicum hallii CGTTACTGCACGAAAGTCGATGCAA-AACTGGCCGA--ACTGGTGCCA-TTAAC-GCAC--AAGTTC-GCTAAACAAAGTCGCGT-CGGAA-T 155
Panicum capillare CGTTACTGCACGAAAGTCGGTGCAA-AACGGGCCA--ACTGGTGCCA-TTAAC-GCAC--AAGTTC-GCTAAACGAAGTTGCGT-CGGAA-- 155
Panicum virgatum CGTAACTGCATGAAACTCGGTGCAA-AACGGGGTGA--ACTGGTGCAA-TTAAT-GCAA--AAGTTCGTGC-CACGAAGTCACGT-CGGGA-T 175
Zea mays CGTTGCCGCACGTCACCCATTCCGAAAACG-GGTAT--CGGGGGCAT-ACAAA-GCAC--GAGTTTTTGCCACCGGAACAATTT-CTTCG-T 156
Zea luxurians CGTTGCCGCACGTCACCCATTCCGAAAACG-GGTGT--CGGGGGCAT-ACAAA-GCAC--GAGTTTTTGCCACCGGAACCATTT-CTTCG-T 156

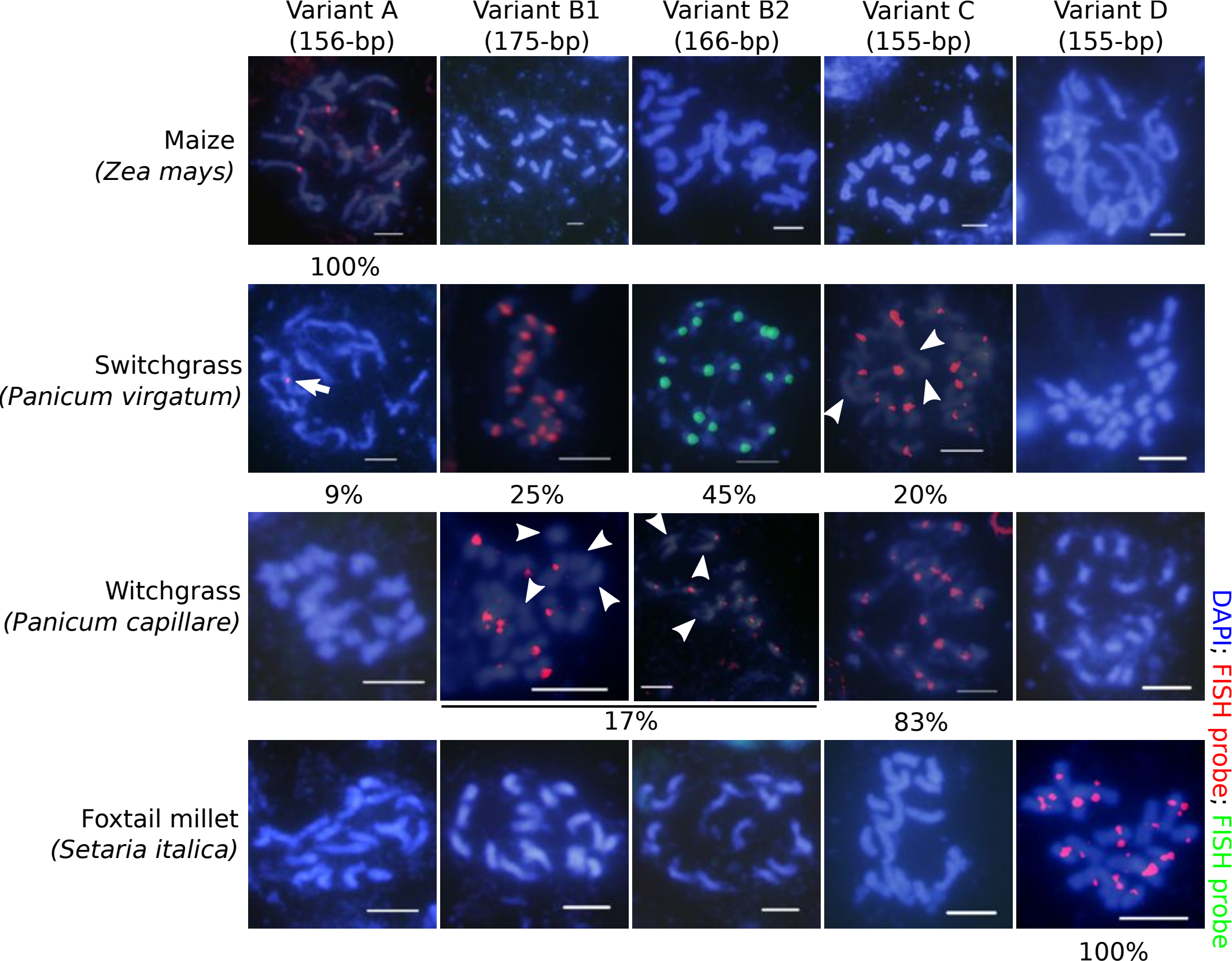

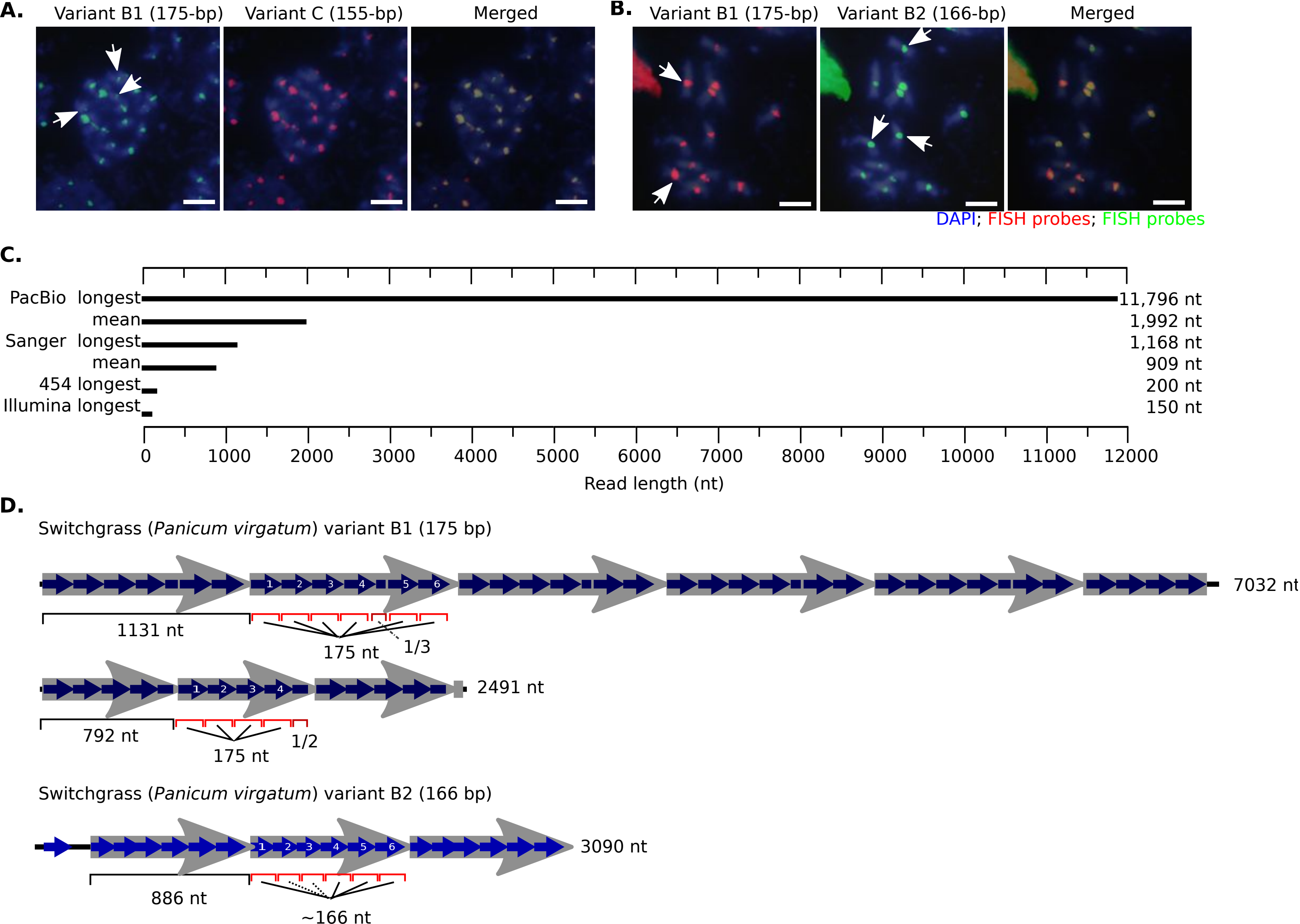

**Supplemental Methods**

# Patterns of Centromere Tandem Repeat Evolution in 282 Animal and Plant Genomes

Daniël P. Melters*[1,2], Keith R. Bradnam*[1], Hugh A. Young[3], Natalie Telis[1,2], Michael R. May[4], J. Graham Ruby[5], Robert Sebra[6], Paul Peluso[6], John Eid[6], David Rank[6], José Fernando Garcia[7], Joseph L. DeRisi[5,8], Timothy Smith[10], Christian Tobias[3], Jeffrey Ross-Ibarra[9], Ian F. Korf#[1] and Simon W.-L. Chan#[2,8]

**Notes**:
1. All (candidate) centromere repeat monomer sequences can be found at the end of this document.
2. All Perl scripts can be downloaded from the Korf Lab website: http://korflab.ucdavis.edu/Datasets/Centromere_data
3. Accession numbers for sequences used in this paper can be found in Supplementary Table S1.

## Bioinformatics pipeline for tandem repeat clustering

In this Methods section we will use results for gorilla (*Gorilla gorilla*) as an example of how each step works.

### Trace Archive Data files

For Sanger reads, we downloaded up to five randomly chosen Trace Archive sequence files along with their supporting ancillary files for each species from the NCBI Trace Archive (http://www.ncbi.nlm.nih.gov/Traces). E.g.

fasta.gorilla_gorilla.002
fasta.gorilla_gorilla.004
fasta.gorilla_gorilla.007
fasta.gorilla_gorilla.011
fasta.gorilla_gorilla.017
anc.gorilla_gorilla.002
anc.gorilla_gorilla.004
anc.gorilla_gorilla.007
anc.gorilla_gorilla.011
anc.gorilla_gorilla.017

If for any species, fewer than five sequence files were available, we used all available sequences. Each file typically contains 500,000 sequences.

***Gorilla gorilla* input: 2.5 million sequences**

We then filtered these sequences as follows (using information in the ancillary files for steps 1–3):

1) Ignored any sequence which wasn't flagged as either Whole Genome Shotgun (WGS) or Whole Chromosome Shotgun (WCS)
2) Clipped sequences for quality and/or vector contamination
3) Ignored sequences with clipping information that referred to coordinates longer than the actual sequence
4) After clipping, removed any sequence which contains >5% Ns
5) After clipping, removed any sequence which was <100 nt
6) Ran DUST filter on resulting sequences
7) Removed sequences that contained >5% Ns after DUSTing

**Resulting number of sequences in *Gorilla gorilla* after filtering: ~2.3 million**

**Tandem Repeats Finder (TRF) parameters**

A command-line version of TRF (version 4.04) was run using the following parameters:

Match = 1
Mismatch = 1
Indel = 2
Probability of match = 80
Probability of indel = 5
Min score = 200
Max period = 750

From the output of TRF, we we only kept tandem repeats that contained a minimum of 2 repeats (TRF will identity tandems from less than 2 complete repeats), and which had a minimum length of 50 bp. The processed TRF output consists of a FASTA file containing the *consensus* repeat sequences that TRF produces.

**Number of tandem repeats in *Gorilla gorilla* after running TRF: 69,041**

**Clustering of tandem repeats**

Clustering similar tandem repeats was a CPU intensive step and not practical for those species in which we had identified tens of thousands of different tandem repeats. At this point we chose to select a random sample of sequence reads for each species, and then find all of the corresponding tandem repeats that we had already identified in those reads. For this step we chose up to 200,000 reads from each species (fewer if tandem repeats were more common in that species).

**Number of reads randomly selected in *Gorilla gorilla*: 100,000**
**Number of tandem repeats present in those reads: 3,489**

We chose to sample reads first, in order to record their lengths for a calculation that occurred in a later step. At this stage we also tracked how many repeat units were present in each tandem repeat that occurred in the selected reads, along with the total length of the tandem repeat array.

We then took the monomer units from the sampled tandem repeats and made tandem versions of them. Next, we used WU-BLASTN to search each monomer repeat sequence against a database of all the tandem-ized version of those repeats (WU-BLAST settings: M=1, N=-1, R=3, Q=3, W=10). We only retained matches that had at least 75% identity.

**Number of BLAST HSPs in *Gorilla gorilla*: 12,509,094**

At this point we processed the BLAST output in two ways. In 'global' mode, we only retained BLAST matches between repeats if the match spanned the majority of the monomer repeat length. In 'local' mode, we did not require this retriction and so partial matches of a repeat monomer to a tandem repeat could occur.

For both global and local mode, we then found the tandem repeat that had the highest 'tandem repeat mass'. This was calculated as follows:

1) Count how many matches each repeat has to all other repeats (excluding self-matches)
2) For each match, multiply the length of the matching repeat by its copy number in the read where it was originally detected
3) Sum these multipled lengths to produce 'tandem repeat mass'

All repeats that matched the individual repeat with the highest mass were then removed from the analysis (along with the high mass repeat) and the process was repeated. Each iteration of this process defines a cluster of related repeats. In local mode, clusters grow bigger in size and 'swallow up' more of the repeats that match only partially.

For each cluster we can record:

1) Number of sequences
2) GC% of sequences
3) Length of representative cluster member
4) Mean pairwise identity between sequences in cluster
5) Cluster 'depth' – total number of matches to top hit in cluster, including matches to the matches
6) Cluster 'mass' – the summed tandem repeat mass of all sequences in cluster
7) Genomic fraction – the tandem repeat mass as a fraction of the total length of the sample sequences

We typically focus our analysis on just the top global and top local cluster, and these typically account for the majority of the total amount of tandem repeat mass. Exceptions occur when there is higher order repeat structure. E.g. the results below for *Gorilla gorilla* show that — in global mode — the top cluster finds the known centromere repeat (length 171 bp) and this accounts for ~13% of the sample reads (inferred to be representative of the genome as a whole). However, the second cluster finds a similar number of sequences which are double the length of the top cluster, and which are presumed to represent the same sequence as the centromere repeat, but in a tandemized version. In contrast, the top local cluster contains a mix of both 171 bp and 340 bp sequences and therefore accounts for a higher genomic fraction (25%).

### Example cluster output for *Gorilla gorilla*

```
Summary for Gorilla gorilla tandem repeat analysis
100000 reads, 3489 tandem repeats, 71072791 bp, 0.4232 GC

Repeats (global mode) - using mass threshold %0.1
```

| Cluster | n | id# | length | gc | mean | stdev | depth | mass | frac |
|---|---|---|---|---|---|---|---|---|---|
| 0 | 1230 | 50980 | 171 | 38 | 87.10 | 6.00 | 5399 | 920589 | 0.012953 |
| 1 | 1057 | 61930 | 340 | 39 | 90.06 | 5.17 | 2437 | 827458 | 0.011642 |
| 2 | 745 | 72354 | 70 | 54 | 99.16 | 1.25 | 7716 | 540129 | 0.007600 |
| 3 | 22 | 33954 | 171 | 39 | 87.29 | 5.44 | 100 | 17016 | 0.000239 |
| 4 | 20 | 47128 | 171 | 38 | 85.17 | 4.26 | 87 | 14891 | 0.000210 |
| 5 | 16 | 14196 | 342 | 34 | 79.77 | 4.19 | 37 | 12680 | 0.000178 |
| 6 | 14 | 6017 | 76 | 50 | 93.72 | 3.08 | 123 | 9322 | 0.000131 |
| 7 | 12 | 27715 | 349 | 53 | 85.31 | 3.27 | 25 | 9073 | 0.000128 |
| 8 | 16 | 33244 | 96 | 37 | 94.10 | 3.76 | 91 | 8813 | 0.000124 |
| 9 | 12 | 58454 | 64 | 37 | 96.93 | 2.44 | 130 | 8316 | 0.000117 |
| 10 | 11 | 36485 | 279 | 53 | 88.40 | 4.60 | 29 | 8226 | 0.000116 |
| 11 | 8 | 728 | 341 | 36 | 79.59 | 3.54 | 17 | 6125 | 0.000086 |
| 12 | 9 | 56545 | 209 | 52 | 93.46 | 2.28 | 28 | 5996 | 0.000084 |
| 13 | 9 | 35159 | 98 | 69 | 97.96 | 1.32 | 54 | 5299 | 0.000075 |
| 14 | 7 | 32585 | 168 | 58 | 89.05 | 5.39 | 25 | 4352 | 0.000061 |
| 15 | 5 | 70673 | 84 | 38 | 89.00 | 3.08 | 44 | 3779 | 0.000053 |
| 16 | 5 | 14582 | 170 | 37 | 88.28 | 7.16 | 21 | 3663 | 0.000052 |
| 17 | 8 | 62437 | 161 | 40 | 84.42 | 4.82 | 22 | 3569 | 0.000050 |
| 18 | 5 | 45011 | 298 | 51 | 80.19 | 3.19 | 11 | 3532 | 0.000050 |
| 19 | 6 | 28593 | 140 | 54 | 95.37 | 2.31 | 24 | 3443 | 0.000048 |
| 20 | 8 | 33970 | 125 | 35 | 85.75 | 3.69 | 21 | 2763 | 0.000039 |
| 21 | 5 | 49512 | 175 | 79 | 85.47 | 5.73 | 13 | 2421 | 0.000034 |

```
WARNING: Top clusters are similar in size. Cluster #1 has a tandem repeat mass within 50%
of cluster #0
NOTE: 193 clusters are not reported due to having a tandem repeat mass less than %0.1 of
cluster #0,
or the cluster contained less than 5 sequences


Repeats (local mode) - using mass threshold %0.1
```

| Cluster | n | id# | length | gc | depth | mass | frac |
|---|---|---|---|---|---|---|---|
| 0 | 2339 | 17934 | 171 | 36 | 8026 | 1787776 | 0.025154 |
| 1 | 787 | 15282 | 348 | 50 | 7849 | 569187 | 0.008009 |
| 2 | 54 | 33244 | 96 | 37 | 298 | 31018 | 0.000436 |
| 3 | 24 | 62336 | 151 | 46 | 157 | 16369 | 0.000230 |
| 4 | 22 | 17637 | 200 | 58 | 118 | 14271 | 0.000201 |
| 5 | 18 | 47128 | 171 | 38 | 68 | 13270 | 0.000187 |
| 6 | 23 | 32502 | 127 | 78 | 79 | 11472 | 0.000161 |
| 7 | 17 | 45011 | 298 | 51 | 48 | 11028 | 0.000155 |
| 8 | 14 | 19227 | 339 | 39 | 48 | 10418 | 0.000147 |
| 9 | 15 | 35159 | 98 | 69 | 76 | 8880 | 0.000125 |
| 10 | 7 | 36688 | 63 | 42 | 47 | 4469 | 0.000063 |
| 11 | 5 | 70673 | 84 | 38 | 44 | 3779 | 0.000053 |
| 12 | 5 | 31667 | 84 | 67 | 21 | 3451 | 0.000049 |
| 13 | 6 | 21535 | 161 | 69 | 37 | 3283 | 0.000046 |
| 14 | 5 | 72797 | 71 | 60 | 30 | 2335 | 0.000033 |

```
NOTE: 107 clusters are not reported due to having a tandem repeat mass less than %0.1 of
cluster #0,
or the cluster contained less than 5 sequences
```

**Graphing of results**

We produced XYZ graphs to further study the sequences within the top clusters (X-axis = repeat length, Y-axis = repeat GC content, Z-axis = genomic fraction). Multiple peaks with different length, but the same GC percentage would be separated by steps less than 50 bp. These peaks would represent various multimers of shorter repeat than the cutoff of 50 bp. In these instances, TRF and the subsequent steps were re-run but with a minimum repeat length of 20 bp. This was done for ~40 species.

**Sequence Read Archive files**

For Illumina and 454 reads, we downloaded either *.lite.sra or *.fastq.bz2 files for species from the DDBJ Sequence Read Archive (http://trace.ddbj.nig.ac.jp/DRASearch/), e.g.

| | |
|---|---|
| Species | : *Gorilla gorilla gorilla* |
| Submission | : ERA013840 |
| Study | : ERP000004 |
| Experiment | : ERX007929 |
| Run | : ERR019516 |
| File | : ERR019516.lite.sra (2.1 Gb) |
| Unpacked files | : ERR019516_1.fastq and ERR019516_2.fastq |

For each FASTQ file two were made: one with 1.000.000 randomly selected reads and one with 100.000 randomly selected reads. This means that per assembly run either 2.000.000 or 200.000 reads were used.
The read length for Illumina data ranged from 36 to 152 bp, whereas for 454 data the read length ranged around 200 bp.

Because Ilumina and 454 reads are often too short to contain at least two copies of a tandem repeat, an assembly step was necessary to find tandem repeats.

**PRICE parameters**

The command-line used for PRICE (version 0.6) was:

-fpp Gorilla_gorilla_100k_1.fastq Gorilla_gorilla_100k_2.fastq 475 90 (input files)
-nc 25 (number of cycles)
-mpi 85
-MPI 95
-tpi 85
-TPI 95
-logf Gorilla_gorilla_run1 (log file)
-o Gorilla_gorilla_run1.fasta (output file)
-picf 20000 Gorilla_gorilla_1M_1.fastq 500 2 25 (seed file)

For a detailed manual go the PRICE website (http://derisilab.ucsf.edu/software/price/index.html) or on the command-line type "PriceTI -h"

Per cycle a FASTA file was created containing the assembled contigs. These FASTA files were fed into the TRF pipeline as described earlier for Sanger reads.

**Tandem Repeats Finder (TRF) parameters**

A command-line version of TRF (version 4.04) was run using the following parameters:

Match = 1
Mismatch = 1
Indel = 2
Probability of match = 80
Probability of indel = 5
Min score = 200
Max period = 2000

The only difference between the TRF parameters for Sanger data vs Illumina or 454 data is that the maximum tandem repeat length is not 700 bp but 2000 bp, the upperlimit of TRF (Benson, 1999).

**Genomic fraction determination of PRICE derived tandem repeats**

Tandem repeats derived from PRICE contigs were doubled in length. 2.000.000 reads (both 1.000.000 files were used) were aligned to the doubled tandem repeat monomers with WU-BLASTn. Only the top hits (hspmax = 1) was used for determining the genomic fraction. The minimum BLAST score was dependent on the length of the reads. The number of reads that significantly aligned to the tandem repeat were counted and the fraction of these reads over all the reads determined the genomic fraction. The average of each file was used as the genomic fraction for that specific tandem repeat. The most abundant tandem repeat was predicted to be the candidate centromere tandem repeat.

**Pacific Biosciences data**

PacBio reads were processed the same way as Sanger reads, as described earlier, with the exception that the maximum tandem repeat length was set to 2000 (Max period = 2000). Please note, that only the insert sequence was used for analysis, not the complete PacBio read. Because two loops are ligated to a double stranded DNA fragment, the polymerase can read multiple times over the same DNA molecule. For determining tandem repeat structures, only a single pass over the dsDNA molecule will suffice.
For the grasses (*Panicum virgatum, Panicum capillare, Zea mays, and Zea luxurians*) the sequences longer than 1,000 bp were used. For the bovine species (*Bos taurus taurus,*

*Bos taurus indicus, Bos grunniens, Bison bison, Bubalus bubalis*) sequences longer than 3,000 bp were used. This was to guarantee that the genomic fraction would not be diluted with sequences which would be too short to contain at least two copies of the described centromere repeat sequences (see Supplementary Table S2).

**Supplementary Table S5 - Number of PacBio reads per species.**

| Species | >1 kbp | >2 kbp | >5 kbp | >10 kbp |
|---|---|---|---|---|
| *Panicum virgatum* | 27702 | 15699 | 2124 | 3 |
| *Panicum capillare* | 14624 | 6596 | 216 | 0 |
| *Zea mays* | 137989 | 43145 | 6829 | 175 |
| *Zea luxurians* | 87448 | 25063 | 2031 | 51 |
| *Bos taurus taurus* (Hereford) | nd | 59294 | 16319 | 593 |
| *Bos taurus indicus* (Nellore) | 120588 | 54463 | 7408 | 35 |
| *Bos grunniens* | nd | 74043 | 9014 | 263 |
| *Bison bison* | nd | 107023 | 29945 | 1093 |
| *Bubalus bubalis* | nd | 158768 | 16941 | 501 |

**Plant material and growth conditions**

Seeds from the inbred line B73 of *Zea mays* were provided by the Hake lab (PGEC, Albany, CA). *Panicum capillare* seeds were obtained from Tom Juenger (UT Austin, TX). Foxtail millet (*Setaria italica*) seeds were obtained from the National Genetic Resources Program (NGRP, Beltsville, MD), whereas dihaploid switchgrass plants (*Panicum virgatum*) were already available (Young et al., 2010; Young et al., 2011). Seeds from all species were germinated in appropriate soil and plants were maintained in the greenhouse at 21-28°C, under supplemental light (16:8 photoperiod), watered as needed, and fertilized weekly with general purpose 20-20-20 fertilizer.

**Switchgrass (*Panicum virgatum*) DNA isolation**

Switchgrass (*Panicum virgatum* AP13 (tetraploid)) genomic DNA was isolated using a modified protocol from Chen and Ronald (Chen and Ronald, 1999). In summary, fresh leaf tissue was ground thoroughly in liquid nitrogen and the powdered tissue was added to 4.0 mL of extraction buffer (1.42 M NaCl, 100 nM Tris-Cl (pH 8.0), 2% (w/v) polyvinylpyrrolidone (PVP-40), 20 nM EDTA and 2% (w/v) CTAB), 5 mM ascorbic acid, 4 nM diethyldithiocarbamicacid (DIECA), and 0.2 mg/mL RNase (DNase free). Next, the solution was incubated at 65°C for 15 minutes, after which 3.0 mL of chloroform/isoamyl alcohol (24:1 ratio) was added. This was centrifuged at 3,500 rpm for 10 minutes at room temperature. The aqueous phase was transferred to a new 15 mL tube and 0.7 volumes of isopropanol was added, followed by 30 minutes of centrifugation at 3,500 rpm at room temperature. The supernatant was removed and the pellet was washed with 5 mL of cold 70% ethanol for 30 minutes on ice. All ethanol was removed and the pellet was airdried overnight. The next day the pellet was resuspended in 100 μL TE buffer and stored at 4°C overnight. The sample was centrifuged for 10 minutes at 3,500 rpm at room temperature and the supernatant containing DNA was transferred to an eppendorf tube.

**Fluorescence in situ hybridization (FISH) experiments**

Mitotic chromosome spreads were generated following a protocol by Zhang and Friebe (Zhang and Friebe, 2009; Zhang et al., 2010) with a few modifications. Actively growing root tips were excised from greenhouse grown plants, pretreated in ice cold water for 18-24 h, and then fixed in 3:1 ratio of 95% ethanol and glacial acetic acid at 4°C, overnight. Root material was either used immediately for slide preparation or stored in fixative at -20°C for up to several months. For slide preparation, the root tips (0.5-1.0 cm) were washed twice for 5 min each in 0.01M citrate buffer and digested in an enzyme mixture of 50 mg/ml Onozuka R-10 cellulase and 30 mg/ml Macerozyme (Phytotechnology Labs, Shawnee Mission, KS) at 37°C. Digestion times varied from 30 min to 2.5 h, depending on the thickness and degree of lignification in the root tip. Softened root tips were then washed for 5 min in 0.01M citrate buffer and transferred to a slide. Forceps and a scalpel were used to carefully excise the white tissue just behind the root cap containing actively dividing mitotic cells. All other root tissues were removed and the remaining cells were macerated in a few drops of 1% acetocarmine stain. A coverslip was placed over the stained tissue and even pressure was applied to generate mitotic chromosome spreads. Slides were viewed under phase-contrast microscopy to identify spreads optimal for use in FISH analysis.

Plasmid vectors containing a single copy of each repeat sequence were synthesized by Bio Basic Inc. (Ontario, Canada) and used as probes for FISH analyses. Plasmid vectors were labeled with either digoxigenin-11-dUTP or biotin-16-dUTP using a nick translation protocol, in accordance with manufacturer instructions (Nick Translation Kit; Roche Applied Sciences, Indianapolis, IN). Hybridization and post-hybridization wash procedures were performed as previously described  (Jenkins and Hasterok, 2007). Probe hybridization signals were detected using anti-digoxigenin (dig) conjugated FITC (green), anti-dig conjugated Rhodamine (red), or Streptavidin conjugated Rhodamine (red) antibodies (Roche Applied Sciences). Dual probe labeling of chromosomes was conducted using non-competing dig and biotin antibodies to allow for simultaneous red and green imaging. Chromosomes were counter-stained with 4',6-diamidino-2-phenylindole (DAPI).

Digital images were recorded using an Olympus BX51 epifluorescence microscope (Olympus Corporation, Center Valley, PA) with a DP70 CCD (charge coupled device) camera and suitable monochrome filter sets (Chroma Technology, Rockingham, VT). Images were processed using GIMP 2.6 (GNU Image Manipulation Program) for Linux. Analysis of FISH data was conducted by overlaying a probe signal image on top of the corresponding DAPI stain image. Adjustments were made to the transparency of the top (FISH signal) layer to demonstrate signal and chromosome alignment.

**Repeat variant sequences:**

**Variant A (*Zea mays*):**
**5'-**
CTTTAGGTCCAAAACTCATGTTTGGGGTGATTTCGCGCAATTTCGTTGCCGC
ACGTCACCCATTCCGAAAACGGGTATCGGGGTGCATACAAAGCACGAGTTTT
TGCCACCGGAACAATTTCTTCGTTTTTCGCAACGAACATGCCCAATCCACTA-
3'
**Variant B1 (*Panicum virgatum*):**
**5'-**
TTCGTTGCGAAAAATTCCGATGCGACTTCATGGCACGAACTTTTGCATTAAT
TGCACCAGTTCAGCCCATTTTGCACCGAGTTTCCTGAAGTAACGAAACGATGC
CAAATGCACCCAAACACTACGAAACGCACCAAAACATGAGTTTAGGGTCCAA
TGGGGTGGATCGGGTGCG-3'
**Variant B2 (*Panicum virgatum*):**
**5'-**TTCGTTGCGAAAAATTCCGACGCGACTTCGTGGCACGAACTTTTGCACTAG
TTCGGCCCGTTTTGCACCGAGTTTCGTAAGGCAACCAAACGGTCCCGAATGC
ACCCAAATAGTACAAAACGCACCAAAACGTCAGTTTAGGGTCCAATGGGGTG
GATCGGGTGCG-3'
**Variant C (*Panicum capillare*):**
**5'-**
TTCGTTGCGAAAAATTCCGACGCAACTTCGTTTAGCGAACTTGTGCGTTAAT
GGCACCAGTTCGGCCCGTTTTGCACCGACTTTCGTGCAGTAACGAAACGGTC
CGAAACGCCCAAAAACATGAGTTTTGGGTCCAATGGAGTGGATCGGGTGCG-
3'
**Variant D (*Setara italica*):**
**5'-**TTCGTTGCGAAAAAATCCACCCGAGTTTCGCTACCCGGAAATAGTGCATTC

GGGTGCCGAAATGCACCCGTTTTGCATCATTTTTCGTGCCGGAACCGAATGCC
CAAAAACACTCCCAAACATGTTCTAGTGTATATTTAGGAAGATTGCATGCG-3’

**All (candidate) centromere repeat monomers for the 282 species in this study:**

```
>Cynocephalus_volans
CATTCTGCTGCTTCCTCCCTTAGAGTTAAACTAGCTTCTGAAAGCTCAACCTCCGGGGAG
TCTGTTTGGCAGAGATATAGAACACACTATTCCAAGACTGTTGGAAGGCCTTTGCAGTGT
CATTTCGCATCTAATTTCCCAAACCAAACTGCTTGCAGCTCTAACTCCCTTCTAACTACC
ACTTCTTCGGTTCAGTGAGATATGCTGGATTCTAAACGGAGAAGCTTCTCTGACGCTTTT
TCACAGAGATTTTGAAACTGAAATCTCCCATGCCGTTTCCACCATAGGCAGCAATGGGCT
TCCCCATAGAACTTGCCAACCTGAATTCCTACAGGCTACATGTGAGAGGCA
>acropora_millepora
TACTTTTTGCAACATTTTTCTAAAAATTGGGTCAAAACCCTAGTGCACGGTACTTTGCAC
AAAAAGTTGCTTATCTCGAGGAGATCGACAAGTTTTGGTGGTTTTTCAGCAAATGCCC
>acropora_palmata
TGGTGAAAAACCACCAAAACGTGTCGATCTCCTCGAAATAAGCAACTTTTTGTGCAAAGT
ACCGTGCACTAGGGTTTTGACCCAATTTTGAGCAAAATGTTGCAAAAAGTATGGCATT
>acyrthosiphon_pisum
AAACGTACACCAAAACTTAGAGTAGATTAAACCCTATCTTTTGGTAAGAAAATTATGAAA
AAATATTAAGTATTGGAGAAACGGGACGGGATTCCACCGCCACAGGTGCTGAAATATAGT
CCAAATTGTAGACGGGACGAGCTGTAGTTAATTCAAAAATGAACTAAGACTTATG
>aedes_aegypti
TTTTGGCAGAGTGAATTGACAAATCCAGCACACGGAAAGTTAGCTGGGATCCAGACGAAG
CGATTGATGCTAAGAAATCGAAAATCCGTTGATATTTACTCAACTTTTTAGGAAATCTAC
GGATTTTTATAATTTCATTGGATTTTCAAGCATGGGTCCCACTAAAAGTTGGAGGCTAAG
G
>ancylostoma_caninum
TATCGATTTATCGATTTTATTGATTTTTCGCTATTTTTGCTCAGGGGTCGGTGGGGACAT
GTCTACTGGGGTTTCTCAATACGCTGAACACGAATATGGCAACCGCTTCGCACGAAAA
>anolis_carolinensis
GACCCGCCGGGCTTGGAAAATGATGAAATTTTCCAGAAAAAGTGACCTCTGCCAAATCTG
AAAAACTTGCCAAATTTTCATTTTTCT
>anopheles_gambiae
CTATTTGAATGGCCTTGAAATTTGATGTTGAATTTTGTTGCCATCTATTATCGTTGTTAC
GGCCATGTTTCATCCAAATGAAGCCATCTTGCA
>anopheles_gambiae_M
AAATTCAACATCAAATTTCAAGGCCATTCAAATAGTGCAAGATGGCTTCATTTGGATGAA
ACATGGCCGTAACAACGATAATAGATGGCAACA
>anopheles_gambiae_S
ATGGCCGTAACAACGATAATAGATGGCAACAAAATTCAACATCAAATTTCAAGGCCATTC
AAATAGTGCAAGATGGCTTCATTTGGATGAAAC
>apis_mellifera
CGCTATGGTCGTCCAAAGAAGTGGAAATTTTAGAATCTTTAAAATAGTGCATTATTAAAA
CGGCTGTATCTCCGAAGATAAGAGAAAAAAATACGATTTCCGCAGCTCGTTGCAAGCGGG
AAAGATTGCTCTTTGACCTCATGTTTGGTTGGGTTCTGGACGACAAGCGGTTCATA
>aplysia_californica
ACAAGCATTCACGCGAGACAGAATGCAAACTAGCATTCAAGCAAGAAAGAATGCAAACGA
GCATTCAAACAAGACAGAATGAAAGCATGCATTCAAGCAAGACAGAATCCAAGCAAGTAC
TGAAGACGGAGAGAATGCAAATATGCATTTAAAATAATCCGAATGCAA
>aquilegia_coerulea
CTCGGTTTTCGGTCCGAACTTGGAAATGGAGAAACTCACTTGTTTTGGACTCTAAAACCT
CAAGATCGACAAGACTTAGCCGGGACGTTTTAAGGCCAATACAGAACATTTGCACCATTG
GAACGGGATTTTGTCCGGACATTCGGCTTGGTATGTACTCGATTTTTGTTCCGAACTTAT
```

GAGTGGACGAAACACACATGTTTTCGCATGGAAAACGATACACCTAAAAGACTAGCCACT
CCATTACTTAACACGGGGCGTTCTAAGGGGAAAAACAACGTTCCGTCCGTAAAAACGTGA
TTTATGAACGTCCGTTCGGTCGGCATTTA
>arabidopsis_lyrata
CGGGATCCGGTTGCGGCTCTAGTTCTTATACCCAATCATAAACACGAGATCTAGTCATAT
TTGACTCCAAAAACACTAACCAAGCTTCTTATTGCTTCTCAAATCTTTGTGGGTGTGGCC
GAAGTCCTATGAGTTTTCGGTTTTGGAGCTTCTAAACGGAAAAACACTACTTTAGCTTT
>ascaris_lumbricoides
ATTGAATGGTCACCAGATCATTGGTGAGTTACGACTCATGATGATTATTCTATATCACTA
AGTTAAT
>ateles_geoffroyi
ACCCTTTCTGTGCAGAGATGCAAAACTGTCATTTCCAGCCAAATCCAGGCACGTCAGAAA
AAGCAGTAATAGTGCGTCTAAAACACAGAAGAAACGCATCATTCAAACTGCTCTGCTGTG
TGTTCGTTTCAACTAAGGGAGCTGAAGCGGAGTTTCATTCAGCGAGTTAGCAACACGTTT
CTTCTAGGAACTGCGTTTGGATTTTCCAGAGCGAAAGGGAGCATTTGCTGTACAGAGGAA
TATCTCGCTCTAAAACCAAAACGGAGCAATCTAACTGAATGCTTGTCAATGTGTGCATTC
AACTTACAGAGTTAAACTCGTGTGTGTTTGCAGCAGTTTAGAA
>biomphalaria_glabrata
CAGCTACTTCCTTGACAATATCCCAGAAGTTATCTTGTTGCTCGGCACTTCTGCCAACGA
GAAGGGGAAGGGCTACTC
>brachypdium_distachyon
GGATTCGCTAGGCATGGTCCCTCACTAGGACGGACAATTCGGGCAAATCGGAGTCGGTGC
TAAACTTTGATCCGGAACGTGTCCATTTGAGGTCCTAGGATGAGCACACCAAACAAAGCT
ACTCATCGAAGCGAAACAAGACTAGACCACTTGTAC
>branchiostoma_floridae
ACCAAGCTTGCCAGGCCGGAACTTGAAGCCTAAGCTACAATCGCGTTATTAAGTCACCCT
CCTACCCGGTGACGTGACAGGCGGTGTTTCAGAAAAAGCAGCTAGCGGTTGCGCCTAAAC
AGGGTCGGTATACTCAAGAATATCTCGAGAAGGAAGTATCCCAACATTTTGCGGTTTTCA
TCTTTCCATTCCTTGCTAAGGAACCTTTCCAACCAT
>brassica_oleracea
TCTCCACTACTTTATGTATCCAAATACAGCTTCTTACATCGCGATTCATCCTGGTTTGAT
CAGAATGACGAGGAAGTTGTCATATTCCCAAACAGGAAAACTGGGATCACCTGATTTGAA
AGTGGGATAACTTCTTCATCCTAACTCCTATGAGATTTATTCAACTTCCTGGTGAT
>brugia_malayi
TTCATTAACTCATACCATTTCTCTACAGATATAACAATATCACTAGAAGACATTTTGATT
AA
>caenorhabditis_briggsae
AATTTATCACATACGAGTTTTGTGAAAAGGCAGAACAAGCAAAGTTTAAAACAATGATAT
TATTATGGCTAATATTTTACAAAAGAATTTGTGTGAATTGGCTCATTCAGAAAAAAGTTG
GAAGCCAAAATCTCGATTTTAGGGCTCAAAAACCGATTTTTCG
>caenorhabditis_japonica
CGTTCAATTTTCAAAACACTAAGGCTCGGCCAATTTTTGATGAAAATTGCTCGTTTTTGG
ATCAAAATGAAGCTTTCAACCTGATTAATCTGATTATGGTTCGTATTGCGCATTCCAATC
GTGTTGTATCATCAAAACACAAGTCTGTATTCACCGAACTGTCTGAGTCTCTTCAAATAG
AACTGTGTTGG
>caenorhabditis_remanei
CCTTGAAAGCACAGTCTTGGAAGTTCTCATATAGACTTTTGTGAAACTTCAAGCGTGTGC
GGCCTTGTTCAGTTTGAACGTTATTGGCTGAAAATTGGCACAGTAGTAGACAATCCATTG
CCAATTCGAATGCACTTTACTGTTTTCAGATATCTCATCCAGAAAAAAAGTTAGAGACAA
TACA
>callithrix_jacchus
AGAGTGGAGCACGTTCCTTCTAGGAGCTGCGTTTTGCTATTCCAGCGTGAATGGGAGTAA
ACGCTGTATTGAGAAATATCTGCTTCTAGAACCAAAACGCAGCTACCTAACAGAACGGTC
TTCAATGTGTGCATTCAACTTACAGAGTTAAACTGATATGCGTTTGCAGGAGTTTGTAAA
CCCTTTCTGTGAAGCAGCGGAAAACGCATTTTTCCAGCCAAATCCAAGCATTTCAGAAAT
AGCTGTAAATCTCCGTCTAAAACACAAAACGAACGTATCCTTCAAACTGCTCTCCTGTGC
GTTCGCTTAACTAAGGGAGTTGAATCCGCGTTTAGATTCAG

>callorhinchus_milii
CCGGTGGCTCAATACCGTCACTGCAGCCGCTATAATGAAGCGCAGCAACAGCGCCGCCCG
GTGGCTCAATACCGTCACTGCAGCCGCTATAGTGAAGCGCAGCAACAGCGCCGCCCGGTG
GCTCAATACCGTCACTGCAGCCGCTATAACGTAGCGCTGAGCAACAGCGCCGCCCGGTGG
CTCAATACCGTCAGTGCAGCCGCTATGTTGAAGCGCAGCAACAGCGCCGCGCCGTGGCTC
AATACAGTCACGGCAGACGCGATAGTGAAGCGCAGCAACAGCGCCGT
>canis_latrans
TGGATACTTGGTCCACATCTGGGCCTCACTGGAACTACGTGACTCTGGTCCATCTCTGGG
CCTCACCCTGACTGGTGACTCTGGTCCCCACCTGGGCATCACTTGGACCTGGTGACTCTG
GTCCTACCTGGGCCTCACCTGGACCTGGTGACTCTGGTCACCAACTGGGCCTCACCTGGA
CCTGGTGAATCTGCTCCCACCTGGGCCTCACCTGGAACTTGTGACTGGTTCCCACCCGGA
CCTCACTGGACCTGGTGAATCTGGTCCCACATGGGCTCATAGGACC
>capitella_sp
TATTGAATACCTAGAGAATCGATTCTGTGAGCTTATTTCATTCGAACAGAACCATCTGAA
GCAATTCGTCTTGACTTCAATGCATTACAGAATCAATTCTGTTGAACTATAAGATCCACT
ACAGAATCAATTCTGTGAGCT
>capsella_rubella
ACAATGAATTTGATTGATTACAAGTGCTAGAGATGCATGAAGAATGTTTAAAAGTGAGAA
GAAAGACTTGTTGATATTTGGTCCCAAATGGGATAAGAACCCAAAACCATTGCTTTGAGG
CAGTGAATGGCTTGTATAAGTTATTTTGGGTTAGAATATGTTATAAT
>cavia_porcellus
TTTCACCACACATGCACATTGTAAAATACATGAAAGAATTCACACTGGAGAGAAACCCTA
TGTATGTAAGCAATGTGGGAAAGC
>chlamydomonas_reinhardtii
AATGCAGACTCGAGCAGGGAGCCATGTTGCCAGCCCTCACAGTGCCTTCAGTGCCCCTGC
ACGCCTGGACAAGGCGGGTGGGGTCCCTGCCGCCCAGCCATCACCAAACACCCCACCTGG
CACAACCACCCTTGTGCACTGTTGTTTCACATTTTCATATGTGCATGTTGCCTGACCTAT
TTGC
>choloepus_hoffmanni
CTGGGCTGAAAAACAACTTTTTTGAAAGCAGTCACTTGGAACTCTGTGAAAGCAGTATGT
TACACAGTTTTGAAAACTAAGTTGAATACAGCTTTCATTCACTTCTATGTGTTTTCC
>ciona_intestinalis
GTGTAAAACGACCTAGGATGGTCGGGTCGGCAATCGGTCGACCTCCGACCGAGCGACTTC
GATCACGACCTCGGGGATCTGGCTCCGCGGCAAGGGGCACCTCGGCACCCCGCGGAGAGC
TCGGGAAGCCCGGCGCAATTGTACAACGGCCGGAGGCGGACAGCTTCCACCACAACGAAG
CATACGTCGGGCCGACGTTTTCGCTCA
>ciona_savignyi
CTTTAAAACGCTAAAGGAACGTGGTACTAGGTCGGTTCTACGACCTTAGCGTGGCACGAC
TCGCCCAACTCGTCGCCGGCACGAGAAAACGAAACTCGCGCGCGCCGGGAGAAAATTCGA
ACGCGAAAATTAGGGGGGTAGATCCGCGCGCGCGCCGAAACGGCCCCTTTGCCGACGTCG
CATCG
>citrus_clementina
TCCGTGTGCCAAAAATAGGCCGCGGGCAAAGCCGCGCCCAAAAATAACCGCCCGAAGGCC
GGGGCGCCAAGTTCGTCCAGCGGAAAAAGGCCACAAAAACGGGTGGGCTATAGCCTTGGG
GGGTGGGGCTGGCCAACTTCGTCCGCCGGACTCGGAATGCCGCGAGACTTTGCGAGGGGC
C
>citrus_sinensis
GGGCGCGGCTGTGCCCGTGGCCTATTTTTGGCACACGGAGGCCGCCCGCAAAGTCTGGGG
CCATTCCGAGTCCGGCGGACGAACTTCGCCCACGCCCCCCACCAAGGCTATAGCCCACCC
GATTTTTGGCCATTTTTCCGCTGGCGAGTCTTGGCGCCCCGACCTTCGGGCGCTCATTTT
T
>culux_pipiens
TACATTTAATTGAGATCCGGCTTCAAAAAAGTACATAAATATCACTTAAGTGGTCATAAC
TCGAGACAGGGTTGCCAGATCTTCAATGTTGTGGACTTGTTGGAAAGGTCTTTTGATTAC
CTAACTAACGATGGGTCGGATGATGGATCCGGACATCATTTACA
>cynocephalus_variegatus
CTCCCTGGAGGTTCAGCTTCTGCAAGCCACTTTAACTTTAAGAAATGAGAGAGCAGAATA

TGCTTCTCACATCTAGCATGCAGGAATTCAGGTTTGCAAGTTCTATGTGAAAGCCCATTG
AAGCCTATGGGGGAAACGGCTTCTGAGACTTCTGTTTCAAAATCTCTGTCAAAAAGCATC
AGAGAAGCTTCCCCGCTTAGAATCCAGCATAACTCACTGAACCGAAGATGTGGCAGTTAC
AAGAGAAGTTGAGCTAAGAGAAGTTTGTTTTCTGAAATTAGAAGCAAAATCACACTGCAA
ACGGCATTCAAGCAGGCTTGAAAAGTGTGGTGCATATCTCTGCCAAAAACA
>danio_rerio
TAAAACGATCCAGCCATAAAATGCATCATTCTTTTTTGTTTTAGACAACAATTCATGCAC
TGTTAAACATGTTAAAGCAAGTTGCAAGTGAAAATCTATGTCTCTGACTGAGTTTGCATT
ACTGTGATTTGACCTCTCTGCTGGCTGAGATAAGCTCATTTTCAACGTCCAATTCAGAAA
GTGA
>daphnia_pulex
TTCAAATCCATTGAAGTTGTGTTCACTTCCTGGAAGATCTCTATCAACCCTTATTTCTAT
TTCTTGTCCATCATGGATGTCTTAAGAGATTCATTGATTTTAAAGACTTTAAATAGTGTA
TTGGACATCGAATTAACATATCGTGGAACAAGAAATCTGTTTAAACTTGTAAAAGATTTC
ATATCGTCAT
>dasypus_novemcinctus
AGGAAAGGAGATAGCTGCAAAATCTCTGCACACAATGCTTAGTTCTCAGAAGTAAATCAT
ATGTGCCATGTTGTTTCATTTAGGTTTAAAGAGTTAGGTGCTATACCCTCGGTTATGCTT
TCCTGTATAGGAAATGCTAGTTCTGGAGATACGAATGTTAGCACAAAAACTGA
>dictyocaulus_viviparus
TTGGCATCGAGAAATGTTCGCATAATTCAACATTACTAGATACCAAAATTTCAAATTTTG
CGCTTCGACCACTCTTGATCGTATTACTCATGGATGAATTTAGCAGATTTCTACAGCGGT
TTTCGGCATTTGTCAGTAGGAATATGGAGATCCACCA
>dipodomys_ordii
CCAGCCTCACAGAGCCCTGTCCAAGCCAGCACCACCTGGACCGGCCTGGG
>drosophila_ananassae
ATCGCTAATTTAGCGATAAAAACGGCTAGAAAATCATCACTGTTAGCGCGTAAAACCGCT
AGAAAATC
>drosophila_erecta
ATTTAGTTAATAAATGTGTTCATGTTTGTGTTTGCGCACGAAAAGTGGTTTCATGTGGTG
CGCAGATAAACAATCTACATCCAGAAAGAAGAAAATATAAACTCTAAACTCTAGACCAAG
TCATCGGTAATTGTAATTAAAAACTGGTGCACATAGTGTTCAAAAATAATTCCCAACTTG
TTATGGCTTATATTTCATTATACGTTCCCTCTAACAGCCTATAAAGTAGTGGACAGGAAG
TTCCGTGA
>drosophila_grimshawi
AGAAATTCTAACTTCTGACAATTATTGGCGAAGATACATCCGTTTGAATTTGGAAACAGA
AACTTAAAACGTCAATTTCTCGAAATGACTTCGAGACTACATATTTTCAGCTTTTGATGC
AGAGTCAAAATATATAAAAAGGGCTTCATAATGTTGAA
>drosophila_mojavensis
AATAGTCCTGTTATTGCCTGTTATACATAGAAAAAAGTCAATTTTCTGATCTTTTACCGC
ATTTCAGGGGATAATAATGTAAATATTTCAAAAATGTTGGCCCATTGAAAATCTGAATAC
GTCAATGGATGGGACTCTCAAAGACGCTTCCAACGAAGTAAGTCGTAATGAAATCGGCCC
AGA
>drosophila_persimilis
ATGTATACTGAACCACCGCACCTATATAGCACCCACGCTCCAGGAAATGTGGCAGGATCG
CACATCTTGGGCATAAGCACCAATATGGCACTATAGCCCCAAAAAAGTGGAACTCCGCAA
CTATATGGCACCTGCACTTATATAGAACCCTTCCCTATCCAGCACTGGC
>drosophila_pseudoobscura
AATACGTCTCTCCGCTGGCGGAAAACTTATCTCCGGTGGCGGAAAATTTATATCCCCTGC
CGGAAAACTTATCTCCGCTGACGA
>drosophila_sechellia
TTTGTGCAAAATTTTTGGATTTTTCGATTTTAGATACCAGGCGATGATAATCAGTAGCGG
GTGTCTACTGAAAACCAACTAATCGTTGGTCACCTTCTGGAATTCTTGTTCGCCTGGTAA
TTTAAACCGAAAAATCTCTCAATTTGCAACAAAATGCGTATT
>drosophila_simulans
CCTACAAAATGCGCATTTTTGTGCAAAATTTTTGGATCTTTCGATTTTAGATACCAGGCG
ATGATAATCAGTAGCGGGTGTCTACAGAAAACCACTTATCGTTGGTCACCTTCTGGAATT

CTTGTTCGCCTGGTAGTTTAAACCGAAAAATCTCTCAATTTGG
>drosophila_virilis
ATATCTTGACCAAACTCGGCATTTATTAGTTTTACTATACTCCTCATATATATGCAAAAT
CCTATTAAGATCGGACCACTATATCATATAGCTGCCATAGGAACGATCGGTCGAAAATTA
AGTTTTTGTATGAAAAACATTTTGTTTTTCAAG
>drosophila_willistoni
CAGCCGCCATTTTGAGTCTTCTACCCGCCATTTTGAGTCTTTTAGCCGCCATCTTTAATA
TAGAAGCCATACGAGCTGCTAGAACCATTCTA
>drosophila_yakuba
AGTATATACATGATGCTTGTAAATAATTCCCAACTTGTTCTGGTTATTTTTCCTTATATG
TACTCTCTACCTGCCCATAAACTAGAGAATGTAATGGCCTCTATAAACGTTGATGCAGGC
GATGGGGATTCAGGAGCTGCGCGGGTGTAAAATCTGCATTCAGGACGCCGTGAACCAGGA
CCAAACCAAGCCAGGTTAGTAGTGGAATGAACTATATTTTTTAAAACTATATGTGTTCAT
TGAATAATACACAATTAAAGTCCATCGACATTTAATATTAAAAAAACTGAATTACTGTCA
GCTTACACCAAATATGACATGTCAACTGATTCCCGAAAATATACACAGTCAAAACAAGTT
AAGTTTGGCCACGTAATTGTTAATGATTGTTTGAAA
>dunaliella_salina
CAGTGCACACCCTCCAGGAAATGTGCAGGTGCACCTTGGTGAAGGAAAAGGCTCTAGCAA
CACAACTACTCAGTTCCAAAATGCCACCAGCAGCATTCTGAAGTTCTGAGTACTATAGTA
TTCAGCTACAAAAGACAGCATGAGAAACTGCTGTGTCTCTGGAAAGATATTCACTGCTCT
GTT
>echinops_telfairi
CCTGCAAGCCAGGCAAGGGACTTCCGGGCAGGCTGGGGAAGCGAGGCTTTGCCTGGCGTC
TC
>equus_caballus
TTGCTTCCAGCTCTTTGGGAAGCTAAGAAACAACTCGCTCTGCACAGTGCTCTTACAGCC
TACCGGGAACATCTCTTACAAAGGCCTGTGAAACGCAGTTTCTTTGCAGGCCTGATCTGG
CTGAAGTAGAACTTCTGCTCTGCGTGTGCTACATTGCCCTAGAGCTGAAAGGCACAACGT
GCAGGCTTCTTTCCAAAGGGCGAGATGGGCCCCAAAGCATC
>erinanceus_europaeus
CTCTCTCAGCACCAGGCACAGCACCCAGGGAGCCCATGGAGGGTGGGCAGGGCTGTGGGG
TCTCTC
>eucalyptus_grandis
CGGATTTAATATTTTGGAGATCCCGATACCTTGAAAGTCGGCTGAAATCAAGAAACGGAG
ACTCGGCACAAAAAATTCCCTTGTGTTCCATTAAAAATTCCTTTTCGGTATTTTCACTAT
AGACCCCGGCAGGCGGAAACGGCATGGATAAGTGGGCTCCAAAAGCTTTATTTCGAT
>felis_catus
GCGGGCACTGGGTTCACTGGAGGCTGCAGTGCC
>felis_silvestris
GCACTGGGTTCACTGGAGGCTGCAGTGCCGCGG
>gasterosteus_aculeatus
TGACACGGTACAGTGGTGTAGTGGTTAGCACTCTCGCCTCCCAGACAGATGCCTCGGGTT
CAACTCCACCCAGTGGTGCCTCTGTGTTAGGTGGTGTAGGCTAAGCCTGCATGGGATTCA
CTCCATACTTGTAAAAGTAATGGAACAATTTTTTAAAACTTCGCATCTGAAAAGAATCCC
CAAATATTTTGCAAAAAGTCATGAAATTCTATCTTTAGTCATATGAAGGGGGACGCCGGT
GAGAAAACGTACTTTTGAATAAAAATCTGAATATTTTCTGAATTGGAGTCATAAGACATT
CTAATGGCAGCAGT
>globodera_pallida
ATGCGTGAGACCAATATGAAGAAAGTTATGGCGCTCGGCTTGCATGGATTTGGTAGGGTG
CACCTTAATTTGTTGGGGCACTTTAAATTCATAATATTCCATCATATTATATACCGTTCG
ATTCCTGCCAACAAGCTGAGCACGTATATCTAAACGCTGG
>glossina_morsitans
AAATTTCAGAAAATTGGTTTTTACGGCTCTCAGAACCCACACGTACAGACATGATAAATT
CGCACAATGTGCTGCAAATAGCGTTTTTGCAACCACAGCAAGCTATTAGAATGCTAACTA
TTGCTTTTTATGAAGAAAAGCAA
>glycine_max
TCACTCGGATGTCCGATTCAGGCGCATAATATATCGAGACGCTCGAAATTGAACAACGGA

AGCTCTCGAGAAATTCAAATGATCATAACTTT
>gossypium_raimondii
TTAGGGAGATAAGATCTACAATCTTCAACCTACTCCACTGCTGCTCAGGGAGATAGGACT
GGTGGCTTAAATCTGCTTCCTACTATCTCGGGAAGATAAGATTCGCCGTCTTCGATCTGC
TCCACTACTGC
>haemonchus_contortus
TGGACCTAGAGCGTTCAAATTTGGTAAGAGTACAGGAGACAGGATGGAAACACTCATGGT
TGGTGGCAGGACCGGACTCCGCCCACAAGGGGCGGGGTTTGTGCAGAAAATCATAGCTGC
AGTTCCCG
>heliconius_melpomene
TTGAAACTGGTGTCGATCGTTCGGGCTCCGAGGGGCCTTCAAGGTGGTATACAACCCGCG
GGAATGCGCGGGAAGGTGCGGGCGGTAAGTGGCGAAAACCGATTTTTTCAGTTTTCCCCG
CTCGGATTCAG
>helobdella_robusta
GTCATACTAAACAGCTACAGAAACGACAGCTCAAACTACTGAAACCACTGGTCCAATGAC
GACGCAAGAAACGACTGTCGAGGCTCCAAGTAAAAATTATTTTAATCGAAACGTGTCTGC
CAACAAATTGATTTGATTTTCACACCTGACAACATTGTTTCGTTTTAATCGTATTTCAGC
TACTGCTTCTCCACCTACTGAAGCTCCGACAACTGTTACAACCAAGGAAATCGAAACATC
AAGTCCAGTGTCAACGACAGAAACAGAGACAACAGTTGTTCCTCCAAGTCAGCTTAATTA
GAACTATTAAAATAAATCTTAATTGATTAAATATAGCTTCAACTCAGTAAAACATATTTA
AATTTATTTGACC
>heterorhabditis_bacteriophora
CCTATGTCGGTAATGGATGCACTGTAGGTGCTGCAGTGAGACTCCCTATGCGAGATAAGA
TCATATCGATAGAGAATTTGATGACGATCATTCTGACATATCTAGAATCTCTGTACGATG
AGTAGATCAAAAGTTATGCCCGAAAAACACGTTTTAACGCATTTTGCA
>homo_sapiens
AATATCTTCACATAAAAACTAGACAGAAGCATTCTGAGAAACTTCTTTGTGATGTGTGCA
TTCAACTCACAGAGTTGAACCTTTCTTTTGATTGAGCAGTTTTGAAACACTCTTTTTGTA
GAATCTGCATGTGGATATTTGGAGCGCTTTGAGGCCTATGGTGGAAAAGGA
>hydra_magnipapillata
CCTGAGTGAATCAAAGGAACCTAAGTTTTTTACGACAAAACACTTTTTCTATATAATAAC
TTATCTAAAATTAAAAAATAAGACAATTGCGTCAAAAAAACAGGTTTCCACTGCTTTCTT
TTGGAATCTTTTGATATCTTTTTTATGATTATGTTGGTTTGTTAAAGAATAAGGTCAAAT
AAA
>hylobates_concolor
ACAGAAGCATTCTGAGAAACTTCTTTGTGATGTGTGCATTCATCTCACAGAGTTGAACCT
TTCTTTTGATTGAGCAGTTTATAAACACTCTTTTTGTAGGATCTGCAAGTGGATAATTGG
AGCACTTTAGGGCCTATAGTGGAAAAGGAAATATCTTCACATAAAAACTAG
>ixodes_scapularis
ACGGTGAACGAATAAAACGGCGCCGAACGTCTCTCGTCAGGCCGTGCTCGCTCCAATAAT
TCGCTTTCACCGAGGCAATAGCCGAGATCGCGGTC
>labeotropheus_fuelleborni
CATAATGAAAACCTATACTTTGTTTCAGGCGAGTTTCCCATTCAAATGCATGTAACAGTG
AGAAACGCATTGTCTTGGCGAAATAAAGCGTTTTTGAACAACTTCATATAAATCGCTGTA
ACTTTTGATAGAAGACTCAGAAACACATGTTTATGGCTTTATCTTATAGAACTCAATATC
CCGTGCTGGGCAAACAGGTTTTGCAGCCGTTTGAGCTAAGATTTTAAAATTATTCA
>lama_glama
CGTTCATCCAGCACAAAGGAAAGGGTTGCAGGAGAAACACTGACTCTGGTTTCTCAGACT
GTTTCCTAGTGCCGGTTTGAAGTGCCTGCCGTTTTCACTGTAAAACCGAGTTGGAGCTTG
TACCTCCCCATATATATGAATGGAGTGGAGTTTGCAACTGAAAAGAAACTTTGTGTTCCC
TTCAAGCTATGTGAAGGACGGCTTTCACACACACTTATCTCTCAGAACTGAGCATGTGGA
GAGACGTT
>latimeria_chalumnae
CAGCAATTTTCAGCACCCCCTTTTTGGGGGGCATCCGGATCAGAAATTTAAAAAGCGCTG
ACT
>lemus_catta
CTTTCTCTAAGTGAATGTAAGCGTTCCTATATATGACTCTCGCACTTCATCCAAACTAGA

GTTTAGCAGGCCTACAAACCACCACCCCATAGTTG
>lottia_gigantea
GCAGACTGCCGACCCCACCCTCTAACTAAAATGGCCACAACTACTCCAATTTTGACGCTA
GGGGAATAATTTTTGTTGCAAACGAGAGTTTAACATGTTCCCTATCTATTTCAGCCGAAA
GCAGAAATTCGGAATTTTTGAAACGTTCGAGAAAATCGACTTTGAAGTGTAACCCAATTT
TCAAGAGGAAAATCGCGTATTTTTGGTTAGGGAAAAATTGAAACCAAAAGTGCTGTAACA
AGCAGAGTTTTCAAGATAGGAACCTGAAATTCGAAACACAACAACTTCCAATAGGGGTGT
ACCTTCAGTGCAAATTTCAACTTTACAGCTCTGATACCCTTTGAACGACAGCCTGTCAAA
GTTTTGGTCTTCACCCTTAACATACAATGCAAAAAGACGCTATAG
>loxodonta_africana
TCGGAAGGAAGGCCTGGTGACTGTACCCAGAGGTACC
>macaca_mulatta
TCCTTTGGCACCATAGCCCTCAAAGGGATCCCAAATATCACTTCGCCGATTCCACAAGAA
CTGGCTAGCGAAAGGCTCCTTGAAAGAAAGATGTAACTCTGTGAGATGAATTCACAGAAC
ACAAAGAAGTTTCTCAGAAAGCTTCTTTCTCTTTTTTATCGGAGGATATT
>macropus_eugenii
TAAAAGCCTTTCCACACTGACTACATTCATAAGGTTTCTCTCCAGTGTGGATTCTCTGAT
GTTCACTAAGACTGGACCTGTCTC
>manis_pentadactyla
GCTTGAAATTAGTATCCCTCACCTTGCATTGGTGGCTCCATCAGAGATTACTGATGTAAA
TTCATTTGCAGCATTTCTGAAAAATTCAAAACATATATGATCAGACTCCAGAGACTTCAG
AGAGTTACACTTTTT
>marcantia_polumorpha
TCACGCATTTTGGAGTTGGGACGCCGAAGTTCACGGATTTCTCAAGTTTCGAGTTGGAGT
TTTCGAGAGTTGGGGTCGAGATCGTTTCTTGCAGAGATTGTTTCTTCGCGTCTGAGTGTC
CACGGAAGCTGAAAATTTGCAGGGGCGTGGACTTGGGTAGCGCGCTCCAGCCTGCAAAAT
T
>melanochromis_auratus
AGTTGTACAAAAATGCTTTATTTCGCCAAGACAGTGCGTTTCTCACTGTTACATGCATTT
GAATGGGAAACTCGCCCGAGACAAAGTATAGGTTTTCATTATGTGAATAATTTTAAAATC
TTAGCTCAAACGGCTGCAAAACCTGTTCGGCCAGCACGGGGATATTGAGTTCTATAAGAT
AAAGCCATGAACATGTCTGTCTGAGTCTACTATCAAAAGTTACAGCGATTTATATGA
>meloidogyne_incognita
GCCTGGGGGAAACCTATTGCCTGAACCCAACCACACTTGGTCAATTGCACTTTGGTCTAC
TTTGCCTGAGGCAATTTTTAATCGGATCGTCACGCACTCTAAACCATGAATGCCAGGTGA
AATTTGATT
>metriaclima_zebra
CACATAATGAAAACCTATACTTTGTCTCGGGCGAGTTCCCCATTCAAATGCATGTAACAG
TGAGAAACGCATTGTCTTGGCGAAAGAAAGCGTTTTTGTACAACTTCATATAAATAGCTG
TAACTTTTGATAGAAGACTCAGATACACATGTTTATGGCTTTATCTTATAAAACTCAAGA
TCTCCGTGCTGGGAAAACAGGTTTTGCAGCCGTTTGAGCTAAGATTTTAAAATTATT
>microcebus_murinus
CGGGCAGGCAGGGCGCAGTGCGGATCTGGCTGTGTCCACTCACCCACGGCAGA
>micromonas_pusilla
GGTGGCCGCCACCCGCGGCATCGTCGCAAGTCCACTCGTCCCACGGAGCGTCGTTCGCGC
GCGCCCACTGCAGCACCTCCA
>micromonas_sp
CACGGGAACCACAGCGGCCGTTTCAATCTGCGCCCCGTCGACGATGACGATCACCGGCTC
GACTCGACTCTGCTCGACCCGTGCGAGGTGTCGATGCGTCGGGCGTGCCGTCGCGGTCGC
GATGAGCGTACGAAGTGCGATTTCCAACGGTTGAATCGACACCAAGCTAGCACGGACCAC
ATGAACCCTCAACATAACCCTC
>monodelphis_domestica
AGCTTACTGTACACCAGAGAATTCATACTGGAGAGAAACCTTATGAATGTAATGAATGTG
GGAAGGCCTTTCACCTGAGGTCAC
>myotis_lucifugus
CAGCTTTCTACATCTAATCTCGGTGATTGAACAAGGCCATAAAGATAGCACGGCTCTATT
TACATCTATTTCAAGTGTTTCTTACACGTTTCTAATCATTTCCCTGACTTATTGCGTATT

TTAAGAGTTTTACCAAAATTAGACAGTTTTCTTAGAAGTATATACATGAATCTTCAGGAT
TTAAACAGTTTCATTCAAAAAAAGCCAGCCAAGTCAGCCTCACTTGTAAATATAAACAGT
ATTCTCCTATTTAAAGGCTTTATGCACAAATTCTGAGCTCTTTAAACTGCAATTCTTAAT
GTTTGAAAGCACATTCTTCGGAAAAAGACCAAGTAAAGTAGCCAGAATTATAAACATTAT
CAGTATTCTCATCACTTCA
>nasonia_giraulti
CGCTTTGGTTTTAGATTTTATACTCGCTTCGCTCGCCTTCCTTCGATGTGCCGAGGTGTT
TTTTAAATAAATTATTGAGCTCGGGGAGCGTGGTGTAGTTGGGTTTAAG
>nasonia_longicornis
TGGGTTTAAGCGCTTTGGTTTTAGATTTTATACTCGCTTCGCTCGCCTTCCTTCGATGTG
CCGAGGTGTTGATTAAATAAATTATTGAGCTCGGGGAGCGTGGTGTAGT
>nasonia_vitripennis
CACATCGGAGGAAGGCGAGCAAAGCGAGTATAAATCAAAAACCAAAGCGCTTAAACCCAA
CTACACCACGCTCCCCGAGCTCAATAATTTATTTAAACAACACCTCGG
>nematostella_vectensis
GGGCTGTAGGCATTTCCTTTTTGTTATCGAAGTACTCACAGGACTGTTATATAATAGATA
GATCTTAAACAGTTCTATCGTCTACGGCCATACCACTTAGAAAGCACCGGTTCTCGTCCG
ATCACCGAAGTTAAGCTCAGTAGGGCGCGGTAAGTACTTGGATGGGTGACCGCCTGGGAA
TACCGT
>nomascus_leucogenys
ATCTGCAAGTGGACATTTGGAGCGCTTTGAGGCCTATGGTGAAAAAGGAAATATCTTCAA
ATAAAAACTACACAGAAGCATTCTGTGAAACTTCTTTCTAATCTGTGCATTCATCTAACA
GAGTTGAACCTTTCTTTTGACTGAGCAGTTTTGAAACACTCTTTTTGTAGA
>ochotona_princeps
AGTGTGAAAACTGAGTTTCCTTGAGAGCTGCTTGGAAGCCTGATGGTGTGTGAAGTTAGT
GAGCCAAATGGAGTTGTTCTCCCACAAGAACCACTCGACTGATTCTTCTTCAATATGCCC
TTCACCAACATGCAGTGTTTCCTTGACAGAAATAGCTCTCACACCAAAATTTGCGTCTAG
AGCCTTTGCATTGAGGGCTATCTGAAAACGGCTTTGGCGCTGAAGTTGTATGTAAATGCG
GAGTCAGAGGCATGTGAAACCCGCATGTGGAGCATTGGTGTGTGTGCGATGCTCACCAAG
CTTCATTCCAGTACAAGGCACTCAAAGGGCATTAAGCATAG
>oesophagostomum_dentatum
TTTTACTCGAAAAACGCAGTTAAATGCATGCATACACTTGTAAAAACTCATG
>onchocerca_volvulus
AATTTTTCCTTGCACCGTCGGCCTTAAACTCTTTATCATCATCATCACTACTAGAAGTGC
TGTCAGATGAACTTTCAGATACTCTGCTTGCAGAAACCGCATTTTTTAAATTAGCATTTG
GCGTAGCTGGAACCGTTTTTTCGGG
>oreochromis_niloticus
AATGTGTTTCAGGCGAGAAACGCACTGTCTCGCCGAAATAAGGCGATTTTCACCAAGTCC
ATAAAGACAGCTGTAACTTTTGATAGGAGACTCGGACACACATATTTCAGGCTTGGCCTT
ATAGAATTCAGCATTTCCATGCTGGGGAAATAGGTTTTGCCACTGTTTGAGCTAAGATTT
TCAGTTATTGACTAAATGAAAACCAT
>ornithorhynchus_anatinus
CAGTTTGGGCCCCCCAGCCTTTGCCTTCCAGGTTTGGCCCCGAACCCTTTTCATTC
>oryctolagus_cuniculus
GAATATCCAGAAAACACAAACACACACAGAGGCTGTAGCAAGATCTGGGATTGAGATAGC
TACAAATCACCAGATAGTTTTGTTTTCATCCAGGAGTATCTGGACAAATTGTTGTGGATT
TCAGGTGAAATTTAACCAAAAATCACCGTGAAAGTGGTTTGGAGAAAAGCACAATCTTCC
TGTTCGGCAAATGTATCCAATGTCATGTTCACAGAAAGAAATATCATGTTTTCTCAGGTT
TCTGAAGCTCTCTTCCTAGAAAACTCGAATGTGTGGAGTAGTTGACTCCAGGTGGAAACA
ATTAGCGTTTGCCAGCAGAAAACAAAAAGGGTCCTGCTAAAATGCATTCAAAACCTTAAA
GTCCATAGA
>oryzias_latipes
AACTGCAAATGAGAACTTTAACTTTTGGGTGCATTTTTTGCTAAAAAATCATTTTGTCAG
TCAAAAGTGCCAAAAGTGTCAAAAAAGCGTTTTGGCTCTCAGTATGACTGTTTTGAATTT
TCAACTTACAATGTGACAAAAAATAACACTTTTTTG
>ostreococcus_lucimarinus
CTGCGAACGCAGCGCGTCGCGCTCAGACTCGAGATCGGACACGGCAGCGGACTGCGACGC

GTCAACCTCGGGCACCTGAGAAAGCACGCGCTCGAGCTCCGACTGCAGCGCGTCGCGCTC
AGACTCAAGTGCCGCCTTGGACTCGTCGAGCTCAGAACGCACGCGCTCGATGTCCTCCTC
GCGCTCGCGAAGCTGGGACTCAAGATCCGCGCGCACAGACGCCAACTCAGAGTCGCGAGC
GGACGAAAGGTCGGCGAG
>ostreococcus_sp
ACGAACGACGCCGGCGACGACGCGTCTGGCGCCGACACGCAGTGCGATCCCATCATGTGC
GCGCGAACAAGCACGTGCAGAGCCACGCGTGCGTGGCGTGTCCTGCCGGGACG
>otolemus_garnettii
CAGGCACCCCTGAGGGCAATTAGGAATCCAATTAGAAACACCTGTGGCCAATTAAGAGAA
AG
>pan_paniscus
CAAGTGGACATTTGGAGCGCTTTGAGGCCTACTTTGAAAAAGGAAATATCTTCACATAAA
AACTACACAGAAGCATTCTCAGAAACTTCTTTGTGATGTGTGCTTTCAACTCACAGAGTT
GAACCTTTCTTTTCATAGAGCAGTTTTGAAACACTCTTTTTGTAGAATCTG
>pan_troglodytes
CTAGACAGAAGCATTCTCAGAAACTTCTTTGTGATGTGTGCATTCAACTCACAGAGTTGA
ACCTTTCTTTTGATAGAGCAGTTTTGAAACACTCTTTTTGTAGAATCTGCAAGTGGATAT
TTGGAGCCCTTTGAGGCCTATGGTGGAAAAGGAAATATCTTCACATAAAAA
>pan_troglodytes_schweinfurtii
CCAACATAGGCCTCAAAGCGCTCCAAATATCCACTTGCAGATTCTACAAAAAGAGTGTTT
CAAAACTGCTCTATCAAAAGAAAGGTTCAACTCTGTGAGTTGAATGCACACATCACAAAG
AAGTTTCTGAGAATGCTTCTGTCTAGTTTTTATGTGAAGATATTTCCTTTT
>panicum_capillare
GTTTTGGGTCCATGGAGTGGATCGGGTGCGTTCCTTGCGAAAATTCCGACGCAACTTCGT
TTAGCGAACTTGTGCGTTAATGGCACCAGTTGGCCCGTTTTGCACCGACTTTCGTGCAGT
AACGAAACGGTCCGAAACGCCCAAAACATGA
>panicum_virgatum
GCACGAACTTTTGCATTAATTGCACCAGTTCACCCCGTTTTGCACCGAGTTTCATGCAGT
TACGAAATGATCCCAAATGCACCCAAACACTATGAAACGTACCAAAACATGAGTTTAGGG
TCCAATGGGGTGGATTGGGTGCGTTCGTTGCGAAAAATCCCGACGTGACTTCGTG
>papilio_dardanus
TTTCTAGATTCAAAATATGAACCCCCTTTTTGACCCCCCAAGGTTGAAATTTGTAAAATC
CTTAAATGGCAAAGTTATTTATTCATGCTGAAGAGCTCTCATTCCAATTTTCATGCAATT
CTGACCTAAGGTGTGAAAAATA
>papio_hamadryas
GAGATGAATTCACAGAACACAAAGCAGTTTCTCAGAAAGCTTCTTTCCAGTTTTCATCTG
AGGATATTTCCTTTTTCACCATAGCCCTCAATGGGCTTCCAAATATCACTTTTCAAATTC
CACAAGAAATAGGCTAGCGAAAGGATCCATGAGAAGAAAGATGTAACTCTGT
>pediculus_humanus
ATTAGCGCTATAAACGGCTAATTGGTCTGCTTAAATGTTTAAATTATCCGGTTAATCGGA
TAAATTAGAATCCGATAACGGATTAATTTGTCATTTATGCGCCCAG
>peromyscus_maniculatus
TTCAACTCCGTTTAAGAAGTTGAATTGAACCCTAGTTGGGCCTTACAAGAAAAACACTCT
GCTTTTGAGAAACAGGCACTCTTGCACTCTACTGTGTTTCCTATAGGGCCAGTACAGTGA
GCTAGCTCAGAACAAAAGAACTGTGCTTCCTAAACAAACGGGGCATATCTAGTTCAACTC
AGTATCAGAAACTAAGTAAAACAAGAGCTTCTCTACAGTCCAATACACTGCACTCGCTAG
GAACAGACAAGAATGCTATTTGCACTGTGTATGGACAGAATAGCACAAGAGTCGCCTTCT
TACCCTGAATGAACACTTAATGCTAGAACTTAAAACTGAATGTGA
>petromyzon_marinus
GTGAATAGCAAAAAGCATTCAACCCCATAATACATGTTTTTAATTGGTCGAGCGTCCATT
TCATATCCCATAGCTTAACCACTCATATCTTTGCATTCCGCTCGATGAGGCGAGCAGTAC
GAGTACCACCTTGATGGGTCTCCGACGTTTCCTTCCAGAGTTATTACAAAAAACCACAAA
CATCCCAGAATCCTCTCGATTTGGGGTCAAAATGTTGTCGCTCAATAAACCCCATAAGAA
ACGTCGCATCGAATAAATGTGATTGAATTCTCCCCGTCTGGACGAGCGGGATTGAATGGT
GCAACTGATTTTGCTGTGCTGTGTTAAATAAGATATTTGCATGACTGCTTTTGTTTTATT
TT
>phlebotomus_papatasi

GCTCTAAATGGATGTAAAGTTTGAGGCCTCTATCTCTCATAGTTTCCGAGATAATCAACT
TTAAAGATTTTCAGACACTTTTATGCATTTCTCGTCCAATTTTCTTAATTTTCCAGTGAA
ATTTTGCACATATCAAGCCTGAAAGAG
>physcomitrella_patens
AGATGTAAGGTTGCCCAATTCATTTGGCAATGATGTCAAGCTTGAGCACCATTCTATATC
AAAGGTAGTCAA
>polychrus_marmoratus
AATACGTAGGAAAGAGAATGACCCTTTCTAAATGTATTAGGAAGGATATTGTACCTGTCT
CCCTTTAATATGTAGGAAGGAGAAGGTACCATTCTCATTT
>pongo_abelii
CTCCAAATATCCACTTGCAGATTCTACAAAAAGAGTGTTTCAAAACTGCTCAATCAAAAG
AAAGGTTCAACTCTGTGAGATGAATGCACACATCACAAAGAAGTTTCTCAGAATGCTTCT
GTCTAATTTTTATGTGAAGATATTTCCTTTTCCACCATAGGCCTCAAAGCG
>pongo_pygmaeus
TATGTGAAGATATTTCCTTTTTCACCATAGGCCTCAAAGCGCTCCAAATATCCCTTTGCA
GATTCTACAAAAAGACTGTTTCCAAACTGCTCAATCAAAAGAAAGGTTCAACTCTGTGAG
ATGAATGCACACATCACAAAGAAGTTTCTCAGAATGCTTCTGTCTAGTTTT
>populus_balsamifera
GGTTGGCCGACCAGACCTCCTCGTCACCGAGTAGCGATTCCATAGCTACCCCAAGGAAAA
CGGCATCCGGAAACTCGTGTAAAAATTTGAGCGCGATCCAACGGTCGGATCAAAAGTTAT
GGCCCTTTTGACCCGTTACTCAA
>porites_lobata
GCCTCTAATTCATCGTTTTCAGCTAGTCTTATGGTGCTTATCATCACAGCGACATTTTGG
AGCTCTTACAGTGAAATTCTCACTTAACCCATTACGACCAAAAGATGGTCAATTTGCAAA
ATACTGTCACAC
>pristionchus_pacificus
AATCGAGTTATGATTTATGGAAATGATAGAAGCACCACTTGTTCACGTAGAACAGGCCTG
ACACCTCTTTTGTATCAAAAGTGATACAAAAGATGAGTTCGCATGGAAGAAGGCCTTTGG
CTATGTGACCACCTAGCGCAAATGTCTTCTAA
>procavia_capensis
ACACAGAGAATTCCAAGTATAACTACAAGAAAATAAGAGATACTAAGAAAACTGGGAAAA
CCTACATTGCTGGTAGAGTGGGGAGTACTTAGCATTGTGAAAACAGTACTTTCGCAGAAA
GAAATCTACTGCTCTAAGAAAGCACAACAGTGAACCGAGAATGTATGAGTTTCTTAAGAT
TGGTAACAGATAGAAACAAGAGAAAGTACACAGTGATTCTTCTACCCAGAGAAGTTACAG
AGGTTT
>prunus_persica
GACCCCTCCCACAATATTATGATGTTCCCCAAGTTTTGGACTCAATCCGATATCGATTGG
TCCATGAAATCGGACAATTCAGGTTCGTACGAAGTTCGTTCTGAAATGGAGTGGTTGGTG
TATTGTATAGTTCTAGACATTGGATTTCACTCAAAACCCACCAAAT
>pteropus_vampyrus
CTATGAATGTAATGAATGTGGGAAAACTTTTTGCTAAAATTCACACCTTAGTGAACATCA
GAGAACTCACACAGGGGAGAAACC
>rattus_norvegicus
AAATTTTAACGAACGGAATCTAAGTATCTTGGTGAATTCAGTTAGTTCCCAATAGGACGC
GCTTGTAATAAGTGTACTTTTCAAGATATAGCA
>rhamphochromis_esox
CTGTAACTTTTGATAGAAGACTCAGACAAACATGTTTATGGCTTTATCTTATAGAACTCA
ATGTCCCCGTGCTGGGAAAACAGGATTTGCAGCTGTTTGAGCTAAGATTTTCAAGTTATA
CACATAATGAAAACCTATACTTTGTTTCGGTCGAGTTCCCCATTCAAATGCATGTAACAG
TGAGAAACGCACTGTCTTGGCGAAAGAAAGCGTTTTTGTACAACTTCATAAAAAATCG
>rhodnius_prolixus
TTTCATGACCCAAAAAATTTTCAACCCTTTATATCTTCTCAATTTATGGATCTGTACACA
AATAAACTATAGCTTATAACTGAGGAAAACAAAGTGTATAAGAGAACAAAATTTCATACG
AATATGTTGTTTACAGAGGAAGATATAGAGTGGCCGTAATTTTACAGGA
>ricinus_communis
TATTTGTGAATATGAGAAAACAACCCCGTAAAAGGGGTA
>saccoglossus_kowalevskii

ATTGCAGAATACACCAGTTCTCGTCCGATCACTGAAGTTAAGCTGCATCGGGCGTGGTTA
ATACTTGCATGGGAGACCGGCTGGGAATACCACCTGCAGTAGCTTATCAACATTTTTATT
CCACCGTACATACCAACTTTTTCCCGGGATTTTTTGCCTAGCCGAAGCGTTAAACGACAT
CAATTCCGACGAATATTCAGCCTACTGCCAAACC
>salmo_salar
GAAATTGCACTATGTTTTGGTGGGGACGTGTCCGTAGCATTTGGGTTACCAGGGGTGAAA
TTTTACAACTGTTTTAAATACTTTTAAAGGCCATAACCACCCCCAGGACACCCCTGACAG
CACCCCACTACGGGCCCAAGTCAATGGGAGCAAGAATGACTGATTTATGGAAGTTTGAAA
AAATCCTCAGAAAATATTCTATGTTTTTCTTCTCAGAAAATCGGCTATGTTTTTCTTCTC
A
>schistosoma_mansoni
CTATGAAAATCGTTGTATCTCCGAAACCACTGGACGGATTTTTATGGTGTTTGTTTTAGA
TTATTTGCGAGAGTGTGGGCGTTAATATAAAACAAGAATCATCTCAAATCCGACACAGCC
GTT
>schmidtea_mediterranea
GACGAGCCCGTTATGATTTTTGCAATTTTTAAAATGCTATCATGTAGAGAATCAAAAGTT
ATCGAGAAATATATCAAAGATTATGAAAACAAATAAATATTGACAGAGATACAAGCCTAT
AAAGATTAAAAAATTTCCGTCCAGAATGACTTGTGTTT
>selaginella_meollendorfii
ACATGCACTGGAAGTCTAGCTGGGGGTGTGGAGCCGGGCACCCGAGGTCTGACCTCGGCT
GCCCAACACTCGCATCCCCCACTATCCTTCCAGTGCTCACGGCCTACCACGCCCCAGGGA
GACAAGCGCTTTGGGCTTTCACCCCCACGTGTCCATGTCCGAGCCGGAACCCGAGTCCGA
GATGGAGCCCGAGTCCGACCCCGACATGGTGCACGCGGGGAGGAGCCCAGCAGCAGCCCG
C
>setaria_italica
AAATGCACCTGTTTTGCATCGTTTTTCGTGCCGGAACCGAATGCTCAAAAACACTCCCAA
ACATGTTCTAGGGTATAATTAGGAAGATTGCATGCGTTCGTTGCGAAAAAATCCACCCGA
GTTTCGCTACCCGAAAATAGTGCATTCGGGTGCCG
>solanum_phureja
AGATCACCAAAAAGTCCATGGACTATAGCACACGAAAATCGGCAAAATTGGGGGTTTACC
TGCTCTGGGGCACATTTGACCTTCAAAATGGGATGTTTTGGCCGTGAGGGCTAACTGGCT
CCATAGCTAAGGTCTTAACGGACGTCCATGAAAAATTTTGGCAAAAATTATGTCGGAATT
CC
>solanum_tuberosum
CGTTAAGACCTTAGCTATGGAGCCAGTTAGCCCTCACGGCCAAAACGTCCCATTTTAAAG
GTCAAATGTGCCCCAAATCAGGAAAACCCCAATTTTGTCGATTTTCGTATGCTATAGTCC
ATGGACTTTTTGGTGATCTGGAATTCCGACAAAATTTTGCCAAAATTTTTCGTGGACGTC
>sorex_araneus
CACGCATGCGCAGTGCGCTCTGGCCGC
>sorghum_bicolor
GAGATAGTGTTAATCTTGATGCAAGATAGGTGCACGGTTTGCACGGAACGCACCATAGGC
TAAGAAACCATTTTGGACGCACCCGATGGAACTCCTAGATGAAGTGTGTCAAATGGAAGC
TCGGTTCGGTCTGTTTG
>spermophilus_tridecemlineatus
CTAGCTCTTTGAAACTTGACACAATAAAGCAGTTTCAAGTTCTATCTCTATGGTTTGTTT
GCTGTTGATGTAGCCCTAAGCAGAGACTGGCGAGAATCACTTTCATGGAAAGGGCCTGCC
AAGCTCCAAAAAACTACGAGTTGCCTGAAATCTCAGCAAACACTTTTTCACCCAATCTTG
ATGATACTTCAGGTAATGATCAGTACACCAATCCCACTTGGCTCGCCAAATTTCGTTCTT
CTAGGTTAAACCGTTCGTTGCAGTAACCGGTTTCATTTTGAGGTTTCACTCTGTTTTCTC
CTATAGGGATACATGTATTTGGAACTCTAAATCAAACACGGTTTGTG
>spirodela_polyrhiza
CACCAATTGAATCTGGAAGATATTGTAATTTTGTACTTGACAAATTAAGATATTCTAACT
TTATAAGTGATCCAATAGTACTTGGAAGACTTTGTAATTTTGTTTATGACACATCTAAAT
GTCTTAATTTCACAAGAG
>strongylocentrotus_purpuratus
TTATTTGAGCGCATATCAACCAATGTGCCAGAACAAACCTTAACGTTTAGCGCATATAAC
GTCCAGAGTATATAGAAGGGGGTTGGGGTTAATACATTCGTTGCTGATTTCCCATTTGCT

TTAAATGTGCTGAAACTCACATGAAATGTAGATTATTATATACCTCACCCAATACGTGGT
GCAAAAAATTCTAGCAGCTGACAATAAAAAGGTC
>strongyloides_ratti
AGAAGCCTTTTTAGGGTAAAATATGTGTTGCTGAAACCATAGGAAAATGAAGATTTAGAA
GTCTAACAATGAAACCATTGTAATCTACATGAAAAATTAATCTAGAATCAACTTTTTTAT
TTTAATATATTTAAAGTCATTGATACATAACTGAATCTTTAATATAAGTTAC
>taeniopygia_guttata
ATGTTTGCGCTCCGAGTCCTGTGGCCCTGAGCCGGCAGCTGACTGCAAAAAATCCGAACA
TCGAAGGATAAGTCCCAGACAGGAGCTGCCCCCACCTCATACTTGTGCTGCCCTGACAGT
CCCAGGGGAGCCCTACAAACTGACATCAGGAACCTGGACTGGTGAGTTTTCTTTCCCAGG
AAAAAGAGCCG
>takifugu_rubripes
TGCATTCACCTCTGTTTTGACAAAAATGTGTCTCCTGACCAAAAGTGATGGTTTCCCCAC
GAGAAAACGTCAAAAACGTCATAATGTGACCGCAGCATGAGTTTTCAGATGATCATGT
>tarsius_syrichta
TGACTTAACTAGCTAGCCAGCTGGCTAGCCAGCTGACTTAACTAGCTAGCTACCTAGCTA
GCCAGCTGGCTAGCCAGCTGACTTAACTAGC
>teladorsagia_circumcincta
CACCTTCGGTCCCCCACCAAATATGTTCCCGTCCCACCCCCTACATCACTTGTTGGAACC
AACGCCTCACCGCTGCTGGGACGCGTGCTATGAATTTCGTGAAATTTGCATTCCAGCGCC
TCTTATCGGCTCTTATCAGCCGTTAT
>tetranychus_urticae
CCAGCATTTACTTTACTCAACTATTTGTCACACATGCAAAAATTTGATATTTGATGATTC
ACCTTGATAAGCTTTAAAGAGTGATGTCCAGAAAAATAGTCTCACGCGAAAATAATTTTA
ACCTTTACTTTAACAGTCAAAAGTTTGAAACAAAATCCAAAATTTACACTTTGA
>tetraodon_nigroviridis
AACACTGGAAATCTGTGTGCTGCATTACATTTTGTCAAAAGTGAGCTTTTCTCGCCGAGA
AAACGCTGTTTTTGACAAAATCCATCATTCTGCACCAAACGCAGGTGCCAGAACTGAA
>tribolium_castaneum
TTGTCAAAATTTATCGAAACCAAATTTTCTTCATTATTTTTCAACTATTCGGAGTTATGT
CCTTAAAATTAAGTGTTTCTAACGTAAAATGACCGGCGAAATCCAAATTTGCAAGAATCA
AGTCGCTACGACTAACCGTTCTGAAAATATCGGCAAAAATTTAGCC
>trichinella_spiralis
AATATAGTTAAGCATTGTTAATCATTGCATTTCAGACCGCACCAAGGAATATGAGCGGTT
TATCGGTATGCAAACTCTGGTAAGGAGTTACTCTTACCAATATTCCGGATTTCTACCTGT
GTTTCTCCCCACAGAAGTCAGGTAATGGTTCCTCTCTATGGCCTGTAATACTATCT
>trichoplax_adhaerens
GGATGATTGTCACCAAGTGTTGCTAGTTCAATCTTGAGTGATTTGTTATACATTGATAGA
GCATCATCATACTTGCCTTGATTACAATAGACAAGTCCAATGTTGTTATATGACATAGCA
ACATCA
>trichuris_muris
GGTATCGGCGAAAACGACAGCCATCATCAGTCGACCAGCACGAAAGCAGAGCGCACAGGA
AAGATCGGTTGTGTTTAGGTCTAGGCTTA
>tupaia_belangeri
TTTTGGCTGAAAATCTACATGCTTATCTCACGAAAGTAAGCTTCATTTCAAAACGTCCAA
TGAAATACACTGTAAAACTCCTTCCAAAACTGAAAATTGAAGTAGAGCAACACTGAACAT
CACAAGTTGTTCAAGTCCTCAGTTAAGTAAGTGGAACAACTAGTTGCTGTAGTTTTTTTA
ACATTTCTTGGGAAGAGGGTTTCTTAGTTCTAGTTTGCTTTGAAAGTGTAGTTCATTCTT
AGGCCCAAAACTAAACTTCCAGCTACTACAAAAATCTTTAGGTGTGAAAATAAGCTCCAT
GAGTGGAGTCATGCTGCCAAAGATCTTTCCGACATTCCAAGGAAAAGTACAGA
>tursiops_truncatus
AACCAGGCAGGATTGACCTCACACCAGAGGGCCCACATCACAAAGGGGATGGG
>vitis_vinifera
AGTACCGAAAAAGGGTCGAATCAGTGTGAGTACCGAAAAATGGTAGAATCCGGGCGAGTA
CCGGGAAAAGGTAGAATCCGTGCGAGTATCGAAAAACTGTCCGGGCG
>volvox_carteri
CACCACCCGCGGCGTTCTACTAGTGCAGTAGCCAGGCGCGCTGTGTCCACTGTCCTAGTA

GCCGCGGCGCTGACGTGT
>xenopus_tropicalis
GAGATATGTAACTGTCCCTGGGCTGCAATGGTATCTTCCAGCCTTTGCCCAGGCT
>zea_mays
CTTTAGGTCCAAAACTCATGTTTGGGGTGATTTCGCGCAATTTCGTTGCCGCACGTCACC
CATTCCGAAAACGGGTATCGGGGTGCATACAAAGCACGAGTTTTTGCCACCGGAACAATT
TCTTCGTTTTTCGCAACGAACATGCCCAATCCACTA
>Jatropha_carcus
CCCAAAATCGTAGGTATACCCCTTTGGTCTTAGGAGCCCTGTGTGAAGCCCTGTTCCGGA
GCCGATATCTTCTGTACGCCTGATGGGAAGCTCCTATAGGCTCATCTAGGGGATATTTTA
GGGTCTTATGGTCCAATCGGTATGCCTATCTCGGCTGTTCAATGCTTCATTAAGGATTGC
AAACTCAATATAAATTAATGCGCAAAATCGGTGCCACTCCGGAAGCATTTGGCAGTTTTA
GCCCAATATCTCTCTGTCTATCGCTTGTCAAGCCATTCAAGGCCATAATATTTCCTCTAA
GACATATAGAAAGCATTGTAATCATTTTCGGGCAATTCCGAGCCGGTGCTTCCGGGC
>Mus_musculus_castaneus
GAAAAATGAGAAATACACACTGTAGGACGTGAAATATGGCGAGAAAACTGAAAATCAT
>Bos_taurus_taurus
AATGGAAGATTGGACTTCCTGGGCAACACAAGAAGGCCATCACTGATTTCCCCGTTCGTA
ACTCGAGATTCCGCCGCCAAACTCGAGACACCAACGTGGATTCCCCCGTCATCGCACAAG
ATTGAAGCCCTTTCCGCTACAGCGTCTCAGGAGAAGTCCCACGTTAGGAATTGAGCGGTG
GAACGGTACTTGGCACCCTTGACGGCGGCCCCACAAAGTTCCCCGGCATCCCGTCTCCCT
GGAGAGGAACACCGAGTTTCCGGCACCACTTCCTCTGAGCCCCTTCTACCCTCCTGATCT
CGACAGGGAGGGTCGACTCCCCTGCTTTGTCTGGAAGGGGTTCCCGACCTTCCGGTCGCA
CCTCAGGATGAGGCCGGGCTCACGACGACATTCCAGACGTGGCCTCGTGGGTGCTTCCAC
ATTGCGAAGCACCCCGATTTCCCGGTCCCCTCTTGGTAAGAACCCGATGCCCGGACACCT
CTCGAACTCCACCCTGTGAATGAGGTCAACACGAAGGGGCAGTGCCCGCCCGTGCATCGT
CCGGAAAGAACCCCCAGGTTCCAAATACAGCTCGACAAGTGGCCTCTCTCCCCGGGGACA
CCTCGAGAGGCAAGCGGAGTTCCATGCCTCAACCCAAGACGAGGACCTGACTCTCCTGTC
CCCAGTCTGCAGGGACCCTGCGATTCGGAGTCTGAAATCAGAGGAACCCTGAGGTTCCTG
CCTCAACTGGAGATGGAGGCCCTCTTCCAATGCACCAAAACCCCGTGGAGTCCCGAGAGG
CCCCTCCCACCTCCAGTTCCGCTGATTCTCAGAGCCCACCATGAGAAGCCCCCTGAGGTC
ACCTGCACAAGTCGAGGGAAGCCAAGGGTTCCCTGCCTCAACCCGAGAAAGACCTCGAGA
GACCTTCTTCAACATCGTCTCGAGGCCAGGTTCCCTACCAGGACTCGAGGGCAATGACGC
GCTCCCCCTCGCCACTCGCGTGGAGACCCGACTTCCCTGGCGCCCCACGAGAGGCTCACT
GACCTCGCCGTCGCACCTCGTGAGAAACCGCACCCTGGGGCCGCCGGCTCGAGAACCACC
CGAGACTCCCCCGTCATCGCGGAGATGAGGGCCTTCGTCTCCTGCATGGGCCTAGAGCCA
ACCTCGCGACCTCTCTCCAAACGCCTCAGGAGGCTGACTCCCATTTGTCCACCCAGTGGA
GCTCAAGAGATACCCGTCGCGACTCGGAGGCAGAAAGTTGGGTTCGTTGCTTCCCCTCGA
GATGAATGCCTGTCTCCCCGGGTGCGTCTTGGAATGCCACCCCGAGTCCCGGTCGCCCCT
GGAGAGGAACCTCGGGCTTCTGGGCACAAGCCTAGATGAAGGTCTATCAGGGCCTGCAAG
TCACTCTGGAGCAATCCCCAGCTTTTCTTTCGCAACTCG
>Mus_musculus
AAAACTGAAAATCATGGAAAATGAGAAATATACACTTTAGGACGTGAAATATGGCGAG
>Mus_spretus
GTGTATATCAATGAGTTACAATGAGAAACCTGGAAAATGATAAAAACGACACTGTAGAAC
ATGTTAGATCAATGAGTTACACTGAAAAACAAATTCGTTGGAAACGGGATTTGTAGAACA
>Oryza_sativa
TGCCAATATTGGCATTAATTGACAAAAGTTCGCCGCGCGAATCACGAAGTGAGTTTTTGC
CACGAACGCACCCAATACACTCCAATATGTCCAAAAATCATGTTTTGGCCTTTTTGAACT
TTTTCATTCCGGTAAAAAACATCGCACCCACGTG
>Sus_scrofa
CGCTGCATGGAATCCACTGCATTCAATCACGCTGCATGGAATCCACTGCATTCAATCGGG
ATGCATTTTACCATGCTGCATGGAATCCACTGCATTCAATCACAATGCATGGAATCCACT
GCATTCAATCAGGATGCATTTAAGCCACTGCATTCAATCGCAATGCATTCACCACGCTGC
ATGGAATACACTGCATTCAATCACGCTGCATGGAATCCACTGCATTCAATCGGGATCCAT
TGCGCCA
>Canis_lupus_familiaris

GATCGGAAGAGCGTCGAGAATGATACGGCATACGAGATCGGTCTCGGCATTCCTGCTGAA
CCGCTCTTCCGATCTAATGATACGGCATACGAGATCGGTCTCGGCATTCCTGCTGAACCG
CTCTTCCGATCTTCAATGATACGGCATACGAGATCGGTCTCGGCATTCCTGCTGAACCGC
TCTTCCGATCAGATCGGAAGAGCGTCGTGTAGGGAAAGAGTGTAGATCTCGGTGGTCGCC
GTATCATTA
>Sacrophilus_harrishii
CTTCAATAGAAATCAAACCTTTTCCACGCATTTCTGTCCGCATTGGGTTCTAGATTATTT
TTACGATCTCCAATCCAGAGAACATTTGTTCTAGATTTTTTAGTCTCATTCATTTCCGG
AATCTTTTCTGGAA
>Gorilla_gorilla_graueri
AACCTTTCTTTTGATACAGCAGTTTGAAACACTCTTTTTGTAGAATCTGCAAGTGGATAT
TTGGATACTCTGAAGATTTCGTTGGAAACGGGAATATCTTCATATAAAATCTAGACAGAA
GCATTCTCAGAAACTTCTTTGTGATGTGTGTCCTCAACTAACAGAGTTG
>Mus_famulus
CACACTGTAGGACCTGGAATATGGCAAGAAAACCTGAAAATCGTGGAAAATGAGAAAAT
>Gorilla_gorilla_gorilla
GAAAGGTTAAACTCTGTGAGTGGAACACACACAACACAAAGAAGTTACTGAGAATGATTC
TCTCTAGTCATTAGACGAAGATAATCCCGTTTCCAACGAAAGCCCCAAAGAGCTCCAAAT
ATCCACTTGCAAACTCCACAAAAAGAGGGTTTCAAAACTGCTCTGTCAAAA
>Echinococcus_multilocularis
ACCAAAAAAACCACCAAATCTACTCGAACTACTCCCAAAGTGACTAAGCCTACCGTCACC
>Mus_caroli
TATGAGTGAGTTGCACTGAAAAACTTAGAATATAAGAAACGCACAGTGTAGTACATAGTA
>Hymenolepis_microstoma
ATCCCACACAACCACCTCAACCAAACACCACTTTTAAACCTTATTACACCCCAATACACC
TATTCTTGCCATTTGCACCACCAATTTCTCTTCCATTTAGCCAAAATCTTCGATTCTTCC
AAGTCAACCAATTTTTTCCACCTTCGCCAATTCCATCCCCATTTTCTCCCAC
>Nippostrongylus_brasiliensis
ATCACTTTGGCACTGTTCGCTGAGAAACAAAAATTTTCACTTGAACCGGTGTGTCTGTAG
ATGATCTTCATGGGTCTGCCCATGGAGAACATCTTCTGTAGGAGACTGTCCACGAAGTTC
AGCTCGTTGAGCAC
>Bos_taurus_inducus
AATGGAAGATTGAACTTGCCCTGGGCCAACCAAAAGGCATCCTGACTTCCCCGTCGTAAC
TCAGAATCCCCGCCCGTAACTCGAGAAAAACCACGTGGCTCCCCGTCCATCCGCAAGATG
AAGGCCCTTCCCGCAACAGCCGGCCCTCCAGGAGAAGTCCCACGTAGGATTGGAGGTCGA
AAGGGCACTTGGCGCCCTTGATGCGACCCACAAAGTTCCCCGAAATCCCCGGTCTCCCTC
GAGAGGAACACTGAGGCTTTCGCACCCCCTCCTCTGACCCTTTTCTCCCCCTCCTGAATC
TGGACAGGAGGTCGACTCCCCTGCTTGTCTGGAAGGGGTTCCGACCTTCCGTCCACCTCC
AGGAATGAGGCCGGTCTCAACGAAGACCATTCCAACGTGCCCCTCCGTGGGTGGTTCCCA
CATTCGTAGGACCCCGATTCCCGTCCCCTCCTGGATAAGAACCCGATGCCGGACCACCTC
TCCGAACTCCAACCCTGGAATGAAGTCAACACGAAGGGCCAATTTTTCCGTGCATCGTTC
AGAAAAACCCAGTTCCAAATACAGCTCGACAAGCCGGCTCTTCTTCCCCCGGGACCATCT
CGAGATGCAAGCGGAGTTCCATGCCTCAAACCAAAGACAGCCGACTCTCCTGTCCCAGTC
CTGCAGACCTGCGATCGGAAGTCTGAATCAAGGTACCCTGCGGTTCCGCCCCTCAACTGG
AGATGAGGCCTCTTCCAATGCACCAAAACCCAGTGGGGTCCCGAGAGGCCCTTCCCACCT
CCAGGTTCCCTGCTTCTCAGAGCCACCATGGAGAAGCCCCTGAGGTCACCTGCAACAAGT
CGAGGAACCCAGGGTTTCCTGCCTCAACCGAAAAGACCTCGAGAGACCTTCTCAACACGT
CTCTGAGGCCAATCCCTAACATGGCTCGGGAATCCAGTGACGCGCTCCCCCCCACCACTT
CGCACTGGAGAACCCGACTTCCCTGGCACCCCACCAGAGGCCTCACTGACCTCGCCGCGT
ACCTCCAGTGAAGAAAACACCACCGGGGCCGCCGCTCGAAGAACAAACCCCCGAGAATTC
CACCCTCATCGAGAGATGAGGGCCCTCCGCCCCCTCAGGCCTAGAAGGCCCAATCCTTCC
GCGAACCTCTCCAAAAACGCCTCAGAGCCTGACTCCCTTCGCGAGTCCCACCCAGTGGAG
CCCAAGAGATGAACCGTCGCCTGATTCGAGAGCCAGAGCGGCCCTCTTTGCTTCCACTTC
CGAGGTGAAATGCCTTGTCTCCCCGGTGGTCTGAATGCACCCCGAGATCCTGTCGCCCTG
AGAGAAAACAATTGCTTCTGCACAAAGCCTAGAATGAGTCCTATTGGCCCTCCAGTACAC
TCGATGAGCAATCCCCCAGCTTTCCCTTCGCAACTCCA
>Danio_albolineatus

TTTAAGCAAATTCTGTCACTTGCTCTTTTTGCACTTACCAAGCTCTGAAAAGCAAGTTGC
AGAATTCTAGGAATTGACAATTATGAAACATTATCCTTTTGCTTAATTTTCACTTGCAAG
TTGTTAAAACATGCTTAATACAGCATGCACTGTTGTTTTTAAACTTAAAGAATGATGATT
TTCACTGT
>Danio_nigrofasciatus
GTTTAGAGGGCCCAGAACAGAGAGTGATCAAAACAAAAAGAAATGATGAAAACCATCATT
CTTTTTGTTTTGATCACTAACAGTACAGGCAACTCTAAACACGCTTTTCAGAGCAAGAGA
AAGCTGAAAATAAACAGTATGGCAAACTCAGCCAGATCTGCTGATTGTCACTTGCAACTT
ATTAAAACAC
>Echinococcus_granulosus
CACCACTCAAAAATAACAAAAACAACCAACAACTAAAAATAAGTAAACCTAAAACAATCC
AAACTAATCATAGCACAAAAAAACCCACATTCATAAGAACTA
>Caenorhabditis_elegans
AAACTGAGAATTATGGAGAATATGGAAATTCAAATGTGCACTGGTTTTCAAAATTTTTT
>Ailuropoda_melanoleuca
TGAGTAAGGGAGCATGAATGAAAAAGTTCAAGATTTTCGAGGACTCAGCGCCCTTCTCAT
TGGTAATAAAGCAGAGCGTCCAGGAAAACACCTCCAGCGAACAGCAGCCTTTTCACTGGG
GACCCCGATTCCACATTGGGTGTAGGGGCCTGGGGAGGATCCAGCCCTTAGAACCAAGAA
ACAAAGTGTGAGGGGAAAGCCCTTGAAGTCACGACGGCCTTATGGAAGGTGGAACAGGAA
TCAAGGTGCTTTGGTACCGTGAGTGAAACACGACGAACACAGCACTGAGAAAGACAATTC
GGCCAAAACCCGCTTCCAACGAAATGTTTGCTCCCTGGGGAAATCCTGTCAAAACTGCTG
GCGGCAATGGAACGAAAAGGCCCAAAACACGTCCGTCAGAAACAACACCGTACAGGGACC
GTCTTAGCACTATCATCACCCTTCGTGAATAGGGAAACATGAGGGAAAGTGGTCGCGGAA
GCCTGACGAGAGTGCCCCATAGTGGTCTCTGAGACACCACACAGTCCAGGGAAATATCAA
CGATCATCATCTCTCG
>Bombyx_mori
AGTGACCCCCGATTCCCTACA
>Glycine_soja
TTTCTCGAGAGCTTCCGTTGTTCAATTTCGAGCGTCTCGATATATTATGCGCCTGAATCG
GACATCCGAGTGAAAAGTTATGACCATTTGAA
>Drosophila_melanogaster
TATTC
>Drosophila_albomicans
AAAATTTTGGAAAATAAAAAAATTAAAAAATTGGG
>Bombyx_mandarina
CCACAGTGTACGTCCACCACATAAGTA
>Bubalus_bubalis
CCCAGCAAGCGCCTGCCAGACTGGGCTTCCCTGGAGCTCTCCCCCAGCGTAGAGTCGTGT
GGATGTGGTTTGGGGCCTTCCCCGGTGCGTTTGGGCTAGGGAGAGCACAGCCCCAGCGCA
GGGGAACCAGTCCCTGAGCAGGGAGCTCCGGGCAAGTGCTCTGTGCCCAGGCCTGGAAGC
CCGGTTCCCAGCACGCCGGAGAGAAAGAGAGGGGTTGCGGGCCTGTTACTGGGGAGGACC
GAGCGAGGCTGCCAGAGCGGCGGAGCTCCAGCTCTGCGCGCCCAGGCGCGATAGCGCAGA
GCCCCGGCCGAGAGGCTGCTCTGGCCTGGGCGAGCCTCGCCCAGCGCTCCCCAGCTTCTA
GCGTTCTCGCACCGGGAAGGAGCAGGCTGCGGGAGTGGGGGAGACGTGCCTGGCGCCTGG
CTGGCCTCAGCTCCGCTCTGGCCTTCCCTTTGGACCTAGGAGCCAGGCCTCTTTGCAGAA
GGCTCAGGGGCCGGGGGTGTGCAAACTGCCAAAGCCTCCCGCAAGCAGAGCTCTTTGTCA
GGCAAAGAGAGCGCCCAGGGCTGCATTCCCAAGCGGTTGCTTGGCCGCCCAGACCTAGCG
CTGGGCGGAGGGAATCAGGGTCCAGGGCCCCTGGCTGGGGGAGGCAGCACCACAGCCAGA
GGGAAACTCCGTTTGTG
>Onchocerca_flexuosa
AGGGGAGAAGAATACGAGGAGCAAGAAAAAAAAGCAAAATACAAAACGAGGAGAACACAA
GGGGCGAAAATCAGAGGAGGATCAAAAGTATGCCACTACGTGGAGTAAGAGAAGTGTGAA
AAACCCGAGGCGAAGAAAC
>Ficus_fistulosa
AAAACAAGCTTTGAGAATGATCAAAACACAAAAAAACCAGACCCAATCTGGTTTGGGTCA
AAAAAGTGGCAAAATTAAGCCAGAATCAAGCATAAAGGGACCAAAATGAGCAAATCTCAG
TTCAAGGCAAAGTGGATCATCTAGGCAGCCCCTAAATGACC

>Capra_hircus
TCCCATCGAGGCTTGCCCACGGGGCCTCTCGGGATTCCTCTCCCGTCGATGCCGGGGCCT
AAGACCTTGTGTGGAGTCGGTGCCGGAACCTGAGGATTCCTCTCCAGTGCAGACAGGGAT
AATGGGGTTCCCTGGAGTCGCCTCAGGGGATCAGGCCTCCTCTCGAATGGTGGCACGAAC
CTGGAGTTCCTCTCGCCTTTCCTGTGGAGAGCGCCTCCTCTTGCGATGCGACGGGAATCC
CGGGAATTCTTTCCCTCCACGCAGGCACACGAGCCCTCCCCACGAGCTTACAAGGCGCAA
ACGGGCCACCTCTGGATGTGAGCGCGACCCTCGTGCTTAAACTCGAGTGCAGACAGGGAA
AACAGGGAACTCCTGGAATGGCAGCAAACATCTGAAGGACCCTTGGGAAGGTCCACACGT
AACCTGTGAGTAGCCTCGAGACGCCCCAGCGGAAATGCGCCTCATCTCGACAGGAGACGA
GAACCGCCGGGATTTTCCCGACACGCGGAAGGTCCTCTCGACCTACGACGCGGACAACAG
GGACCCGCTCTGGTGGCCGCAGGAAACGCCAGTCCCCATGCGAGTTGCACACGGGCATCT
CGGGAAACCTCTCCAGTCGAAGCAAGGGCCTAAGACCCTGTGTGAAGTCCCAGACGGAAC
CTGTGATTACCCTCCAGCGCTCACACGGATAATGGGCCACATCTGAGTCTCCCCAGGGGA
GAAACTCCTCGTCTCGAGTGGCGGCATGCGTGCGCTCGACTCACGAGCGGGAACAGCAGG
GACACGCTTCCCGTCGCCTGCAGCAAAGGACCAGTGC
>Lithocarpus_balansea
GAAAAATATTTCCCAGCCAAATTCGTGAGCAAAACTTCAAACTTCCCCTTTTGAGCCCTT
ACATCGGCACGAAATTGGAATCCAAATTTCGGGTAACTCGGTGGCCCACATTCAAACGAG
GATAACTCTCCCGATTTTTATCAAAAAAATACAAAATTTGTGCTCAAATTCAAGCTCAGG
ATGTCTACTTTCTAAAACACGAAGGCCGCGTCAAAGAATTCCTCACGGTTCAAAAGTTAT
TGACGAAACGGTGGCCAAAGCACACTTTTCGTCACACTTCCTAACCGATTAACGCCCGCG
GATAACTACCCACTCACGTATTTTTTCCCCGAAACTTGATTTTGGGAAGATTGATCTCGC
GATATAG
>Lithocarpus_grandilofolius
CCGAAACCCAAAAAATGACCAAAATACCCCCGAAACCCAAAAAATGACCAAAATACCCCC
GAAACCTAAAAATGACCAAAATACCCTCGAAACAAAAAAATTACCAAAATACCT
>Lithocarpus_hancei
AGACTTCTCCTTTTGAGCCCTTATATCGGCACAAAATTGGAATCCAACTTTCGGGTAACT
CAGTGGCCCATATTCAAACGATAATAACTCTCATGATTTTTGTCGAAAAAATACAAAATT
TGTGTTCAAATTCAAGCTCAGGACGTCTACTTTCTAAAACACTTAGGCCGCGTTAAAGAA
TTCCTTTCGGTTCAAAAGTTATTGACGAAACGGTGGCCAAAGGTCACTTTTCGTCTCATT
TCCTAACCGATTAACGCTCGTGGATAACAACCCACTCACATATTTTTTCCCCGAAACTTG
ATTTTGGGAAGATTGATCTCGCGATGTAGGAAAAATATTTCATGCACAAATTCGTGAGCA
AAACCAC
>Lithocarpus_xylocarpus
TTTACGCCTTTATATCGGCACAAAATTGGAGTCCAAATTTCGGGGAACTCGGTGGCCCAT
ATTCAAACGACAATAACTCTCTTGATTTTTATCGAAAAAACACCAAATTTGTGTTCAAAT
TCAAGCTCAGAATGTCTACTTTCTAAAATATTAAGGCCGCGTTGAAGAATTCCCTACGGT
TCAAAGTTATTGTCGAAAGGGCGACCAAAGGTCACTTTTCGACACATTTCCAAACCGATG
AAGGCTCGTGAATAACTACCCACCAAAGTATTTTTTACACAAAACTTCACTTTGAGAGGA
TTGACCTCAAGATATAGAAAAAATATTTCATCGCCAAATCCGAGAAAAAAACCTTAGACA
TCAATG
>Lithocarpus_calolepis
TTTCGGGGGTATTTTGGTCATTTTTTGGGTTTCGGGGGTATTTTGGTCATTTTTTGGGTT
TCGGGAGTATTTTGGTCATTTTTAGGTTTCGGGGGTTATTTTGGTAATTTTTAGG
>Castanopsis_echinocarpa
AAAATTTTAACACATAATTCCATTATTAAATTAAATGTTAATGTCAACCTCGGCTAAGCC
AGGTCCTCGGACCTTCTGCCTTGGGAAAATTTAACATTCAATTTATTCATTTTTTAAGTG
TTAAATGTCAACTGGGGCTACGCCAGGTCCTTATACCTTCTGCCTTAGG
>Castanopsis_indica
GCTGCCAAGGCCAAGGCATGCAGCCAAGGCCAAGGCAGCCCCCCATGGGCGT
>Ficus_altissima
TATATATCAATTTCTAAAGCCTATAGCCTTCTCTAGCAACAGAAAATTGGTCGGGCTCAA
AATTTCTACAAATTTAAATCATTAAAACATGCTCTAAACATAATTCTAACGATCCACTTT
AAGTCGAAAAACACCTTCCGGATCACGATTGACGATACTCA
>Ficus_langkokensis
TTTTGCTGGTTTCTTGACCCAAAGTTGTTAGGATCTGGTTTTTTGTGTTTTGATCCTTGT

CAAACCTTGTTTAGGTCATTTACAGGCCACCTCACTGGTTTACACGACCTTGATCTGATT
ACATTGCATTTTGGGTCACTTTTTGCCTCATTCGGGCTTGA
>Trigonobalanus_doichangensis
GGACAGACCGAATAAGAGTGAAAACAAAATTTCGCACCGGGATCAGTTCTGCCCAAACCG
TGCCAGACATTCCTTAATAAATATTAAAAAATAGAAATAAAATTGAAAAAATTAAAAAAT
ATAATACTTGCATCATTCCTGGCCAAACCGGGCCG
>Myzus_persicae
TCATTTGGTTGAATAGAACCGGAGAACGAGAACTTTGAATATCCCGAAATTCGATTAATT
TTACCCATTTTCAAAATTGTCTCACGACCTGTGTATTTGGGCTAACCAAGTCAGAGTGGT
ATCAAAAGAAAGCTATTAACTTCCTCTTGAGTTTGGTGAAGGTTTGAAG
>Ficus_microcarpa
AAAAGAATATGCAGATGAGGAAAATAGAGGAGGTTAAAAAAAAAAAAAAAAAAAAAAAAA
AAAAAAAAAAAAAAAAAAAAATGGTTGTTGTCCGAAATCTTCAATACTGTAGGGTTTAACT
AAGTTTACTAGGTACTGCTAGAAGACGAAATTTACTTCGGACTACTACTCCTTTTTTTTT
TTATCGTTCGGACCGGTGCTGCAAACTCTGTTAAAAAAAAAATCCAGACCATC
>Ficus_tinctoria
TACCATTTTTAGAACACGGTCAATCCCGTTATGACCGTCTGTTGGGAAGAATCGAGAAAA
ATACAAAAAAAATCATTTCGGGCCATTTTTAACACACGGTCAAAACCGTTATGACCGTCT
GGTCATAACAGTTATGACCACAAATGGAAAAAAATCGAAAAAAATACAAAAAAAATCATT
TCGGGCCATTTTTAACACACGGTCAATCCCGTTATGACCGTCTGGTCATAACAGTTATGA
CCACAAATGGAAAAAAGCGAAAAAAACTGGAAAAAATCTATTTT
>Anas_platythynchos
ACAACTGAACTCCCCAGGGGAGCCTTTACGCAGCCCCACATGCAACTGGACCCTCCGGGT
TCGCTCTTTCACAGCAACCCTAAATATAGCCCTAGCCGCATGGCAGAAGACTGAAAAATC
GGGCCCCAGAAAACGCGTTGCAAGCCCCTTTCGCCCATCCAAAAGTGCCCCCAAAGGAAG
GCTTTGGGT
>Harpegnathos_saltator
TATCGAAGGTTAAAGTTTTCCAAGCTTTCTCCGTCGATATCTCCGCTTCTACTGGGAGTA
GAGAGTCGTCACTTACTCCAAAATGTAGGGTTTTGAAAGGACTTTAATTTGGTACCAGTA
CCAAGCCTGTATCTCTAACCGTTTAGGAGA
>Camponotus_floridanus
AAGCATTCATTTCTAGTTAGTTCAGTAGTTCGCATCATCGCATGATATCCCGGAAAAAAG
CGTTTAATTCTCGTTTGTACGTCGTATAGGCGTGCGTTATTCGGATTAA
>Linepithema_humile
TTCTTCGTATGTTTCTGGTCTCAAATTAATTTATAACTTCAGTACTATCGAAATATATAG
GAAAATGAATTTGTTGCTCGTTTTAAAGTATCGAGAAGCGACACTGTCGTAACAAATTCA
TACAGCACACATTAGAAAAGTTATAATAGTAAAATAATTT
>Carica_papaya
GTCCGCAGGTTTATTCGCAAATGTGAGTCACGGCGCCGGGGCATTTTACTTGATATTAGA
ACGAGAACGAGAGGTGCAGTGTGCACGCACAAGGAAACAAATGACGGGAGGGGGAAGAAA
GGAGAAGGAGTGTATGTATTTGCAGCGCGAAATTAAACAGGTGCGATCATACCAGCACTA
ATGCACCGGATCCCATCAGAACTCCGCAGTTAAGTGTGCTTGGGCGAGAGTAGTACTCGG
ATGGGTGACCTCCTGGGAAGTCCTCGTGTTGCACCAATTTTTTTTCAGATGGGTCGTTCG
ACGTTACGCGCGACTGGAATTATG
>Boechera_holboellii
AGAACCTTGTTCTAGAAGGATAATTAGTAAGAACCTTGTCCGAGAAGGATAACTAGTGAG
AACCTTGTTCTAGAAGGATAACTAGTG
>Mimulus_laciniatus
ACATAAGTAAAAGTACAAAAGCATGCACAAAAGTACTCACGCAAAAGTGCATTCGCAAAA
AGGATCATGAAAACACAAACTACAATCCTAGGACATTAATACAAATCAACAAACATGCAC
AAACACACTCCCTCAAATGGACATGCACAAACGTACATGTAAAAAAACATGCGCAAAAGT
GCATCCGCAACAACGTTCTTAAAAACACGAACTACAATACTATGACTTTAATACGCATCA
ACAAACAACAGCAAAAACACTAGATAAAAAAGATACAATGGAGTGTAGAAATAAAACTCG
ACGTTTACGTTTACGTGGATTTTTGTTTGGATTAATTTATATTTCTGTTTTAATTGGTGG
ATTTGGTATCTTATAGGTACAAATCGAAAATTAGGTGTCCAGGAAGCAAAAATGATCAAA
GGGTAAGAAAAGAGTTGCAATTAAGAAAATAAAATCAAATTAGAAGAGAGAATCCCTCAA
CATCTTTGGTAGTGTCCACGAAGTCGAGCATAACTCTCTCATCCGGACTCCAAATCAACT

GGTTCCGGCGGCATTAGAAAGCTATTTCAGCGGGCTACAATTCCCGTCTAACGTCAAAAT
TCCAAATTCGGACTCGAACATGGTCAAAATTGGATAACAAAAGCGATCATGAAAATACGT
ACTACAATTCAAGGACATGAATGCGCATCAAGAAACGTGCACAAAAACACTCGCTAGAAA
AGATACGC
>Mimulus_guttatus
TTTCTTAATTGCAACTCTTTTCTTACCCTTTGATCATTTTTGCTTCCTGGACACCTAATT
TTCGATTTGTACCTAAAAGATACCAAATCCACCAATAAAAACAGAAATATAAATTAATCC
AAACAAAAATCCACGTAAACGTAAACGTTGAGTGTTATTTCTACACTCCACTGTATCTTT
TGTATCTAGTGTTTTTGCTGTTGTTTGGTTATGCGTATTAAAGTCCTAGTATTGTAGTTC
GTGTTTTTAAGAACGTTGTTGCGAATGCACTTTTGCGCATGTTTTTTACATGTACGTTT
GTGCATGTCCATTTGAGCGAGTGTGTTTGTGCATGTTTGTTGATTTGTATTAATGTCCTA
GGATTGTAGTTTGTGTTTTCATGATCCTTTTTGCGAATGCACTTTTGCGTGAGTACTTTT
GTGCATGCTTTTGAACTTTTACTCATGTGCGTATCTTTTCTAGCGAGTGTTTTTGTGCAC
GTTTCTTGATGCGCATTCATGTCCTCGAACTGTAGCACGTATTTTCATGATCGCTTTTGG
AATCCAATTTTGACCATGTTCGAGTCCGAATTTGGAATTTTGACGTTAGACAGGAATTGT
AGCCAACTGAAATAGCTTTCTAATGCCACCGGAACCAGTCGATTTGGAGTCCGGATGAGA
AAGTTATG
>Mimulus_dentilobus
ATGAGAAAGTTATGCTCGACTTCGTGGACACTACCAAAGATGTTGAGGGATTCTCTCTTC
TAATTTGATTTTATTTTCCTAATTGCAACTCCTTTCTTACTCTTTGATCATTTTTGCTTC
CTCGACACCTAATTTTCGATTTGTACCTATAAGATACCAAATCCACCAATTAAAACAGAA
ATATAAATTAATCCAAAAAAAAATCCACGTAAACGTAAACGTAGAGTGTTATTTCTACAC
TCCACTGTATCTTTTTTATCTAGTGTTTTTGCTGTTGTTTGTTGATGCGTATTAAAGTCC
TAGTATTGTAGTTCGTGATTTTAAGAACGTTGTTGCGAATGCACTTTGGCGCATGTTTTT
TTACATGTACGTTTGTGCATGTCCATTTGAGCGAGTGTGTTTGTGCATGTTTGTTGATTT
GTATTAATGTCCTAGGATTGTAGTTTGTGTTTTCATGATCCTTTTTGCGAATGCACTTTT
GCGTGAGTACTTTTATGCATGTTTTTGTACTTTTACTTATGTGCGTATCTTTTCTAGCGA
GTGTTTTTGTGCACGTTTCATGATGCGCATTCATGTCCTCGAACTGTAGTACGTATTTTC
ATGATCGCTTTTGGAATCCAATTTTGACCATGTTCGAGTCCGAATTTGGAAATTTGCCAT
TAGACGGGAATTGTAGCCAACTGAAATAGCTTTCTAATGCCGCCGGAACCAGGTCATTTG
GAGTCCGG
>Sorghum_propinquum
TCAAATGGAAGCTCGCTTTGGTCTGTTTGGAGACAGTGCTAATCTCGATGCAAGATAGGT
GCACGGTTTGCATGGAACATACCATATGCTAAGAAATCAATTTGGACGCACCCGATGGAA
CTCCTAGATGACGTGTG
>Panicum_hallii
AAAGTCGATGCAAAACTGGCCGAACTGGTGCCATTAACGCACAAGTTCGCTAAACAAAGT
CGCGTCGGAATTTTTCGCAACGAACGCACCCGATCCACTCCATTGGACCCAAAACTCATG
TTTTGGGGCGTTTCGGACCGTTTCGTTACTGCACG
>Staria_viridis
TTGCGAAAAAATCCACCCGAGTTTCGCTACCCGGCAATAGTGCATTCGGGTGCCGAAAT
GCACCCGTTTTGCATCATTTTTCGTGCCGGAACCGAATGCCCAAAAACACTCCCAAACAT
GTTCTAGGGTATAATTAGGAAGATTGCATGCGTTC
>Phaseolus_vulgaris
TAAAAGAAGTATTAAACATATCAAAACACAATTTTAACACATGAAATTGTTTTTCAAAGA
TATCATAAGTGTTTCAATCAATTAATAGACTTAAATTGTAAAAAACATATCAAAACACAA
TTTTAACACATGAAATTGTTTTCAAAGATATTCATAAGTTTTAAAACAATTAATAGACTA
ACATTG
>Arabidopsis_thaliana
GGTTGGTTAGTGTTTTGGAGTCGAATATGACTTGATGTCATGTGTATGATTGAGTATAAG
AACTTAAACCGCAACCCGATCTTAAAAGCCTAAGTAGTGTTTCCTTGTTAGAAGACACAA
AGCCAAAGACTCATATGGACTTTGGCTACACCATGAAAGCTTTGAGAAGCAAGAAGAA
>Cajanus_cajan
ACTTGCTACACCTGGGGAGACTAATAACCAACACAGATGCACAACATAGCATGTAATTGG
TTTACTGTTCATTGGTTCTCTCTAATTCTTCAACTGACTT
>Lemna_gibba
CACACGCACGTATGCACACACATACGCACGCACACACGCACGTACGCACGCACGTACACA

CGTA
>Medicago_truncatula
ATAAGGTCATTTTGAACGGTCGGATTGAACGTGGCTGGTGTCGTTCACGATCTAGGCACG
TTTAGGTCCCCGCAGTGAACTAGTTCCTAAGTTGACTAGTCAATTAGGTGATAGTTCGTC
CGGATGACGTACCTCCGTGAACCCGATCTGAGAAATTCAAGTTTCTGCATCCTTCTATGT
TTG
>Solanum_pimpinellifolium
CCCCAAATGCATCAAACCAACCAATGGGAGACCTACATCTCCTACCATACAATGCTTCAA
ATGGAGCCATATCAATGCTTGAGTGATAGCTATTATTGTATGAAAACTCCGCTAAGGGTA
GGAAGCTATCCCAATGACCACCAAACTCTATCACACACGCACGAAGCATATCCTCCAACA
CTTGAATCGTTCTCCCAGACTGACCATCGGTCTGGGGATGGAACGCAGGACCGCCCCCCC
CCCAAACCCGACCCACCCCCCCCCACCC
>Neolamprologus_brichardi
CGGGCCGCCTGGCCACCCTTTATTCGGGTGTCCGCTCCCCAGGCTCCGAAAGCCTTGCCA
GCCCTTGTGATCGGGTGGCACCGCCCACCAGGCGTGTGAAAAAAAGAAGAAAAAAACCCGG
AC
>Mustela_putorius_furo
AGGTTCAGTTCAGGCTTACGGTTAGGGTCAGAGTTAGGTTCAGGGTCAGGGTCAGCGTTA
GTAGGTTCAGGTTCTGTTTTATGAGTAGGTTCAGGATCAGGGTTAGCTTTAGGATT
>Pundamilia_nyererei
CAGTGAGAAACGCACTGTCTTGGCGAAATAAAGCGTTTTTGTACAACTTCATATAAATCG
CTGTAACTTTTGATAGAAGACTCAGACAAACATGTTTATGGCTTTATCTTATAGAACTCA
ATATCCCCGTGCTGGGCAAACAGGTTTTGCAGCCGTTTGAGCTAAGATTTTAAAATTATT
CACATAATGAAAACCTATACTTTGTTTCAGGCGAGTTTCCCATTCAAATGCATGT
>Zea_luxurians
TCTCCACCAGAAATCCAAGAATGTGATCTATGGCAAGGAAACATATGTGGGGTGAGGTGT
ATGAGCCTCTGGTCGATGATCAATGGCCACACAACCCCCATTTTTGTCAAAAATAGCCAT
GAACGACCATTTACGGATTTTTGACCAAGAAATGG
>Theobroma_cacao
AAAAAATGTAAAAAAATTCAACTTAAAATCAATAATGAGATTATAAAACCATTTAAAGCG
AGTAATAAGGCTTACAAAGGCATTGGTGACTATAAATGGACTTTTCTTGTCAATGGATGG
GATTCGCTTCACGTTTCTCGAGGGAAACATCCAATTATAATCTGTTTTAAC
>Salix_purpurea
AAATCCAAGGCTAACACCCCCTCCAAAAACCAAGTTCAACCATTTTGGGGGGAAAAGTCT
ATTACCCATAAATTTGAAGTCTAATACCAACTAAAAGGGATTGGATATGAACAAGGTTAG
AAATCTAAGGCTAATACTCATTTAAAATGGGGCTCGTCGAGTCAGGACCAATTTGTCCG
>Branta_leucopsis
CAGGGAAAAAAAAAAAAAAAAAAAAAAAAAAAAAAAAAAAAAAAAAAAACGAAGCAAAG
>Aegilops_tauschii
TTCTGACATCATTTGTTATTTTTCAGGCATTTACCGAATTATTTAAGAGCTAAAAGACCC
TAAAATTGAAAAGCACTACAAAATGAACTCTGAAAAGGTTGAAAGTTGGCATGGTATCAT
CATTTCATCCACATAGCATGTGCAAGAAAGTTGAGAGGGTTACGGCAAAAACTGGATGCA
CTTCGTGTACAAAACGGACAATCTCTTTCAAAGTATCAGGATTTCATACGGAAACTCGTC
TGTTACAAAGGGATTTCATTTTTTAAAACTTATTTGAACTCCTGACTTTTTGTGTGTTCA
AAATGCACCATTCAAAGCCACATCATCATTTTTCAATCCT
>Lottia_digitalis
CAATAGGGTTAATCCTTGGTTGATTACGAACACTCACACCAAGTTTCGTCAAAATCCGTC
AAGAATTGTGCTTTCTAGAGTGTTTACGAGCTAAAAATTGAACGCGTTTTCAAGAACAAG
GGCAATAATTCATACTAAATATTTTTCGATTTCGCTGATTTT
>Gallus_gallus
TGATTTTCGGGTTAAATGGGGGATTCTTGAAGAGAAAATGCA
>Mayetiola_hordei
AACTTCATTTCTCACTGATCTAATGACATCGGTTTTGTAATTCTAGACATTTTTCAAGGC
AAATGCTGATTGGTTATCTCTCAGAGAATCATTTCTGAATTATGTATATAGCTATCGGCG
ACCCTGAATTTCCAATCCAATCATATATTGGTCATAGATTTGCTTCAAAAGTTTTGAAAA
TATTCGAAAAACTTCATTTCTCACTGATCTAATGACATCGGTTTTGTAATTCTAGACATT
TTTCAAGGCAAATGCTGATTGGTTATCTCTCAAGAATATAGCCATCAGCGATTCTGAATT

TCCAATCCAATCATATATTGGTCATAGATTTGCTTCAAAAGTTTTGAAAATATTCGAAA
>Eucalyptus_globulus
CAAAAAATGGCCGGAAAGGGAATTTTCTATGGAAAAACAGGGAATTTTCTCTGTGGAATC
TTCGTTCTCTTGTTTCGCCGGGCGTTCAGGGTGCTGTGATGTCCAAAATGATGGGTCGGA
CTGATCTAATGCTTTTGGAAGCCATGTTTCCCAACAGATTTCCAACTGTCGGGGTTTTTC
AGC
>Lottia_scutum
CAAAAAATGGCCGGAAAGGGAATTTTCTATGGAAAAACAGGGAATTTTCTCTGTGGAATC
TTCGTTCTCTTGTTTCGCCGGGCGTTCAGGGTGCTGTGATGTCCAAAATGATGGGTCGGA
CTGATCTAATGCTTTTGGAAGCCATGTTTCCCAACAGATTTCCAACTGTCGGGGTTTTTC
AGC
>Chinchilla_lanigera
CTAAGTTCACTATTCCCAGGAAGCACTCACTGTAGGAAACCTATACCACCCACAGGAAAC
ATTCTGCCTTGCTCAGGTAAGGGGGGATTCCACACATTCTAAGTTGACTATTCCCAGGAA
GCACTCACTGTAGGAAACCTATACCACCCACAGGAAACATTCTGCCTTGCTCAGGTAAGG
GGGGAAACGCACTCACTCTTGTGCTCATAGGCGCCTATGTGAAAGGAATGCCATGGGTGC
ATTGCTCTGTGAAAACAAGCAGGGAGAGCATAGCCAATGTAGGCCCATACTTAGTTTCAC
TAGGGAAGCTCACCACAGCACTTAGAAGTTTGGGAGTGTTTCCGGGAGTTTGGGGCAGAA
AGTTGTTCTTTTCGTTCTAAGTGGGATTTACTGTTTTCTAAGGGAGCATGTATAGGACTG
GTGTGCGGTGTATTCCACCCATT
>Trichechus_manatus_latirostris
TGCTATCACAAATGTTCTCTCAGTGGGAGAAGGCAGTAGAGAGAAGGCCTAGCAGGAATT
GGAATCCCTGTTTCAAAGGGAACTTTGCATAACAGGAATGAAGGTATTGCTAGCTTACAT
AGCCAAACAGGCAGATTGTTTGCCTTCAGCTAGAGAACAGCAGAAGAGGTTCAGTTTTGC
TAACTAAACCTTGTGTGCTTAACACGAGCATGTAGCTTCTTCCCTGCCTGAAAGAGAAAG
CAAGAGCTTGCTGCTCTCAGTCTGGGGATGAAGGGAAAAAGATTCACTTTCTGTGTTTGA
AAATCCCTCTGTGTTACATGCCTAAGAAAGTTGCCTGTGCATTGTCTACCTGGATAAAGT
GTATGACTTTTCTCCCTTTCAAGTAGTTGAAGCCGTTCAGGGTTTGCAGCATTCCGGGAG
GCATTCTGTGCTGCGCGGCTGAGAATACCGCTTTACAGACGGTATCGCTAAAGGAGAAGA
GAAACCGCGTTTCTGCCTTTCTGGAACACGAATGAGGCAAAACCGTAGTTGTGCTGTGTA
AAAACTTTGTCAGCAGCAAGTGTAAGTTCCCCATTCACCACATGAAAGAGAAAGCAAAAT
ACAGCGTTCTCCTTCATGCTGCTTAACTGAATCTACTGAAGTTTTCAGTTTTGTTAGAAG
AAGCAGTCCGCCCACCACAAGC
>Leptonychotus_weddellii
AGAAATATAGAAGATAAATGAGGCAGAAGAGAGAAGC
>Saimiri_boliviensis
ATTCCAGAGTGGATGGGAGAACAGGCTGTAAAGAGAAATACCTTGCTCTAAAACCAAAAC
GGAGCTATCTAGCAGAATGGTTGTCAATGTGTGTATTCAACTTCCAGAGTTAAACTGATG
TGTGTTTGCAGCAGTTCAGAAACCCTTTCTTTGGAGAATGCAGAAACAGTCATTTCCAGC
CCTAGAGACGCATATAAGAGTTAGCAGTAAGAATGCGTTTCAAACACGGAAAAAAGATGT
GTGTAAGCACGCAGTTAAGAAACCCTTTCTTAGGAGAATGTAGAAACAGTCTCAGCGAGG
TAGAAACACGTTTCTTCTAAAAGCTGCGTTTGGAC
>Cricetulus_griseus
TTAGGG
>Drosophila_bipectinata
CCTGCCGAGGCGTGGAGTGCCCGGCCGATCCCCTGCCAGCTGATACCCCGCCAATACGTG
AAGTGCCTGGCCGATCTCCCTGCCAAGGCCTGGAGTGCCCGGCCGATCCCCCTGCCAAGG
CGTGGAGTGCCCGGCCGATTCC
>Drosophila_elegans
GTTTAGTTCTTCGACATATAGCAATGGTTAAATATTTCAGAATTACGGTTTAAATTTCAT
CAAAATCGGACGACTATATCATATAGCTCCCATAGGAACAATCGAAAAATAAATGAAAAA
AATTATAACTTTTCTGTTTTTAACTTTTT
>Drosophila_ficusphila
TTTTCGAATTTTTGGTCAATATTTTGTAATTTTTTATGACCCCCGACCTGTCAAAATTTG
CAAAAAATGGGTTTGCAGAAAAGTGACCAGATCCCAGCACTGCTTAGCCCATAACTTTTG
AAATTTTTACCCGATTTAAAAGTGGAATACCTCTCTGAATTTGTTATTAAAATATCTATC
TAGCTGCATTATTGGTTTA

>Drosophila_kikkawai
ATACTCAATTGTAAATTTGTAAACAAAAAGTTCACAACTCAGCTAATAATGCATGTATCC
ACTTCAGAATTTGTATTTATATTAGTTTTTACAAGGCTAATACATTTGC
>Drosophila_rhopaloa
TTTCCGAGGACTTGTAACAGAGTGACTCAAAATTGGACTTAGAGAGTCCATAACTTTACC
AAACTTAAGCCGATTCCAAAGTGGCATACCTCTAAAGACTTCTGATTCGATTCTCTAAAA
ATCTGCATTCAAATTTTTCAATTTAACGTTGATCAAATTTTTGACCCATTTTCATGTCAG
TTATTTCTAA
>Drosophila_takahashii
ATTGGAATGTCCGATTTAAGCCATAATTGCTGAGTTTATAGGTCTTGATGAGTAGAACAA
AGTTGCCATACAATCCTTCTTTCTATCTCTTGTAGTTTTTCAAAACCGTCTTATATCTGA
AAATAAGGAAAAACATGGTAACGGATGAAAATGAAAAGACAAGTAGCTAAGACGGCCAAT
AACTTTTG
>Drosophila_eugracilis
GGTTTACCATTTTCCTCCTTTAAGTGACAGTATTATGGCTATAACTAAGCCAATAGTAGC
CGAATCTGGAATGGCATAGTCGTTGGATTCTTTATTAATTTAAACTT
>Drosophila_biarmipes
TTTGCCATCGATCTGCATCATTTAAAAGAAAATTTATTTTTTGACCCACTTTTTCATTTT
TTACAAGGAGGTGTGTCTTAATTTTTTAGAATTCGGATTTTAGCCAAAAAACACATTTTC
TTGCGGTTTTTCAATCGTAGTTTATCCAATTTAAGGAGTACAGCTAAAAGACCGACTGTT
TTGGAAAGCTGGCTTAA
>Boechera_stricta
ATTTCATGAACTTATTTTCCTCCTTCTCGATTCTTATGAGTTATCGCTGTTTTCAATGGA
AGGTCACGAAATCTCGAAACGCAATGTCATAATAAAGTTTGATATATTGAAACTTACACC
ACTTATGAGCATATTTATCCCTTAATTCACTTTGGTGGTTTTCAAC
>Cannabis_sativa
TCAAGACAGCTTTCAAACATCCTCAATCAACCAACGTGAACGCACCCAATTGAGGGTGGT
GACACCGGATGATTTCAAATGAGTGTCCCCCATCACTCCGACAATGAAGTCGG
>Clonorchis_sinensis
GGGTTTGTGATGCGGTTGCCACGTGTTGTGTGACCGCGACGAGCG
>Ctenopharyngodon_idella
ATTGCGAGATATAAAGTCAGA
>Acropora_digitifera
GAAAATGCAAAAATGGCCACCATGCAAAGGCTATAGCCCATGCAAAATCGTCACTTTGGG
TCAAAAATTAAAAATGCAAAAAACATGCCAAAAACCATTCTACAAATCATTTACAGTTGT
TCTGTGCAAAAAACCGCTCCAAAAAACACCAAATATTCGAAAAATGAGAGCATTTT
>Daubentonia_madagascariensis
GGTTAATCACTGCTGAAAAACCGCGTTTACACTGCCAGAGAATCACATTCGCAGCAAGGC
GTTAGAAGAGGCAGGAGAAGAAAACATCGGTTTTTCTGCCAGCCAGCCGCTTATGGAGAG
TGCACGGGAGTTTCACAGGAAACACTGCCTTTCCCAGAAAGAAAGCACCTACTTCTTAGA
TTGATTGCAGGCTTGCTCTATTCTGTGAAAGGAAGCAGAATATACACTGCCTTCCTCCTA
GCTTCCTCCTAGGATGAATATTGAAACCTG
>Dirofilaria_immitis
ACAAATTCATCGTATATATATATGATTAGACGGGTCACAATGCATACCTATGATTAGACA
AACACAGACGAGTCATAATCCGTATCTATGATTAGACAAATATATACGACAAATTCATCG
TATATATATATGATTAGACGGGTCACAATGCGTATCTATGAGTTATACAAATATATACG
>Perdix_perdix
TTTGGGGAGAAAAGTGAAGATTCGAGCAGAAAAATGGGGATTTGAGGGGAAAAATGGGGA
TT
>Heliconius_cydno
AAACCATGAAACATGTGTTCAATTAAATACTTTTAAGAGTCTAAATTTCAAATTTTATAA
AAATCGGTGACGATTTAAGAATTTTTCCATACAAACTTTCAACCCCTTATTTGCACCCTT
AGGGGGAGAATTCTAGA
>Heterocephalus_glaber
AGGGGCTTCTGGGCCACCATCCCTACTGCTTCCCGGCCCAGAGGA
>Heterodera_glycines
CGTCATTTTTTTACGGAAGAATAACCGGAATACAAATTAATATTTTCAATTAATTTTTGG

CACTATACTGCTATAGTGGCTTGGCCTTGGCTTGGAATTTTTCTGTAGCTCAGCTCCCTG
ACAGTCGTTTTTTGATTTGAAGGTACCATTCGAAAGAGCTCTTTTTCTAGGCCTAATGGC
TAAGGTGG
>Lepisosteus_oculatus
GGCAGTAAAACCACATGTAACAAAACGGGCATAACAGGCACACTTTTCAAAACGTGTTTT
TCAGCCCGAAATAAACGCGTTTTTTCAGACAACTGAATACCCATGTAGCAAGCAAGGCTC
AGAGAGCCCACACACTCAGATCTTAAAGTGTTTTTCAGCAGAAATAAACATGTTTCTGCA
AGTTTGGTCAACCCCATGTAAGTTATGGGGCCAACAGAAATAAACAAAACGTTTGGTCAA
CCCCAAATAATACGC
>Leucoraja_erinacea
TGGTATGGTCAGACGTGGGGGCAAGCCTGCCGCTGATAAAGGAAAGGACAATCGTGATTA
TCTTGCTGCCGCATTCCCTGAGCTTAATGCTGGGGACCGTGCTGGTGCTCATGCCTCCTC
CGC
>Linum_usitatissimum
GTTCTGTCTATTTATAACTCTTTTATGTCGGTGATATACACTTTACGAGTATGTTATGCG
TCTTTTATTTATGTCTTTGTTACGTCACTTACGATTAGTTGAATTCATCATC
>Lytechinus_variegatus
TTGAATTACCATCATAACTTTGAAAGTTTATGGATCTAGTTCATGAAACTTGGACATAAG
AGTAATCAAGTATCACTGAACATCCTGTGCGAGTTTCAGGTCACATGATCAAGGTCAAAG
GTCATTTAAGGTCAATGAACTTTGGCCGAATTGGGGGTATCTG
>Manihot_esculenta
TTTGGGATGGAAGCTATTAGGCCGAAATCGGGCACCGGATGTAACATGGGTGATAGACTT
GGCAGTT
>Musa_acuminata
GCTGAGAAAAGGGCAGTCAAATGCCAAGAAAACAGAACGAATACCAGGTGCTAAACCATA
GCAGATTCTAAACTTGGGGTTGATCTCTTTAGGGGATCGGCCTCCTTGGAACTCTATAGG
GGGAATTACTCCAAGTGGCTGTCAAAA
>Patiria_miniata
CTGACAAATCGGCAAGGTTCAGACTTGAAACTTTTGGAGTGAGCCCCAGACATAGGTACA
TGATGCCCCCAGTTTCGTTTTTGTGCAGTAAGTCCCATAGCAAGGGCTGTCAGCGCCAGC
CGACATGGCTCTCATGTGTCGCGGTCTAAAGGTACCTTAACATGGATGGATTTTGACAAA
CTATAACTTGTTGGATGGCGGATGAATGGCTCAACAAG
>Phoenix_dactylifera
TCGACTCGCCTCTCAGAGACGAAAATACCGTCTTCTGTCTTTTTGGAGTTGCGCTGTCTC
GGAGTCGACTCGAACTTTGCGGGAGTCGACTCGGCTCTCAGTGTCTGAAATTGGCTCTCT
GACTTTTTGCTTGTATTCCTCTGGGAGTCGACTCTAGCTTCCTTTGGAGTCGACCTGCCA
ACCATCGGAGTCGACTCGAGTTCCTCAGGAGTCGACTCGGCTCTCAGAGTCCAAAATAGC
TCTCTGACTTTTGCCTTGTTCTCCCTAGGAGTCGACTCGTACCACACTGGAGTCGACTCG
AAAAGCATCGGAG
>Pogonomyrmex_barbatus
ACCTGACGGTCTCGGCGGTACCTCCGCGCGCGGCGTGTTGCTTTCGCTCGCGGCGGT
>Radix_balthica
AATAATCTAAAAAATCAATAATTTAATTACGGTACATGCGATCGATCTGAAATTTTGATA
CATTGTTTTTATCATCGGTACGAAACTGCCTGCAAAATTTCAGCTCGCTAGCTTACGAGC
AAGTGGGTTAAAAATCGATCAAAAGTTTTGGCGACAAGCAAGCAAGCAAGCAAGGTCGGC
CGCAGACACCACGAGTTAATATAAAGGATTTAATAAAAAAAACCCGTAAAAAACCAAGTT
TTTGACTATAAATCCTTTATGTACGAGACGTAACTAAAAGAATGTTCATTATAATACGCA
GCGCAAGACCATCCAAAATACAAAATTGCAGTTGTGTCGTAAAACATGTTTAATATAATG
CGCACCGATCAGGAATATGTTTATAAAAAGAGCTGTCAATAAAATACGCACCTTATTGCG
CGACAGAAAGGTTGTTCAATATAATACGCACCGCGACGCTCTATCTCAAAAAATAGCATG
ATACATAGGAATAAATAA
>Ovis_aries
TCCACAGGAAAGGCGAGAGGAACTCCAGGGTCGTGCCACCATTCCAAGAGTCCCCCAGAT
GTGTCAGTCCATTCCAGAGGAACCTGTTTTCCCTGCACTGCCTTGACGTTCAAGCCGAGG
ATCGACTCCCACCACGTGTGCACGTGGGACAGCCCTGTGGGAAAGCCTCGTGGGAAAGCC
TCGTGGGAAAGACACGAGGGAAAACCATAGATGCTTTGATCCACGGGGCGGACTGCGTGA
CACTGCTGCTACCGCTCTGGAGGAAAGCGCAAGTGCATGCCCGCATTCGAGACGAGGACT

GACTCCCCTGGGGAGACTCCAGAAGTACCCCAAGATCCATGTCAGCACTGGAGAGGAATC
CTCAGGTTCCGGCCCTGAGTTCACACAAGGTCTTAGGCCCCGGCATCAGCGGGAGAGGAA
TTCCGAAAGGCCCCCGAGGAACTCGCATGGGGACTGGCCTTTCCTGAGGCCACCAGAGCG
GGTCCCTGAGGGCCCCGTCGTAAGTCGAGAGCACCTGCCGCAACTCGAGAAAATCCAGGA
GGTTTTGCCCTCCAGGCGAGATGAGGCCCATTTCCGCTGAGGCTTCTCGAGGCTAATCAC
ATCTAACCCCTGGAACTTCCAAAGGGTCCTTCACACCCTTGCTGCAACTCAAGAAGTTCC
CGACACACCCGTCTCCACTCGAGAGGAAGCACGAGGGTCCCGAACACATCCAGGGGAGCC
CCGTTTCCGCCTCCGAGCTCGAGATGAGGGATCCTTTCCCTGTTTCGTAGGGAAAGAATT
CCCGGCGTTCCCGTCGCATCTCAAGAGGAGGCGCTC
>Eucidaris_tribuloides
TGGTTTTCTACCATATATGCTTGCATCTGACAGTATTAAGAAGAACACCAGACCGCAATC
AAACAAAACGAATGCATCACTATCTGATGAAACGAAAAAATATGTTTACACCCACTGGGA
AGTAGATTTCAGCGAGATATTGTCATTGGTAGAAGATAGTTAGGGATAGGGTAGAGTTAG
ACGTTAGCATGTAGATTAGGGTTAGCGACAAGATATGCTTGCATCTGACAGTCCATTGAA
GAAGATTATCAGACCGCAATGAAACAAAACGGATGCATTACTATCTGATGAAACGAAAAA
ATATGTTTACACCCACTGGGAAGTAGATTTCAGCGAGATATTGTCATTGGTAGGGTTAGG
GTTAGGGTTAGGGTTAGGGTTAGGGTTAGGGTTAGGGTT
>Oryza_brachyantha
TTGTATGAATGGTGAAGTTGTTGATGGAATCCAAGTGTACCCTTCGTGCTCGAAAATTTC
GCTACCGAAGTTGTGTTCGAGCACGAAGTGTAAACCAAGTGGATACACAATTTGTGTAGT
TATCCCAAAGTTCATAGTCGAATCTTGGTAGACT
>Anopheles_merus
CTTTCCCATCCATCCATATGGCCATGGTTGAAATTTTGACGAAAATGCAAGGTCAACCAT
TATCGTTCTTATATCATTGAGGCTTCTTGAATAACCAAACCAAAGCCCTGGAACACAATA
ATAAGGATCTTGGCGATGAAAAAGTCTCAAAATGGGTCATAGTACCCTTTATACACATGC
TAAAATTGATAGGTTTTCCCGAGTACAATGCATTCAGAGGCCATTTACTAGGCTACCAGT
GAGTCTTATCTTTAAACCTATGTGCGGATCTAGAAGACCACTTCAATACCTTTCCAAAAC
TGTCTTAATCGTCTCGCTAGGACACATGGTTACGAAGTTATGGCCATTTGAAGGAATGTC
CAA
>Zostera_marina
CGCACTAGAGTGCATTCCTTGGTCGGGATCACACGAATTGCACGCATGAAGTACTTAGTC
GGGAT
>Miscanthus_giganteus
ACACTAACACTATCTCCAAACGGACCGAAACGAGATTCCACATGACCCACGTCACCTAGG
AGTTCCATCGGGTGCGTCCAAAATGATTTCTGAGCCTATGGTACGTTTGGCGCAAACCGT
GCACCTATCTTGCATCA
>Hordeum_vulgare
TCATGCCTCTAGTTGGCTAGCCAGTTGATCAAAGATAGTCAAGGTCTTCTGACTATGAAC
AAGGTGTTGTTGCTTGATAACTGGATCACGTCATTAGGAGAATCACGTGATGGACTAGAC
CCAAACTAATAGACGTA
>Macaca_fascicularis
AAGAAGCTTTCTGAGAAACTGCTTTGTGTTCTGTGAAATCATCTCACAGAGTTACAGCTT
TCCCCTCAAGAAGCCTTTGCTAAGACAGTTCTTGTGGAATTGGAAAAGAGATATTTCGAA
GCCCATAGAAGACTATAGGGAAAAAGGAAATATCCTCAGATAAAAAAGAGA
>Thellungiella_parvula
TTTGAGAAGATACTAAAATCTAGAAACTAAAAGACCAAGAGTGAGAACATGATCATGAAA
GGAGTCTCTTAGAGATCCTAAAGGGAGTTTTAAATCCCAAAATCCTTACACACATACTTT
TGTTGAATCAAACAAACTAAATGTATGGACACAACTAGGCAAAAGGATTAGGGAATCCTA
GATGCATTAGGACATGCATATCTAGTTGGTACTCATTGT
>Bison_bison
AATGGAAGATTGGACTTCCCTGGGCCAAACACAAGAGGCATCCTGAATTCCCCGTCGTAA
TTCGAGAATCCTGCCGACACTCGAGAAAATCCACGTGTTCCCCCGTCATCGCAAGATGAA
GCCCTTTCCCGCTACAGTGTCTCAGGAGAAGTCCCACGTTAGGTATTGGAGGTCGAAACG
GTACTTGGCACCTTGATGCGACCACAAAGTGCCCCGACATCCCGGTCCTCCCTCGAGAGG
AACACCGAAGTTTTCCGGCACCACTTCCTCTGAGCCCCTTCTACCCTCCTGATCTGGACA
GGAGGGTCGACTCCCCTGCTTTGTCTGGAAGGGGTTCCCGACCTTCCGGTCGCACCTCAG
GATGAGGCCGGGTCTCACGACGACATTTCAGACGTGGCCTCGTGGGTGGTTCCACATTCC

GAAGGACCCCGATTTCCCGGTCCCCTCTTGATAAGAACCCGATGCCCGGACACCTCTTCG
AACCTCCACCCTGTGAATGAAGTCAACACGAAAGGGCAGTGACTCGCCCGTGCATCGTCG
GGAAAAAACCCCCAGGTTCCAAATACCGCTCGACAAGTGGCCTGTCTCCCCGGGAAACAC
CTCGAGAGGCAAGCGGAGTTCCATGCCTCACCCAAGACGAGGCCTGACTCTCCCTGTCCC
AAGTCTGCAGGGACCTTGCGATCAGAGTCTGAAGTCAGAGGAACCCTGAGGTTCCTGCCT
CAACTGGAGATGAGGCCCTCTTCCAATGCACCAAACCCAGTGGGAGTGCCGAGAAGCCCT
CCCACCTCCAGTTTTCCCTGACTTCTCAGAGCCACCATGAGAAGCCCCCTGAGGTCACCT
GCACAAGTCAAGGGAAGCCAGGTTTTCTGCCTCAACCCGAGAAAGACCTCGAGAGACCTT
CTTCAACACGTCTCGAGCCAGATTCCCCTAACAGTGACTCGAGAGCAATGACGCGCTCCC
CCTCGCCATTGCGCATGGAGACCCCGACTTCCCTGGCGCCCACGAGAGCTCACTGACCTC
GCCGTCGTACACTAGGAAAAACCGCACACTGTGGCGCCAGCTCGAGAACAACCCTGAGCC
CTCCCCATCATCGCGAGTTGAGGGCCTTCGTCTCCTGTATGGCCTAGAGACCAATCTCCT
CGACTCTCTCTCAAACGCCTCAGGAGGCTTGACTCCCTTTAGTCCACCCAGTGAGCTCCA
AGAGATACCCGTCGCGACTCGAGAGCAGAGCGGGGTTCTTTGCTTCCACTCGACATGAAT
GCTGTCTCCCCGGGTGCGTCTGGAATTGCAACCCTGAGATCCCTTTCGCCCCCTGGAGAG
GAACACTGGCTTCTGGACACGAAGCCTAGATGAGGTCTATTGGCCCTGCAGTCACTCGAG
AGCAATCCCCAGCTTTCCTTCGCAACTCG
>Bos_grunniens
GGTGGCCAAGTACCGTTTCGACCTCCAATTCCTAACGTGGGACTTCTCCTGAGACGCTGT
AGCGGGAAAGGGCTTCATCTTGCGATGACGGGGGAACCACGTGGTTTTTCTCGAGTTGCG
GCGGGATTCTCGAGTTACGACGGGGAATTCAGGATGCCTCTTGTGTTGGCCCAGGGAAGT
CCAATCTTCCATTCGAGTTGCGAAGGAAAGCTGGGGATTGCTCTCGAGTGACTGCAGGGC
CAATAGACATCATCTAGGCTTGTGTCCAGAAGCCAATGTTCCTCTCCAGGGCCGACAGGG
ATCTCGGGTTGCATTCCAGACGCACCCGGGGAGACAGGCATTCATTCTCGAGTGGAAGCA
AAGAACCCCCGCCTGCTCTCGAATCGCGACGGGTATCTCTTGGAGCTCACTGGGTGGACT
AAAGGGAGTCAAGCCTCCTGAGGCGCTTGGAGAGAGGTCGCGAGATTTGGTCTCTAGGCC
ATGCAGGAGACGAAGGCCCCTCATCTCGCGATGACGGGGGAGTCTCGGGGTTGTTCTCGA
GCGGCGGCCCCAGGGTGCGGTTTCTCACGAGGTACGACGGCGAGGTCAGTGAGCCTCTCG
TGGGGCCCAGGGAAGTCGGGTCTCCATGCAAGTGGCGAGGGGGAGCGCGTCATTGCTCCT
CGGAGTCATGGTAGGGGAATCTGGCCTCGAGACGTGTTGAAGAAGGTCTCTCGAGGGGGT
CTTTCTCGGTTGAGGCAGGAAACACTGGTTCCCTCGACTTGTGCAGGTGACCTCAGGGGG
CTTCTCATGGTGGCTCTGAGAAGTCAGGGAAACTGGAGGTTGGGAGGGGCCTCTCGGGAC
TCCATGGGGTTGGTGCATTGGAAGAGGGCCTCATCTCCAGTTGAGGCAGGAACCTCAGGG
TTCCTCTGATTTCAGACCGATCGCAGGGTCCCTGCAGACTTGGGACAGGAGAGTCAGGCC
TCGTCTTGGGTTGAGGCATGGAACTCCGCTTGCCTCTCGAGGTGTCCCCGGGGAGAGAGG
CCCATTGTCGAGCTGTATTTGGAACCTGGGGGTTTTCCCCGAACGATGCACGGGCGAGTC
ACTGCCCCTTCGTGTTGACTTCATTCACAGGGTGGAGTTCGGAAGAGGTGTCCGGGCATC
GGGTTCTTATCAAGAGGGGACCGGGAAATCGGGTGTCTTACGCAATGTGGAACCACCCAC
GAGGCCACGTCCTGGAATGTCGTCGTGAGACCGGCCTCATCCTGAGGTGCGACCGGAAGG
TCGGGAACCCCTTCCAGAAAAAGCAGGGGAGTCGACCCTCCTGTCCAGATCAGGAGGGTA
GAAGGGGCTCAGAGGAAGTGGTGGCCGGAAAACCTCAGGGGTTCCTCTCCGAGGGAGACC
GGGATGTCGGGAACTTTGTGGGTCGCATCAAG

## References for Supplementary Table S2

| |
|---|
| (Wade et al., 2009; Hasson et al., 2011; Alkan et al., 2011) |
| (Matzke et al., 1990; Shang et al., 2010) |
| (Tek et al., 2010) |
| (Lee et al., 1997; Alkan et al., 2007; Hasson et al., 2011) |
| (Alkan et al., 2011) |
| (Alkan et al., 2011) |
| (Haaf et al., 1995; Alkan et al., 2007; Locke et al., 2011) |
| (Zhong et al., 2002; Nagaki et al., 2004) |
| (Cheng et al., 2002; Lee et al., 2005) |
| (Heslop-Harrison et al., 1999; Hall et al., 2003; Zhang et al., 2008) |
| (Krzywinski et al., 2005) |
| (Tarès et al., 1993; Beye and Moritz, 1995) |
| (Cellamare et al., 2009) |
| (Alkan et al., 2011) |
| (Abad et al., 1992; Abad et al., 2000; Sun et al., 2003) |
| (Haaf et al., 1995; Samonte et al., 1997; Alkan et al., 2007) |
| (Musich et al., 1980; Alkan et al., 2007) |
| (Fishman and Saunders, 2008) |
| (Guenatri et al., 2004; Kuznetsova et al., 2006) |
| (Cellamare et al., 2009) |
| (Alkan et al., 2011) |
| (Haaf et al., 1995; Samonte et al., 1997; Alkan et al., 2007) |
| (Musich et al., 1980; Alkan et al., 2007) |
| (Viñas et al., 2004) |
| (Tek and Jiang, 2004) |
| (Jiang et al., 1996; Zwick et al., 2000) |
| (Fischer et al., 2000; Roest Crollius et al., 2000) |
| (Ugarković et al., 1996) |
| (Harrison and Heslop-Harrison, 1995; Lim et al., 2007; Koo et al., 2011) |
| (Teschke et al., 1991; Niedermaier and Moritz, 2000) |
| (Alkan et al., 2011) |
| (Meyer et al., 2010) |
| (Ekes et al., 2004) |
| (Kikuchi et al., 2011) |
| (Tanaka et al., 1999; Pathak et al., 2006) |
| (Kipling et al., 1995) |
| (Spence et al., 1998) |
| (Jantsch et al., 1990; Bratanich et al., 2000) |
| (Gaillard et al., 1981; Płucienniczak et al., 1982; Taparowsky and Gerbi, 1982) |
| (Sternes and Vig, 1995) |
| (Buckland, 1983; Novak, 1984) |
| (Argout et al., 2011) |
| (Lee et al., 2011) |
| (Carone et al., 2009) |
| (Samonte et al., 1997; Alkan et al., 2007) |

**Supplementary Figures.**

**Figure S1** - group alignment (from Figure 2 - The Tree):
For each of the 26 groups of species with sequence similarity as depicted by the red bars in Figure 2, the alignment is shown in top-to-bottom order.

**Figure S2** - primates and cichlids:
The colored bars represent the lenght of the centromere repeat sequences and the colors represent sequence similarity.
*A* There were 21 primates in the 282 species. All the Old World Monkeys (macaques and baboons) and apes (gibbons, gorillas, chimpanzees, bonobos, oerangutans, and humans) have similar 171 bp centromere repeat sequences. The New World Monkeys (spidermonkey, marmoset, squirrelmonkey) have a 343 bp repeat. This 343 bp repeat is a doubled version of the 171 bp repeat (ligher blue bars, Figure 5D). The (candidate) centromere repeat sequences of the more basal tarsiers and prosimians did not show sequence similarity with the apes or monkeys or between themselves.
*B* The centromere repeat sequences of the Lake Malawi cichlids (*Metriclima zebra, Melanochromis auratus, Labeotrophus fuelleborni, Rhamphochromis esox,* and *Pundamilia nyererei*) and the nile tilapia (*Oreochromis niloticus*) showed sequence similarity, eventhough the nile tilapia centromere repeat is 20 bp shorter than the 237 bp repeat found in the Lake Malawi cichlids. Surprisingly, the centromere repeat sequence of the Lake Tanganyika cichlid (*Neopamprologus brichardi*) did not show sequence similarity with either the Lake Malawi cichlids or the nile tilapia.

**Figure S3** - correlations between karyotype, genome size and our found parameters (length, GC content, genomic fraction):
Genome size and karyotype were correlated to tandem repeat length, GC content and genomic fraction, but no correlation was observed.

**Figure S4** - Nile tilapia (*Oreochromis niloticus*) tandem repeat alignment:
In Nile tilapai (*Oreochromis niloticus*) is a close relative to the cichlids and the centromere repeat sequence is similar to the candidate centromere repeat sequence in Malawi cichlids, except that the most abundant tandem repeat in Nile tilapia has a 206 rather than a 237 bp repeat. In addition, a 237 bp repeat is also present in the Nile tilapia genome, but is less abundant. The alignment shows a 20 bp indel.

**Figure S5 -** Bovideae fractions of short (680 bp) and long repeat (1410 bp):
In Hereford cattle (*Bos taurus taurus*), Nellore cattle (*Bos taurus indicus*), yak (*Bos grunniens*), bison (*Bison bison*), and waterbuffalo (*Bubalus bubalis*) two predominant tandem repeats were found. One was around ~680 bp in length (named short) and one was around 1410 bp (named long). No sequence similarity was found between these two repeats, but all five species had these repeats, albeit in different ratios. In waterbuffalo the short repeat was the most abundant, whereas in the other species the long repeat was the most abundant.

**Table S1** - all species info:
For each of the 282 species the basic phylogenetic, repeat, and genome information is given as well as the accession numbers used. The accession numbers for the NCBI Trace Archive (Sanger data) and DDBJ DRA (Illumina and 454 data) is given seperatly.

**Table S2** - literature comparison:
For 60 of the 282 species literature data was available. For 51 of the 60 species we found the published repeat.

**Table S3** - PRICE vs Sanger discordance explanation:
For 10 of the 37 we found a different repeat between PRICE assembled contigs and tandem repeat directly derived from Sanger sequences. Each case is individually addressed.

**Table S4** - Number of species with data from a particular sequencing technology.

Supplementary Figure S1

Orange

```
Citrus_sinensis       CATTCCGAGTCCGGCGGACGAACTTCGCCCACGCCCCCCACCAAGGCTATAGCCCACCCG 60
Citrus_clementina     CATTCCGAGTCCGGCGGACGAAGTTGGCCAGCCCCACCCCCCAAGGCTATAGCCCACCCG 60
                      ********************* ** ***  * ** *** *******************

Citrus_sinensis       ATTTT-TGGCCATTTTTCCGCTGG-CGAGTCTTGGCGCCCCGACCTTCGGGCGCTCATTT 118
Citrus_clementina     TTTTTGTGGCC-TTTTTCCGCTGGACGAA-CTTGGCGCCCCGGCCTTCGGGCGGTTATTT 118
                       **** ***** *********** ***  *********** ********** * ****

Citrus_sinensis       TTGGGCGCGGCTGTGCCCGTGGCCTATTTTTGGCACACGGAGGCCGCCCGCAAAGTCTGG 178
Citrus_clementina     TTGGGCGCGGCTTTGCCCGCGGCCTATTTTTGGCACACGGAGGCCCCTCGCAAAGTCTCG 178
                      *********** ****** *********************** * ********* *

Citrus_sinensis       GGC 181
Citrus_clementina     CGG 181
                       *
```

Candidate centromere repeat monomer:

```
>Citrus_sinensis
CATTCCGAGTCCGGCGGACGAACTTCGCCCACGCCCCCCACCAAGGCTATAGCCCACCCG
ATTTTTGGCCATTTTTCCGCTGGCGAGTCTTGGCGCCCCGACCTTCGGGCGCTCATTTTT
GGGCGCGGCTGTGCCCGTGGCCTATTTTTGGCACACGGAGGCCGCCCGCAAAGTCTGGGG
C
>Citrus_clementina
CATTCCGAGTCCGGCGGACGAAGTTGGCCAGCCCCACCCCCCAAGGCTATAGCCCACCCG
TTTTTGTGGCCTTTTTCCGCTGGACGAACTTGGCGCCCCGGCCTTCGGGCGGTTATTTTT
GGGCGCGGCTTTGCCCGCGGCCTATTTTTGGCACACGGAGGCCCCTCGCAAAGTCTCGCG
G
```

##########################################################################################

Arabidopsis

```
Arabidopsis_lyrata       TTGCTTCTCAAATCTTTGTGGGTGTGGCCGAAGTCC-TATGAGTTTTCGGTTTTGGAGCT 59
Arabidopsis_thaliana     TTGCTTCTCAAAGCTTTCATGGTGTAGCCAAAGTCCATATGAGTCTTTGGCTTTGTGTCT 60
                         ************ ****   ***** *** ****** ******* ** ** ****   **

Arabidopsis_lyrata       TCTAAACGGAAAAACACTACTTTAGCTTTCGGGATCCGGTTGCGGCTCTAGTTCTTATAC 119
Arabidopsis_thaliana     TCTAA-CAAGGAAACACTACTTAGGCTTTTAAGATCGGGTTGCGGTTTAAGTTCTTATAC 119
                         ***** *    ***********  *****   **** ******** *  ***********

Arabidopsis_lyrata       CCAATCATAAACACGAGATCTAGTCATATTTGACTCCAAAAACACTAACCAAGCTTCTTA 179
Arabidopsis_thaliana     TCAATCATACACATGACATCAAGTCATATTCGACTCCAAAA-CACTAACCAACCTTCTTC 178
                          ******** *** ** *** ********* ********** ********** ******
```

Centromere repeat monomers:

```
>Arabidopsis_lyrata
TTGCTTCTCAAATCTTTGTGGGTGTGGCCGAAGTCCTATGAGTTTTCGGTTTTGGAGCTT
CTAAACGGAAAAACACTACTTTAGCTTTCGGGATCCGGTTGCGGCTCTAGTTCTTATACC
CAATCATAAACACGAGATCTAGTCATATTTGACTCCAAAAACACTAACCAAGCTTCTTA
>Arabidopsis_thaliana
TTGCTTCTCAAAGCTTTCATGGTGTAGCCAAAGTCCATATGAGTCTTTGGCTTTGTGTCT
TCTAACAAGGAAACACTACTTAGGCTTTTAAGATCGGGTTGCGGTTTAAGTTCTTATACT
CAATCATACACATGACATCAAGTCATATTCGACTCCAAAACACTAACCAACCTTCTTC
```

##########################################################################################

Eucalyptus

```
Eucalyptus_grandis      ATTTTGGAGATCCCGATACCTTGAAAGTCGGCTGAAATCAAGAAACGGAGACTCGGCACA 60
Eucalyptus_globulus     ATTTTGGACATCACAGCACCCTGAACGCCCGGCGAAACAAGAGAACGAAGATTCCACAGA 60
                        ******** *** *   *** **** * * *  ****  *   **** *** **  ** *

Eucalyptus_grandis      AAAAATTCCCTTGTGTTCCATTAAAAATTCCTTTTCGG----TATTTTCACT-ATAGACC 115
Eucalyptus_globulus     GAAAATTCCCTGTTTTTCCATAGAAAATTCCCTTTCCGGCCATTTTTTGGCTGAAAAACC 120
                         **********  * ******  ******** **** *     * ****  ** * * ***

Eucalyptus_grandis      CCGGCAGGCGGAAA-CGGCATGGATAAGTGGGCTCCAAAAGCTTTATTTCGATCGGATTT 174
Eucalyptus_globulus     CCGACAGTTGGAAATCTGTTGGGAAACATGGCTTCCAAAAGCATTAGATCAGTCCGACCC 180
                        *** ***  ***** * *   *** *  ***  ********* ***  **  ** **
```

```
Eucalyptus_grandis       AAT 177
Eucalyptus_globulus      ATC 183
                         *
```

Candidate centromere repeat monomer:

```
>Eucalyptus_grandis
ATTTTGGAGATCCCGATACCTTGAAAGTCGGCTGAAATCAAGAAACGGAGACTCGGCACA
AAAAATTCCCTTGTGTTCCATTAAAAATTCCTTTTCGGTATTTTCACTATAGACCCCGGC
AGGCGGAAACGGCATGGATAAGTGGGCTCCAAAAGCTTTATTTCGATCGGATTTAAT
>Eucalyptus_globulus
ATTTTGGACATCACAGCACCCTGAACGCCCGGCGAAACAAGAGAACGAAGATTCCACAGA
GAAAATTCCCTGTTTTTCCATAGAAAATTCCCTTTCCGGCCATTTTTTGGCTGAAAAACC
CCGACAGTTGGAAATCTGTTGGGAAACATGGCTTCCAAAAGCATTAGATCAGTCCGACCC
ATC
```

################################################################################################

Figs

```
Ficus_fistulosa         TTAAGCCAGAATCAAGCATAAAGGGACC-AAAATG-AGCAAATCTCAGTTCAAGGCACAG 58
Ficus_langkokensis      TCAAGCCCGAATGAGGCAAAAAGTGACCCAAAATGCAATGTAA-TCAGATCAAGGTCGTG 59
                        * ***** **** * *** **** **** ****** *    *  **** ******    *

Ficus_fistulosa         TGGATCATCTAGGCAGCCCCTAAATGACCAAAACAAGCTTTGAGAATGATCAAAACACAA 118
Ficus_langkokensis      TAAACCAGTGAGGTGGCCTGTAAATGACCTAAACAAGGTTTGACAAGGATCAAAACACAA 119
                        *  * **   ***  ***  ********* ******* ***** ** *************

Ficus_fistulosa         AAAAACCAGACCCAATCTGGTTTGGGTCAAAAAAGTGGCAAAA 161
Ficus_langkokensis      AAAA-CCAGATCCTAACAACTTTGGGTCAAGAAACCAGCAAAA 161
                        **** ***** ** * *   ********** ***   ******
```

Candidate centromere repeat monomer:

```
>Ficus_fistulosa
TTAAGCCAGAATCAAGCATAAAGGGACCAAAATGAGCAAATCTCAGTTCAAGGCACAGTG
GATCATCTAGGCAGCCCCTAAATGACCAAAACAAGCTTTGAGAATGATCAAAACACAAAA
AAACCAGACCCAATCTGGTTTGGGTCAAAAAAGTGGCAAAA
>Ficus_langkokensis
TCAAGCCCGAATGAGGCAAAAAGTGACCCAAAATGCAATGTAATCAGATCAAGGTCGTGT
AAACCAGTGAGGTGGCCTGTAAATGACCTAAACAAGGTTTGACAAGGATCAAAACACAAA
AAACCAGATCCTAACAACTTTGGGTCAAGAAACCAGCAAAA
```

################################################################################################

Soybean

```
Glycine_max       TCACTCGGATGTCCGATTCAGGCGCATAATATATCGAGACGCTCGAAATTGAACAACGGA 60
Glycine_soja      TCACTCGGATGTCCGATTCAGGCGCATAATATATCGAGACGCTCGAAATTGAACAACGGA 60
                  ************************************************************

Glycine_max       AGCTCTCGAGAAATTCAAATGATCATAACTTT 92
Glycine_soja      AGCTCTCGAGAAATTCAAATGGTCATAACTTT 92
                  ********************* **********
```

Centromere repeat monomer:

```
>Glycine_max
TCACTCGGATGTCCGATTCAGGCGCATAATATATCGAGACGCTCGAAATTGAACAACGGA
AGCTCTCGAGAAATTCAAATGATCATAACTTT
>Glycine_soja
TCACTCGGATGTCCGATTCAGGCGCATAATATATCGAGACGCTCGAAATTGAACAACGGA
AGCTCTCGAGAAATTCAAATGGTCATAACTTT
```

################################################################################################

Beechs 1

```
Lithocarpus_calolepis           TATTTTGGTCATTTTTTGGGTTTCGGGGGTATTTTGGTCATTTTTTGGGT 50
Lithocarpus_grandilofolius      TATTTTGGTAATTTTTTTG-TTTCGAGGGTATTTTGGTCATTTTT-AGGT 48
                                ********* ******* * ***** *******************  ***

Lithocarpus_calolepis           TTCGGGAGTATTTTGGTCATTTTT-AGGTTTCGGGGGTTATTTTGGTAAT 99
Lithocarpus_grandilofolius      TTCGGGGGTATTTTGGTCATTTTTTGGGTTTCGGGGGT-ATTTTGGTCAT 97
                                ****** *****************  ************ ******** **
```

```
Lithocarpus_calolepis           TTTT-AGGTTTCGGGGG 115
Lithocarpus_grandilofolius      TTTTTGGGTTTCGGAGG 114
                                ****  ******** **

Candidate centromere repeat monomer:
>Lithocarpus_calolepis
TATTTTGGTCATTTTTTGGGTTTCGGGGTATTTTGGTCATTTTTTGGGTTTCGGGAGTA
TTTTGGTCATTTTTAGGTTTCGGGGGTTATTTTGGTAATTTTTAGGTTTCGGGGG
>Lithocarpus_grandilofolius
TATTTTGGTAATTTTTTTGTTTCGAGGGTATTTTGGTCATTTTTAGGTTTCGGGGGTATT
TTGGTCATTTTTTGGGTTTCGGGGGTATTTTGGTCATTTTTTGGGTTTCGGAGG

####################################################################################################
Beechs 2

Lithocarpus_balansae        TTTGAGCCCTTACATCGGCACGAAATTGGAATCCAAATTTCGGGTAACTC 50
Lithocarpus_hancei          TTTGAGCCCTTATATCGGCACAAAATTGGAATCCAACTTTCGGGTAACTC 50
Lithocarpus_xylocarpus      TTTACGCCTTTATATCGGCACAAAATTGGAGTCCAAATTTCGGGGAACTC 50
                            ***  *** *** ******** ******** ***** ******* *****

Lithocarpus_balansae        GGTGGCCCACATTCAAACGAGGATAACTCTCCCGATTTTTATCAAAAAAA 100
Lithocarpus_hancei          AGTGGCCCATATTCAAACGATAATAACTCTCATGATTTTTGTCGAAAAAA 100
Lithocarpus_xylocarpus      GGTGGCCCATATTCAAACGACAATAACTCTCTTGATTTTTATCGAAAAAA 100
                             ******** **********  *********  ******* ** ******

Lithocarpus_balansae        TACAAAATTTGTGCTCAAATTCAAGCTCAGGATGTCTACTTTCTAAAACA 150
Lithocarpus_hancei          TACAAAATTTGTGTTCAAATTCAAGCTCAGGACGTCTACTTTCTAAAACA 150
Lithocarpus_xylocarpus      CACCAAATTTGTGTTCAAATTCAAGCTCAGAATGTCTACTTTCTAAAATA 150
                             ** ********* **************** * *************** *

Lithocarpus_balansae        CGAAGGCCGCGTCAAAGAATTCCTCACGGTTCAAAAGTTATTGACGAAAC 200
Lithocarpus_hancei          CTTAGGCCGCGTTAAAGAATTCCTTTCGGTTCAAAAGTTATTGACGAAAC 200
Lithocarpus_xylocarpus      TTAAGGCCGCGTTGAAGAATTCCCTACGGTTCAAA-GTTATTGTCGAAAG 199
                               *********  *********   ********* ******* *****

Lithocarpus_balansae        GGTGGCCAAAGCACACTTTTCGTCACACTTCCTAACCGATTAACGCCCGC 250
Lithocarpus_hancei          GGTGGCCAAAGGTCACTTTTCGTCTCATTTCCTAACCGATTAACGCTCGT 250
Lithocarpus_xylocarpus      GGCGACCAAAGGTCACTTTTCGACACATTTCCAAACCGATGAAGGCTCGT 249
                            ** * ******  ********* * ** **** ******* ** ** **

Lithocarpus_balansae        GGATAACTACCCACTCACGTATTTTTTCCCCGAAACTTGATTTTGGGAAG 300
Lithocarpus_hancei          GGATAACAACCCACTCACATATTTTTTCCCCGAAACTTGATTTTGGGAAG 300
Lithocarpus_xylocarpus      GAATAACTACCCACCAAAGTATTTTTTACACAAAACTTCACTTTGAGAGG 299
                            * ***** ******  *  ******** * * ****** * **** ** *

Lithocarpus_balansae        ATTGATCTCGCGATATAGGAAAAATATTTCCCAGCCAAATTCGTGAGCAA 350
Lithocarpus_hancei          ATTGATCTCGCGATGTAGGAAAAATATTTCATGCACAAATTCGTGAGCAA 350
Lithocarpus_xylocarpus      ATTGACCTCAAGATATAGAAAAAATATTTCATCGCCAAATCCGAGAAAAA 349
                            ***** ***  *** *** ***********     ***** ** **  **

Lithocarpus_balansae        AACTTCAAACTTCCCCT 367
Lithocarpus_hancei          AACCACAGACTTCTCCT 367
Lithocarpus_xylocarpus      AACCTTAGACATCAATG 366
                            ***   * ** **

Candidate centromere repeat monomer:
>Lithocarpus_balansae
TTTGAGCCCTTACATCGGCACGAAATTGGAATCCAAATTTCGGGTAACTCGGTGGCCCAC
ATTCAAACGAGGATAACTCTCCCGATTTTTATCAAAAAAATACAAAATTTGTGCTCAAAT
TCAAGCTCAGGATGTCTACTTTCTAAAACACGAAGGCCGCGTCAAAGAATTCCTCACGGT
TCAAAAGTTATTGACGAAACGGTGGCCAAAGCACACTTTTCGTCACACTTCCTAACCGAT
TAACGCCCGCGGATAACTACCCACTCACGTATTTTTTCCCCGAAACTTGATTTTGGGAAG
ATTGATCTCGCGATATAGGAAAAATATTTCCCAGCCAAATTCGTGAGCAAAACTTCAAAC
TTCCCCT
>Lithocarpus_hancei
TTTGAGCCCTTATATCGGCACAAAATTGGAATCCAACTTTCGGGTAACTCAGTGGCCCAT
ATTCAAACGATAATAACTCTCATGATTTTTGTCGAAAAAATACAAAATTTGTGTTCAAAT
TCAAGCTCAGGACGTCTACTTTCTAAAACACTTAGGCCGCGTTAAAGAATTCCTTTCGGT
TCAAAAGTTATTGACGAAACGGTGGCCAAAGGTCACTTTTCGTCTCATTTCCTAACCGAT
TAACGCTCGTGGATAACAACCCACTCACATATTTTTTCCCCGAAACTTGATTTTGGGAAG
ATTGATCTCGCGATGTAGGAAAAATATTTCATGCACAAATTCGTGAGCAAAACCACAGAC
```

```
TTCTCCT
>Lithocarpus_xylocarpus
TTTACGCCTTTATATCGGCACAAAATTGGAGTCCAAATTTCGGGGAACTCGGTGGCCCAT
ATTCAAACGACAATAACTCTCTTGATTTTTATCGAAAAAACACCAAATTTGTGTTCAAAT
TCAAGCTCAGAATGTCTACTTTCTAAAATATTAAGGCCGCGTTGAAGAATTCCCTACGGT
TCAAAGTTATTGTCGAAAGGGCGACCAAAGGTCACTTTTCGACACATTTCCAAACCGATG
AAGGCTCGTGAATAACTACCCACCAAAGTATTTTTTACACAAAACTTCACTTTGAGAGGA
TTGACCTCAAGATATAGAAAAAATATTTCATCGCCAAATCCGAGAAAAAAAACCTTAGACA
TCAATG
```

```
#################################################################################################
Potato

Solanum_tuberosum      CGTTAAGACCTTAGCTATGGAGCCAGTTAGCCCTCACGGCCAAAACGTCCCATTTTAAAG 60
Solanum_phureja        CGTTAAGACCTTAGCTATGGAGCCAGTTAGCCCTCACGGCCAAAACATCCCATTTTGAAG 60
                       ********************************************** ********* ***

Solanum_tuberosum      GTCAAATGTGCCCCAAATCAGGAAAACCCC-AATTTTGTCGATTTTCGTATGCTATAGTC 119
Solanum_phureja        GTCAAATGTGCCCCAGAGCAGGTAAACCCCCAATTTTGCCGATTTTCGTGTGCTATAGTC 120
                       *************** * **** ******* ******* ********** **********

Solanum_tuberosum      CATGGACTTTTTGGTGATCTGGAATTCCGACAAAATTTT-GCCAAAATTTTTCGTGGACG 178
Solanum_phureja        CATGGACTTTTTGGTGATCTGGAATTCCGACATAATTTTTGCCAAAATTTTTCATGGACG 180
                       ******************************** ****** ************* ******

Solanum_tuberosum      TC 180
Solanum_phureja        TC 182
                       **

Candidate centromere repeat monomer:
>Solanum_tuberosum
CGTTAAGACCTTAGCTATGGAGCCAGTTAGCCCTCACGGCCAAAACGTCCCATTTTAAAG
GTCAAATGTGCCCCAAATCAGGAAAACCCCAATTTTGTCGATTTTCGTATGCTATAGTCC
ATGGACTTTTTGGTGATCTGGAATTCCGACAAAATTTTGCCAAAATTTTTCGTGGACGTC
>Solanum_phureja
CGTTAAGACCTTAGCTATGGAGCCAGTTAGCCCTCACGGCCAAAACATCCCATTTTGAAG
GTCAAATGTGCCCCAGAGCAGGTAAACCCCCAATTTTGCCGATTTTCGTGTGCTATAGTC
CATGGACTTTTTGGTGATCTGGAATTCCGACATAATTTTTGCCAAAATTTTTCATGGACG
TC
```

```
#################################################################################################
Monkeyflower

Mimulus_guttatus        CTTTGATCATTTTTGCTTCCTGGACACCTAATTTTCGATTTGTACCTAAAAGATACCAAA 60
Mimulus_laciniatus      CTTTGATCATTTTTGCTTCCTGGACACCTAATTTTCGATTTGTACCTATAAGATACCAAA 60
Mimulus_dentilobus      CTTTGATCATTTTTGCTTCCTCGACACCTAATTTTCGATTTGTACCTATAAGATACCAAA 60
                        ********************* ************************** ***********

Mimulus_guttatus        TCCACCAATAAAAACAGAAATATAAATTAATCCAAACAAAAATCCACGTAAACGTAAACG 120
Mimulus_laciniatus      TCCACCAATTAAAACAGAAATATAAATTAATCCAAACAAAAATCCACGTAAACGTAAACG 120
Mimulus_dentilobus      TCCACCAATTAAAACAGAAATATAAATTAATCCAAAAAAAAATCCACGTAAACGTAAACG 120
                        ********* ************************** ***********************

Mimulus_guttatus        TTGAGTGTTATTTCTACACTCCACTGTATCTTTTGTATCTAGTGTTTTTGCTGTTGTTTG 180
Mimulus_laciniatus      TCGAGTTTTATTTCTACACTCCATTGTATCTTTTTTATCTAGTGTTTTTGCTGTTGTTTG 180
Mimulus_dentilobus      TAGAGTGTTATTTCTACACTCCACTGTATCTTTTTTATCTAGTGTTTTTGCTGTTGTTTG 180
                        * **** **************** ********** *************************

Mimulus_guttatus        GTTATGCGTATTAAAGTCCTAGTATTGTAGTTCGTGTTTTTAAGAACGTTGTTGCGAATG 240
Mimulus_laciniatus      TTGATGCGTATTAAAGTCATAGTATTGTAGTTCGTGTTTTTAAGAACGTTGTTGCGGATG 240
Mimulus_dentilobus      TTGATGCGTATTAAAGTCCTAGTATTGTAGTTCGTGATTTTAAGAACGTTGTTGCGAATG 240
                         * *************** ***************** ******************* ***

Mimulus_guttatus        CACTTTTGCGCATGTTTTTTTACATGTACGTTTGTGCATGTCCATTTGAGCGAGTGTGTT 300
Mimulus_laciniatus      CACTTTTGCGCATGTTTTTTTACATGTACGTTTGTGCATGTCCATTTGAGGGAGTGTGTT 300
Mimulus_dentilobus      CACTTTGGCGCATGTTTTTTTACATGTACGTTTGTGCATGTCCATTTGAGCGAGTGTGTT 300
                        ****** ******************************************* *********
```

```
Mimulus_guttatus        TGTGCATGTTTGTTGATTTGTATTAATGTCCTAGGATTGTAGTTTGTGTTTTCATGATCC 360
Mimulus_laciniatus      TGTGCATGTTTGTTGATTTGTATTAATGTCCTAGGATTGTAGTTTGTGTTTTCATGATCC 360
Mimulus_dentilobus      TGTGCATGTTTGTTGATTTGTATTAATGTCCTAGGATTGTAGTTTGTGTTTTCATGATCC 360
                        ************************************************************

Mimulus_guttatus        TTTTTGCGAATGCACTTTTGCGTGAGTACTTTTGTGCATGCTTTTGAACTTTTACTCATG 420
Mimulus_laciniatus      TTTTTGCGAATGCACTTTTGCGTGAGTACTTTTGTGCATGCTTTTGTACTTTTACTTATG 420
Mimulus_dentilobus      TTTTTGCGAATGCACTTTTGCGTGAGTACTTTTATGCATGTTTTTGTACTTTTACTTATG 420
                        ********************************* ****** ***** ********* ***

Mimulus_guttatus        TGCGTATCTTTTCTAGCGAGTGTTTTTGTGCACGTTTCTTGATGCGCATTCATGTCCTCG 480
Mimulus_laciniatus      TGCGTATCTTTTCTAGCGAGTGTTTTTGTGCACGTTTCTTGATGCGCATTCATGTCCTTG 480
Mimulus_dentilobus      TGCGTATCTTTTCTAGCGAGTGTTTTTGTGCACGTTTCATGATGCGCATTCATGTCCTCG 480
                        ************************************** ******************* *

Mimulus_guttatus        AACTGTAGCACGTATTTTCATGATCGCTTTTGGAATCCAATTTTGACCATGTTCGAGTCC 540
Mimulus_laciniatus      AATTGTAGTACGTATTTTCATGATCGCTTTTGTTATCCAATTTTGACCATGTTCGAGTCC 540
Mimulus_dentilobus      AACTGTAGTACGTATTTTCATGATCGCTTTTGGAATCCAATTTTGACCATGTTCGAGTCC 540
                        ** ***** **********************  ***************************

Mimulus_guttatus        GAATTTGGAATTTTGACGTTAGACAGGAATTGTAGCCAACTGAAATAGCTTTCTAATGCC 600
Mimulus_laciniatus      GAATTTGGAATTTTGACGTTAGACGGGAATTGTAGCCCGCTGAAATAGCTTTCTAATGCC 600
Mimulus_dentilobus      GAATTTGGAAATTTGCCATTAGACGGGAATTGTAGCCAACTGAAATAGCTTTCTAATGCC 600
                        ********** **** * ****** ************  *********************

Mimulus_guttatus        ACCGGAACCAGTCGATTTGGAGTCCGGATGAGAAAGTTATG------------------- 641
Mimulus_laciniatus      GCCGGAACCAGTTGATTTGGAGTCCGGATGAGAGAGTTATGCTCGACTTCGTGGACACTA 660
Mimulus_dentilobus      GCCGGAACCAGGTCATTTGGAGTCCGGATGAGAAAGTTATGCTCGACTTCGTGGACACTA 660
                         **********   ******************* *******

Mimulus_guttatus        -----------------------------------------TTTCTTAATTGCAACTCTT 660
Mimulus_laciniatus      CCAAAGATGTTGAGGGATTCTCTCTTCTAATTTGATTTTATTTTCTTAATTGCAACTCTT 720
Mimulus_dentilobus      CCAAAGATGTTGAGGGATTCTCTCTTCTAATTTGATTTTATTTTCCTAATTGCAACTCCT 720
                                                                 **** ************ *

Mimulus_guttatus        TTCTTACC 668
Mimulus_laciniatus      TTCTTACC 728
Mimulus_dentilobus      TTCTTACT 728
                        *******
```

Candidate centromere repeat monomer:

```
>Mimulus_guttatus
CTTTGATCATTTTTGCTTCCTGGACACCTAATTTTCGATTTGTACCTAAAAGATACCAAA
TCCACCAATAAAAACAGAAATATAAATTAATCCAAACAAAAATCCACGTAAACGTAAACG
TTGAGTGTTATTTCTACACTCCACTGTATCTTTTGTATCTAGTGTTTTTGCTGTTGTTTG
GTTATGCGTATTAAAGTCCTAGTATTGTAGTTCGTGTTTTTAAGAACGTTGTTGCGAATG
CACTTTTGCGCATGTTTTTTTACATGTACGTTTGTGCATGTCCATTTGAGCGAGTGTGTT
TGTGCATGTTTGTTGATTTGTATTAATGTCCTAGGATTGTAGTTTGTGTTTTCATGATCC
TTTTTGCGAATGCACTTTTGCGTGAGTACTTTTGTGCATGCTTTTGAACTTTTACTCATG
TGCGTATCTTTTCTAGCGAGTGTTTTTGTGCACGTTTCTTGATGCGCATTCATGTCCTCG
AACTGTAGCACGTATTTTCATGATCGCTTTTGGAATCCAATTTTGACCATGTTCGAGTCC
GAATTTGGAATTTTGACGTTAGACAGGAATTGTAGCCAACTGAAATAGCTTTCTAATGCC
ACCGGAACCAGTCGATTTGGAGTCCGGATGAGAAAGTTATGTTTCTTAATTGCAACTCTT
TTCTTACC
>Mimulus_dentilobus
CTTTGATCATTTTTGCTTCCTCGACACCTAATTTTCGATTTGTACCTATAAGATACCAAA
TCCACCAATTAAAACAGAAATATAAATTAATCCAAAAAAAAATCCACGTAAACGTAAACG
TAGAGTGTTATTTCTACACTCCACTGTATCTTTTTTATCTAGTGTTTTTGCTGTTGTTTG
TTGATGCGTATTAAAGTCCTAGTATTGTAGTTCGTGATTTTAAGAACGTTGTTGCGAATG
CACTTTGGCGCATGTTTTTTTACATGTACGTTTGTGCATGTCCATTTGAGCGAGTGTGTT
TGTGCATGTTTGTTGATTTGTATTAATGTCCTAGGATTGTAGTTTGTGTTTTCATGATCC
TTTTTGCGAATGCACTTTTGCGTGAGTACTTTTATGCATGTTTTTGTACTTTTACTTATG
TGCGTATCTTTTCTAGCGAGTGTTTTTGTGCACGTTTCATGATGCGCATTCATGTCCTCG
AACTGTAGTACGTATTTTCATGATCGCTTTTGGAATCCAATTTTGACCATGTTCGAGTCC
GAATTTGGAAATTTGCCATTAGACGGGAATTGTAGCCAACTGAAATAGCTTTCTAATGCC
GCCGGAACCAGGTCATTTGGAGTCCGGATGAGAAAGTTATGCTCGACTTCGTGGACACTA
CCAAAGATGTTGAGGGATTCTCTCTTCTAATTTGATTTTATTTTCCTAATTGCAACTCCT
TTCTTACT
>Mimulus_laciniatus
CTTTGATCATTTTTGCTTCCTGGACACCTAATTTTCGATTTGTACCTATAAGATACCAAA
```

```
TCCACCAATTAAAACAGAAATATAAATTAATCCAAACAAAAATCCACGTAAACGTAAACG
TCGAGTTTTATTTCTACACTCCATTGTATCTTTTTTATCTAGTGTTTTTGCTGTTGTTTG
TTGATGCGTATTAAAGTCATAGTATTGTAGTTCGTGTTTTTAAGAACGTTGTTGCGGATG
CACTTTTGCGCATGTTTTTTTACATGTACGTTTGTGCATGTCCATTTGAGGGAGTGTGTT
TGTGCATGTTTGTTGATTTGTATTAATGTCCTAGGATTGTAGTTTGTGTTTTCATGATCC
TTTTTGCGAATGCACTTTTGCGTGAGTACTTTTGTGCATGCTTTTGTACTTTTACTTATG
TGCGTATCTTTTCTAGCGAGTGTTTTTGTGCACGTTTCTTGATGCGCATTCATGTCCTTG
AATTGTAGTACGTATTTTCATGATCGCTTTTGTTATCCAATTTTGACCATGTTCGAGTCC
GAATTTGGAATTTTGACGTTAGACGGGAATTGTAGCCCGCTGAAATAGCTTTCTAATGCC
GCCGGAACCAGTTGATTTGGAGTCCGGATGAGAGAGTTATGCTCGACTTCGTGGACACTA
CCAAAGATGTTGAGGGATTCTCTCTTCTAATTTGATTTTATTTTCTTAATTGCAACTCTT
TTCTTACC

##################################################################################################
Grasses

Zea_mays               -TTTTTCGCAACGAACATGCCCAATCCACTACTTTAGGTCCAAAACTCATGTTTGGG-GT 58
Zea_luxurians          -TTTTTCGCAACGAACATGCCCAATCCACTACTTTAGGTCCAAAACTCATGTTTGGG-GT 58
Setaria_italica        TTTTTTCGCAACGAACGCATGCAATCTTCCTAATTATACCCTAGAA-CATGTTTGGGAGT 59
Setaria_viridis        TTTTTTCGCAACGAACGCATGCAATCTTCCTAATTATACCCTAGAA-CATGTTTGGGAGT 59
Panicum_capillare      --TTTTCGCAAGGAACGCACCCGATCCACTCCAT-GGACCCAAAACTCATGTTTTGGG-C 56
Panicum_hallii         -TTTTTCGCAACGAACGCACCCGATCCACTCCATTGGACCCAAAACTCATGTTTTGGGGC 59
Panicum_virgatum       -TTTTTCGCAACGAACGCACCCAATCCACCCCATTGGACCCTAAACTCATGTTTTGGTAC 59
Oryza_sativa           --TTTTTGCCACGAACGCACCCAATACACTCCAATATGTCCAAAAATCATGTTTTGG--C 56
                         **** ** * ****     * **   *          ** * *  ******* **

Zea_mays               GATTTC-------------------GCGCAATTTCGTTGCCGCACGTCACCCATTCCGAA 99
Zea_luxurians          GGTTTC-------------------GCGCAATTTCGTTGCCGCACGTCACCCATTCCGAA 99
Setaria_italica        GTTTTT-------------------GAGCA-TTCGGTTCCGGCACGAAAAACGATGC-AA 98
Setaria_viridis        GTTTTT-------------------GGGCA-TTCGGTTCCGGCACGAAAAATGATGC-AA 98
Panicum_capillare      GTTTC--------------------GGACCGTTTCGTTACTGCACGAAAGTCGGTGC-AA 95
Panicum_hallii         GTTTC--------------------GGACCGTTTCGTTACTGCACGAAAGTCGATGC-AA 98
Panicum_virgatum       GTTTCATAGTGTTTGGGTGCATTTGGGATCATTTCGTAACTGCATGAAACTCGGTGC-AA 118
Oryza_sativa           CTTTTT-------------------GAACTTTTTCATTCCGGTAAAAAACATCGCACCCA 97
                         **                     *     **   *  * * *    *       *  *

Zea_mays               AACGGGTA----TCGGGGTGCATACAAA---GCACGAGTTTTTGCCACCGGAACAATTTC 152
Zea_luxurians          AACGGGTG----TCGGGGTGCATACAAA---GCACGAGTTTTTGCCACCGGAACCATTTC 152
Setaria_italica        AACAGGTGCATTTCGG----CACCCGAAT--GCACTATTTTCGGGTAGCGAAAC--TCGG 150
Setaria_viridis        AACGGGTGCATTTCGG----CACCCGAAT--GCACTATTGCCGGGTAGCGAAAC--TCGG 150
Panicum_capillare      AACGGGCC-AA-CTGGTGC-CATT--AAC--GCACAAGT-TCGCTAAACGAAG---TTGC 144
Panicum_hallii         AACTGGCCGAA-CTGGTGC-CATT--AAC--GCACAAGT-TCGCTAAACAAAG---TCGC 148
Panicum_virgatum       AACGGGGTGAA-CTGGTGC-AATT--AAT--GCAAAAGT-TCGTGCCACGAAG---TCAC 168
Oryza_sativa           CGTGTGCCAATATTGG----CATT--AATTGACAAAAGT-TCGCCGCGCGAA----TCAC 146
                            *        **     *    **    **  * *         *  *    *

Zea_mays               TTCG---- 156
Zea_luxurians          TTCG---- 156
Setaria_italica        GTGGA--- 155
Setaria_viridis        GTGGA--- 155
Panicum_capillare      GTCGGAA- 151
Panicum_hallii         GTCGGAA- 155
Panicum_virgatum       GTCGGGA- 175
Oryza_sativa           GAAGTGAG 154
                          *

Centromere repeat monomer:
>Zea_mays
TTTTTCGCAACGAACATGCCCAATCCACTACTTTAGGTCCAAAACTCATGTTTGGGGTGA
TTTCGCGCAATTTCGTTGCCGCACGTCACCCATTCCGAAAACGGGTATCGGGGTGCATAC
AAAGCACGAGTTTTTGCCACCGGAACAATTTCTTCG
>Zea_luxurians
TTTTTCGCAACGAACATGCCCAATCCACTACTTTAGGTCCAAAACTCATGTTTGGGGTGG
TTTCGCGCAATTTCGTTGCCGCACGTCACCCATTCCGAAAACGGGTGTCGGGGTGCATAC
AAAGCACGAGTTTTTGCCACCGGAACCATTTCTTCG
>Setaria_italica
TTTTTTCGCAACGAACGCATGCAATCTTCCTAATTATACCCTAGAACATGTTTGGGAGTG
TTTTTGAGCATTCGGTTCCGGCACGAAAAACGATGCAAAACAGGTGCATTTCGGCACCCG
AATGCACTATTTTCGGGTAGCGAAACTCGGGTGGA
>Setaria_viridis
TTTTTTCGCAACGAACGCATGCAATCTTCCTAATTATACCCTAGAACATGTTTGGGAGTG
```

```
TTTTTGGGCATTCGGTTCCGGCACGAAAAATGATGCAAAACGGGTGCATTTCGGCACCCG
AATGCACTATTGCCGGGTAGCGAAACTCGGGTGGA
>Panicum_virgatum
TTTTTCGCAACGAACGCACCCAATCCACCCCATTGGACCCTAAACTCATGTTTTGGTACG
TTTCATAGTGTTTGGGTGCATTTGGGATCATTTCGTAACTGCATGAAACTCGGTGCAAAA
CGGGGTGAACTGGTGCAATTAATGCAAAAGTTCGTGCCACGAAGTCACGTCGGGA
>Panicum_capillare
TTTTCGCAAGGAACGCACCCGATCCACTCCATGGACCCAAAACTCATGTTTTGGGCGTTT
CGGACCGTTTCGTTACTGCACGAAAGTCGGTGCAAAACGGGCCAACTGGTGCCATTAACG
CACAAGTTCGCTAAACGAAGTTGCGTCGGAA
>Panicum_hallii
TTTTTCGCAACGAACGCACCCGATCCACTCCATTGGACCCAAAACTCATGTTTTGGGGCG
TTTCGGACCGTTTCGTTACTGCACGAAAGTCGATGCAAAACTGGCCGAACTGGTGCCATT
AACGCACAAGTTCGCTAAACAAAGTCGCGTCGGAA
>Oryza_sativa
TTTTTGCCACGAACGCACCCAATACACTCCAATATGTCCAAAAATCATGTTTTGGCCTTT
TTGAACTTTTTCATTCCGGTAAAAAACATCGCACCCACGTGTGCCAATATTGGCATTAAT
TGACAAAAGTTCGCCGCGCGAATCACGAAGTGAG


#################################################################################################
Sorghum

Sorghum_bicolor          GATGCAAGATAGGTGCACGGTTTGCACGGAACGCACCATAGGCTAAGAAACCATTTTGGA 60
Sorghum_propinquum       GATGCAAGATAGGTGCACGGTTTGCATGGAACATACCATATGCTAAGAAATCAATTTGGA 60
Miscanthus_giganteus     GATGCAAGATAGGTGCACGGTTTGCGCCAAACGTACCATAGGCTCAGAAATCATTTTGGA 60
                         *************************    ***  ****** *** ***** ** ******

Sorghum_bicolor          CGCACCCGATGGAACTCCTAGATGAAGTGTGTCAAATGGAAGCTCGGTTCGGTCTGTTTG 120
Sorghum_propinquum       CGCACCCGATGGAACTCCTAGATGACGTGTGTCAAATGGAAGCTCGCTTTGGTCTGTTTG 120
Miscanthus_giganteus     CGCACCCGATGGAACTCCTAGGTGACGTGGGTCATGTGGAATCTCGTTTCGGTCCGTTTG 120
                         ********************* *** *** ****  ***** **** ** **** *****

Sorghum_bicolor          GAGATAGTGTTAATCTT 137
Sorghum_propinquum       GAGACAGTGCTAATCTC 137
Miscanthus_giganteus     GAGATAGTGTTAGTGTT 137
                         **** **** ** * *

--ACTUAL READS---
>Sorghum_bicolor
GATGCAAGATAGGTGCACGGTTTGCACGGAACGCACCATAGGCTAAGAAACCATTTTGGA
CGCACCCGATGGAACTCCTAGATGAAGTGTGTCAAATGGAAGCTCGGTTCGGTCTGTTTG
GAGATAGTGTTAATCTT
>Sorghum_propinquum
GATGCAAGATAGGTGCACGGTTTGCATGGAACATACCATATGCTAAGAAATCAATTTGGA
CGCACCCGATGGAACTCCTAGATGACGTGTGTCAAATGGAAGCTCGCTTTGGTCTGTTTG
GAGACAGTGCTAATCTC
>Miscanthus_giganteus
GATGCAAGATAGGTGCACGGTTTGCGCCAAACGTACCATAGGCTCAGAAATCATTTTGGA
CGCACCCGATGGAACTCCTAGGTGACGTGGGTCATGTGGAATCTCGTTTCGGTCCGTTTG
GAGATAGTGTTAGTGTT

#################################################################################################
Corals

Acropora_millepora      CAAAACCCTAGTGCACGGTACTTTGCACAAAAAGTTGCTTATCTCGAGGAG-ATCG---A 56
Acropora_palmata        CAAAAC----GTGT-CGAT-CTCCTCGAAATAAGCAACTTTTTGTGCAAAGTACCGTGCA 54
                        ******    ***  ** * **   *  ** ***   *** *   *   ** * **   *

Acropora_millepora      CAAG--TTTTGGTG-GTTTTTCAGCAAA-TGCCCTACTTTTTGCAACATTTTTCTAAAAA 112
Acropora_palmata        CTAGGGTTTTGACCCAATTTTGAGCAAAATGTTGCAAAAAGTATGGCATTTGGTGAAAAA 114
                        * **  *****      **** ****** **    *     *    *****    *****

Acropora_millepora      TTGGGT 118
Acropora_palmata        CCAC-- 118

Candidate centromere repeat monomer:
>Acropora_millepora
CAAAACCCTAGTGCACGGTACTTTGCACAAAAAGTTGCTTATCTCGAGGAGATCGACAAG
TTTTGGTGGTTTTTCAGCAAATGCCCTACTTTTTGCAACATTTTTCTAAAAATTGGGT
```

```
>Acropora_palmata
CAAAACGTGTCGATCTCCTCGAAATAAGCAACTTTTTGTGCAAAGTACCGTGCACTAGGG
TTTTGACCCAATTTTGAGCAAAATGTTGCAAAAAGTATGGCATTTGGTGAAAAACCAC

#####################################################################################
Mosquitos
anopheles_gambiae         ATGGCCGTAACAACGATAATAGATGGCAACAAAATTCAACATCAAATTTCAAGGCCATTC 60
anopheles_gambiae_S       ATGGCCGTAACAACGATAATAGATGGCAACAAAATTCAACATCAAATTTCAAGGCCATTC 60
anopheles_gambiae_M       ATGGCCGTAACAACGATAATAGATGGCAACAAAATTCAACATCAAATTTCAAGGCCATTC 60
                          ************************************************************

anopheles_gambiae         AAATAGTGCAAGATGGCTTCATTTGGATGAAAC 93
anopheles_gambiae_S       AAATAGTGCAAGATGGCTTCATTTGGATGAAAC 93
anopheles_gambiae_M       AAATAGTGCAAGATGGCTTCATTTGGATGAAAC 93
                          *********************************

Candidate centromere repeat monomer:
>anopheles_gambiae
ATGGCCGTAACAACGATAATAGATGGCAACAAAATTCAACATCAAATTTCAAGGCCATTC
AAATAGTGCAAGATGGCTTCATTTGGATGAAAC
>anopheles_gambiae_M
ATGGCCGTAACAACGATAATAGATGGCAACAAAATTCAACATCAAATTTCAAGGCCATTC
AAATAGTGCAAGATGGCTTCATTTGGATGAAAC
>anopheles_gambiae_S
ATGGCCGTAACAACGATAATAGATGGCAACAAAATTCAACATCAAATTTCAAGGCCATTC
AAATAGTGCAAGATGGCTTCATTTGGATGAAAC

#####################################################################################
Drosophila 1

Drosophila_erecta       TAATTCCCAACTTGTTATGGCTTATATTTCATTATACGTTCCCTCTAACAGCCTATAAAG 60
Drosophila_yakuba       TAATTCCCAACTTGTTCTGG-TTATTTTTCCTTATATGTACTCTCTACCTGCCCATAAAC 59
                        **************** *** **** **** ***** ** * ***** * *** *****

Drosophila_erecta       TAGTGGACAGGAAG--TTCCGTGAAT-TTAGT-TAATAAATGTG--TTCATG-------- 106
Drosophila_yakuba       TAGAGAATGTAATGGCCTCTATAAACGTTGATGCAGGCGATGGGGATTCAGGAGCTGCGC 119
                        *** * *    * *   **  * **  **  *  *    *** *  **** *

Drosophila_erecta       ----------TTTGTGTTTG---CGC-----------ACGAAA------------AGTGG 130
Drosophila_yakuba       GGGTGTAAAATCTGCATTCAGGACGCCGTGAACCAGGACCAAACCAAGCCAGGTTAGTAG 179
                                  * **  **     ***           ** ***            *** *

Drosophila_erecta       T--------------------------------TTCAT---GTGGTGCGCAGATAAA--C 153
Drosophila_yakuba       TGGAATGAACTATATTTTTTAAAACTATATGTGTTCATTGAATAATACACAATTAAAGTC 239
                        *                                *****    *  * * **  ****  *

Drosophila_erecta       AATCTACATCCAG----AAAGAA---GAA-------------------AATATAA----- 182
Drosophila_yakuba       CATCGACATTTAATATTAAAAAAACTGAATTACTGTCAGCTTACACCAAATATGACATGT 299
                         *** ****  *     *** **   ***                   ***** *

Drosophila_erecta       ------ACTCT--AAACTCTAGAC--------CAAGTCA---TCGG-----TAATTGTAA 218
Drosophila_yakuba       CAACTGATTCCCGAAAATATACACAGTCAAAACAAGTTAAGTTTGGCCACGTAATTGTTA 359
                              * **   *** * ** **        ***** *   * **     ******* *

Drosophila_erecta       --------TTAAAAACTGGTGCACATAGTGTTCAAAAA 248
Drosophila_yakuba       ATGATTGTTTGAAAAGTA-TATACATGATGCTTGTAAA 396
                                ** **** *  *  ****  ** *   ***

Candidate centromere repeat monomer:
>Drosophila_erecta
TAATTCCCAACTTGTTATGGCTTATATTTCATTATACGTTCCCTCTAACAGCCTATAAAG
TAGTGGACAGGAAGTTCCGTGAATTTAGTTAATAAATGTGTTCATGTTTGTGTTTGCGCA
CGAAAAGTGGTTTCATGTGGTGCGCAGATAAACAATCTACATCCAGAAAGAAGAAAATAT
AAACTCTAAACTCTAGACCAAGTCATCGGTAATTGTAATTAAAAACTGGTGCACATAGTG
TTCAAAAA
>Drosophila_yakuba
TAATTCCCAACTTGTTCTGGTTATTTTTCCTTATATGTACTCTCTACCTGCCCATAAACT
AGAGAATGTAATGGCCTCTATAAACGTTGATGCAGGCGATGGGGATTCAGGAGCTGCGCG
GGTGTAAAATCTGCATTCAGGACGCCGTGAACCAGGACCAAACCAAGCCAGGTTAGTAGT
GGAATGAACTATATTTTTTAAAACTATATGTGTTCATTGAATAATACACAATTAAAGTCC
ATCGACATTTAATATTAAAAAAACTGAATTACTGTCAGCTTACACCAAATATGACATGTC
```

```
AACTGATTCCCGAAAATATACACAGTCAAAACAAGTTAAGTTTGGCCACGTAATTGTTAA
TGATTGTTTGAAAAGTATATACATGATGCTTGTAAA

################################################################################################
Drosophila 2

Drosophila_sechellia      TTTGTGCAAAATTTTTGGATTTTTCGATTTTAGATACCAGGCGATGATAATCAGTAGCGG 60
Drosophila_simulans       TTTGTGCAAAATTTTTGGATCTTTCGATTTTAGATACCAGGCGATGATAATCAGTAGCGG 60
                          ******************** ***************************************

Drosophila_sechellia      GTGTCTACTGAAAACCAACTAATCGTTGGTCACCTTCTGGAATTCTTGTTCGCCTGGTAA 120
Drosophila_simulans       GTGTCTACAGAAAACCA-CTTATCGTTGGTCACCTTCTGGAATTCTTGTTCGCCTGGTAG 119
                          ******** ******** ** **************************************

Drosophila_sechellia      TTTAAACCGAAAAATCTCTCAATTTG--CAACAAAATGCGTATT 162
Drosophila_simulans       TTTAAACCGAAAAATCTCTCAATTTGGCCTACAAAATGCGCATT 163
                          **************************  * ********** ***

Candidate centromere repeat monomer:
>Drosophila_sechellia
TTTGTGCAAAATTTTTGGATTTTTCGATTTTAGATACCAGGCGATGATAATCAGTAGCGG
GTGTCTACTGAAAACCAACTAATCGTTGGTCACCTTCTGGAATTCTTGTTCGCCTGGTAA
TTTAAACCGAAAAATCTCTCAATTTGCAACAAAATGCGTATT
>Drosophila_simulans
TTTGTGCAAAATTTTTGGATCTTTCGATTTTAGATACCAGGCGATGATAATCAGTAGCGG
GTGTCTACAGAAAACCACTTATCGTTGGTCACCTTCTGGAATTCTTGTTCGCCTGGTAGT
TTAAACCGAAAAATCTCTCAATTTGGCCTACAAAATGCGCATT

################################################################################################
Drosophila 3

Drosophila_ficusphila      CCATAACTTTTGAAATTTTTACCCGATTTAAAAGTGGAATACCTCTCTGA 50
Drosophila_rhopaloa        CCATAACTTTACCAAACTTAAGCCGATTCCAAAGTGGCATACCTCTAAAG 50
                           **********   **  ** * ******  ******* ********

Drosophila_ficusphila      ATTTGTTATTAAAATATCTATCTAGCTGCATTATTGGTTTATTTTCGAAT 100
Drosophila_rhopaloa        ACTTCTGATTCGATTCTCTAAAAATCTGCATT-----CAAATTTTTCAAT 95
                           * ** * ***  * * ****   * *******        *****  ***

Drosophila_ficusphila      TT----TTGGTCAATATTTTGTAATTTTTTATGACCCCCGACCTGTCAAA 146
Drosophila_rhopaloa        TTAACGTTGATCAA-ATTTT-TGACCCATTTTCATGTCAGTTATTTCTAA 143
                           **    *** **** ***** * *    ** * *   * *   * ** **

Drosophila_ficusphila      ATTTGCAAAAAATGGGTTTGCAGAAAAGTGACCAGATCCCAGCACTG--C 194
Drosophila_rhopaloa        -TTTCCGA-----GGACTTGTAACAGAGTGACT------CAAAATTGGAC 181
                            *** * *     **  *** *  * ******       **  * **  *

Drosophila_ficusphila      TTAGC---- 199
Drosophila_rhopaloa        TTAGAGAGT 190
                           ****

Candidate centromere repeat monomer:
>Drosophila_ficusphila
CCATAACTTTTGAAATTTTTACCCGATTTAAAAGTGGAATACCTCTCTGAATTTGTTATT
AAAATATCTATCTAGCTGCATTATTGGTTTATTTTCGAATTTTTGGTCAATATTTTGTAA
TTTTTTATGACCCCCGACCTGTCAAAATTTGCAAAAAATGGGTTTGCAGAAAAGTGACCA
GATCCCAGCACTGCTTAGC
>Drosophila_rhopaloa
CCATAACTTTACCAAACTTAAGCCGATTCCAAAGTGGCATACCTCTAAAGACTTCTGATT
CGATTCTCTAAAAATCTGCATTCAAATTTTTCAATTTAACGTTGATCAAATTTTTGACCC
ATTTTCATGTCAGTTATTTCTAATTTCCGAGGACTTGTAACAGAGTGACTCAAAATTGGA
CTTAGAGAGT

################################################################################################
Drosophila 4

Drosophila_virilis      ACTATATCATATAGCTGCCATAGGAACGATCGGTCGAAAATTAAGTTTTTGTATGAAA-- 58
Drosophila_elegans      ACTATATCATATAGCTCCCATAGGAACAA----TCGAAAAATAAAT----GAAAAAAATT 52
                        **************** ********** *    ******* *** *    * *  ***

Drosophila_virilis      --AACATTT-TGTTTTTCAAGATATCTTGACCAAAC--TCGGCATTTATTAGTTTTACTA 113
```

```
Drosophila_elegans      ATAACTTTTCTGTTTTT---AACTTTTTGTTTAGTTCTTCGACATATAGCAATGGTTAAA 109
                          *** *** *******    *  * ***   *     *** *** **  * *  *   *

Drosophila_virilis      TACTCCTCATATATATGCAAAATCCTATTAAGATCGGACC 153
Drosophila_elegans      TATTTCAGAATTACGGTTTAAATTTCATCAAAATCGGACG 149
                        ** * *  *  **      ****   ** ** *******
Candidate centromere repeat monomer:
>Drosophila_virilis
ACTATATCATATAGCTGCCATAGGAACGATCGGTCGAAAATTAAGTTTTTGTATGAAAAA
CATTTTGTTTTTCAAGATATCTTGACCAAACTCGGCATTTATTAGTTTTACTATACTCCT
CATATATATGCAAAATCCTATTAAGATCGGACC
>Drosophila_elegans
ACTATATCATATAGCTCCCATAGGAACAATCGAAAAATAAATGAAAAAAATTATAACTTT
TCTGTTTTTAACTTTTTGTTTAGTTCTTCGACATATAGCAATGGTTAAATATTTCAGAAT
TACGGTTTAAATTTCATCAAAATCGGACG

###################################################################################################
Wasps

Nasonia_giraulti         CGCTTTGGTTTTAGATTTTATACTCGCTTCGCTCGCCTTCCTTCGATGTGCCGAGGTGTT 60
Nasonia_longicornis      CGCTTTGGTTTTAGATTTTATACTCGCTTCGCTCGCCTTCCTTCGATGTGCCGAGGTGTT 60
Nasonia_vitripennis      CGCTTTGGTTTTTGATTT-ATACTCGCTTTGCTCGCCTTCCTCCGATGTGCCGAGGTGTT 59
                         ************ ***** ********** ************ *****************

Nasonia_giraulti         TTTTAAATAAATTATTGAGCTCGGGGAGCGTGGTGTAGTTGGGTTTAAG 109
Nasonia_longicornis      GATTAAATAAATTATTGAGCTCGGGGAGCGTGGTGTAGTTGGGTTTAAG 109
Nasonia_vitripennis      GTTTAAATAAATTATTGAGCTCGGGGAGCGTGGTGTAGTTGGGTTTAAG 108
                           ***********************************************

Candidate centromere repeat monomer:
>Nasonia_giraulti
CGCTTTGGTTTTAGATTTTATACTCGCTTCGCTCGCCTTCCTTCGATGTGCCGAGGTGTT
TTTTAAATAAATTATTGAGCTCGGGGAGCGTGGTGTAGTTGGGTTTAAG
>Nasonia_longicornis
CGCTTTGGTTTTAGATTTTATACTCGCTTCGCTCGCCTTCCTTCGATGTGCCGAGGTGTT
GATTAAATAAATTATTGAGCTCGGGGAGCGTGGTGTAGTTGGGTTTAAG
>Nasonia_vitripennis
CGCTTTGGTTTTTGATTTATACTCGCTTTGCTCGCCTTCCTCCGATGTGCCGAGGTGTTG
TTTAAATAAATTATTGAGCTCGGGGAGCGTGGTGTAGTTGGGTTTAAG

###################################################################################################
Zebrafish

Danio_rerio               ATTTTCACTTGCAACTTGCTTTAA----CATGTTTAACAGTGCATGAATTGTT---GTCT 53
Danio_nigrofasciatus      ATTTTCAGCTTTCTCTTGCTCTGAAAAGCGTGTTTAGAGTTGCCTGTACTGTTAGTGATC 60
Danio_albolineatus        ATTTTCACTTGCAAGTTGTTAAAA----CATGCTTAATACAGCATGCACTGTTG--TTTT 54
                          *******  *     *** *   *    * ** ***     ** ** * ****

Danio_rerio               AAAACAAAAAAGAATGATGCATTTTATGGCTGGATC--GTTTT-ATCACTTTCTGAATTG 110
Danio_nigrofasciatus      AAAACAAAAA-GAATGATGGTTTTCATCATTTCTTTTTGTTTTGATCACTCTCTGTTCTG 119
Danio_albolineatus        TAAACTTAAA-GAATGATGATTTTCACTGTTTTAAGCAAATTCTGTCACTTGCTCTTTTT 113
                           ****  *** ********  *** *    *         **    *****  **    *

Danio_rerio               G-ACGTTGAAAATGAGCTTATCTCAG---CCAGCAGAGAGGTCA----AATCACAGTAAT 162
Danio_nigrofasciatus      G-GCCCTCTAAACG----TGTTTTAA---TAAGTTGCAAGTGAC----AATCA--GCAGA 165
Danio_albolineatus        GCACTTACCAAGC-------TCTGAAAAGCAAGTTGCAGAATTCTAGGAATTG--ACAAT 164
                          *  *     **         * * *      **  *            ***      *

Danio_rerio               GCAAACTCAGTCAGAGACATAG---- 184
Danio_nigrofasciatus      TCTGGCTGAGTTTGCCATACTGTTT- 190
Danio_albolineatus        TATGAAACA--TTATCCTTTTGCTTA 188
                                  *            *

Candidate centromere repeat monomer:
>danio_rerio
TAAAACGATCCAGCCATAAAATGCATCATTCTTTTTTGTTTTAGACAACAATTCATGCAC
TGTTAAACATGTTAAAGCAAGTTGCAAGTGAAAATCTATGTCTCTGACTGAGTTTGCATT
ACTGTGATTTGACCTCTCTGCTGGCTGAGATAAGCTCATTTTCAACGTCCAATTCAGAAA
GTGA
>Danio_albolineatus
TTTAAGCAAATTCTGTCACTTGCTCTTTTTGCACTTACCAAGCTCTGAAAAGCAAGTTGC
```

```
AGAATTCTAGGAATTGACAATTATGAAACATTATCCTTTTGCTTAATTTTCACTTGCAAG
TTGTTAAAACATGCTTAATACAGCATGCACTGTTGTTTTTAAACTTAAAGAATGATGATT
TTCACTGT
>Danio_nigrofasciatus
GTTTAGAGGGCCCAGAACAGAGAGTGATCAAAACAAAAAGAAATGATGAAAACCATCATT
CTTTTTGTTTTGATCACTAACAGTACAGGCAACTCTAAACACGCTTTTCAGAGCAAGAGA
AAGCTGAAAATAAACAGTATGGCAAACTCAGCCAGATCTGCTGATTGTCACTTGCAACTT
ATTAAAACAC

####################################################################################################
Cichlids

Pundamilia_nyererei           GTAACTTTTGATAGAAGACTCAGACAAACATGTTTATGGCTTTATCTTAT 50
melanochromis_auratus         GTAACTTTTGATAGTAGACTCAGACAGACATGTTCATGGCTTTATCTTAT 50
labeotropheus_fuelleborni     GTAACTTTTGATAGAAGACTCAGAAACACATGTTTATGGCTTTATCTTAT 50
metriaclima_zebra             GTAACTTTTGATAGAAGACTCAGATACACATGTTTATGGCTTTATCTTAT 50
rhamphochromis_esox           GTAACTTTTGATAGAAGACTCAGACAAACATGTTTATGGCTTTATCTTAT 50
oreochromis_niloticus         GTAACTTTTGATAGGAGACTCGGACACACATATTTCAGGCTTGGCCTTAT 50
                              ************** ****** ** * **** **   *****   *****

Pundamilia_nyererei           AGAACTCAATATCCCCGTGCTGGGCAAACAGGTTTTGCAGCCGTTTGAGC 100
melanochromis_auratus         AGAACTCAATATCCCCGTGCTGGCCGAACAGGTTTTGCAGCCGTTTGAGC 100
labeotropheus_fuelleborni     AGAACTCAATATCCC-GTGCTGGGCAAACAGGTTTTGCAGCCGTTTGAGC 99
metriaclima_zebra             AAAACTCAAGATCTCCGTGCTGGGAAAACAGGTTTTGCAGCCGTTTGAGC 100
rhamphochromis_esox           AGAACTCAATGTCCCCGTGCTGGGAAAACAGGATTTGCAGCTGTTTGAGC 100
oreochromis_niloticus         AGAATTCAGCATTTCCATGCTGGGGAAATAGGTTTTGCCACTGTTTGAGC 100
                              * ** ***   *  *  ******   ** *** *****  * ********

Pundamilia_nyererei           TAAGATTTTAAAATTATTCACATAATGAAAACCTATACTTTGTTTCAGGC 150
melanochromis_auratus         TAAGATTTTAAAATTATTCACATAATGAAAACCTATACTTTGTCTCGGGC 150
labeotropheus_fuelleborni     TAAGATTTTAAAATTATTCACATAATGAAAACCTATACTTTGTTTCAGGC 149
metriaclima_zebra             TAAGATTTTAAAATTATTCACATAATGAAAACCTATACTTTGTCTCGGGC 150
rhamphochromis_esox           TAAGATTTTCAAGTTATACACATAATGAAAACCTATACTTTGTTTCGGTC 150
oreochromis_niloticus         TAAGATTTTCA-GTTATTGACTAAATGAAAACC-ATAATGTGTTTCAGGC 148
                              ********* *  ****  **  ********** *** * *** ** * *

Pundamilia_nyererei           GAGTTTCCCATTCAAATGCATGT--CAGTGAGAAACGCACTGTCTTGGCG 198
melanochromis_auratus         GAGTTTCCCATTCAAATGCATGTAACAGTGAGAAACGCACTGTCTTGGCG 200
labeotropheus_fuelleborni     GAGTTTCCCATTCAAATGCATGTAACAGTGAGAAACGCATTGTCTTGGCG 199
metriaclima_zebra             GAGTTCCCCATTCAAATGCATGTAACAGTGAGAAACGCATTGTCTTGGCG 200
rhamphochromis_esox           GAGTTCCCCATTCAAATGCATGTAACAGTGAGAAACGCACTGTCTTGGCG 200
oreochromis_niloticus         GAG-----------------------------AAACGCACTGTCTCGCCG 169
                              ***                             ******* ***** * **


Pundamilia_nyererei           AAATAAAGCGTTTTTGTACAACTTCATATAAATCGCT 235
melanochromis_auratus         AAATAAAGCATTTTTGTACAACTTCATATAAATCGCT 237
labeotropheus_fuelleborni     AAATAAAGCGTTTTTGAACAACTTCATATAAATCGCT 236
metriaclima_zebra             AAAGAAAGCGTTTTTGTACAACTTCATATAAATAGCT 237
rhamphochromis_esox           AAAGAAAGCGTTTTTGTACAACTTCATAAAAATCGCT 237
oreochromis_niloticus         AAATAAGGCGATTTTCACCAAGTCCATAAAGACAGCT 206
                              *** ** **  ****   *** * **** * *  ***

Candidate centromere repeat monomer:
>pundamilia_nyererei
GTAACTTTTGATAGAAGACTCAGACAAACATGTTTATGGCTTTATCTTATAGAACTCAAT
ATCCCCGTGCTGGGCAAACAGGTTTTGCAGCCGTTTGAGCTAAGATTTTAAAATTATTCA
CATAATGAAAACCTATACTTTGTTTCAGGCGAGTTTCCCATTCAAATGCATGTCAGTGAG
AAACGCACTGTCTTGGCGAAATAAAGCGTTTTTGTACAACTTCATATAAATCGCT
>melanochromis_auratus
GTAACTTTTGATAGTAGACTCAGACAGACATGTTCATGGCTTTATCTTATAGAACTCAAT
ATCCCCGTGCTGGCCGAACAGGTTTTGCAGCCGTTTGAGCTAAGATTTTAAAATTATTCA
CATAATGAAAACCTATACTTTGTCTCGGGCGAGTTTCCCATTCAAATGCATGTAACAGTG
AGAAACGCACTGTCTTGGCGAAATAAAGCATTTTTGTACAACTTCATATAAATCGCT
>labeotropheus_fuelleborni
GTAACTTTTGATAGAAGACTCAGAAACACATGTTTATGGCTTTATCTTATAGAACTCAAT
ATCCCGTGCTGGGCAAACAGGTTTTGCAGCCGTTTGAGCTAAGATTTTAAAATTATTCAC
ATAATGAAAACCTATACTTTGTTTCAGGCGAGTTTCCCATTCAAATGCATGTAACAGTGA
GAAACGCATTGTCTTGGCGAAATAAAGCGTTTTTGAACAACTTCATATAAATCGCT
>metriaclima_zebra
GTAACTTTTGATAGAAGACTCAGATACACATGTTTATGGCTTTATCTTATAAAACTCAAG
```

```
ATCTCCGTGCTGGGAAAACAGGTTTTGCAGCCGTTTGAGCTAAGATTTTAAAATTATTCA
CATAATGAAAACCTATACTTTGTCTCGGGCGAGTTCCCCATTCAAATGCATGTAACAGTG
AGAAACGCATTGTCTTGGCGAAAGAAAGCGTTTTTGTACAACTTCATATAAATAGCT
>rhamphochromis_esox
GTAACTTTTGATAGAAGACTCAGACAAACATGTTTATGGCTTTATCTTATAGAACTCAAT
GTCCCCGTGCTGGGAAAACAGGATTTGCAGCTGTTTGAGCTAAGATTTTCAAGTTATACA
CATAATGAAAACCTATACTTTGTTTCGGTCGAGTTCCCCATTCAAATGCATGTAACAGTG
AGAAACGCACTGTCTTGGCGAAAGAAAGCGTTTTTGTACAACTTCATAAAAATCGCT
>oreochromis_niloticus
GTAACTTTTGATAGGAGACTCGGACACACATATTTCAGGCTTGGCCTTATAGAATTCAGC
ATTTCCATGCTGGGGAAATAGGTTTTGCCACTGTTTGAGCTAAGATTTTCAGTTATTGAC
TAAATGAAAACCATAATGTGTTTCAGGCGAGAAACGCACTGTCTCGCCGAAATAAGGCGA
TTTTCACCAAGTCCATAAAGACAGCT

#################################################################################################
Cats

Felis_catus           GCACTGGGTTCACTGGAGGCTGCAGTGCCGCGG 33
Felis_silvestris      GCACTGGGTTCACTGGAGGCTGCAGTGCCGCGG 33
                      *********************************

Candidate centromere repeat monomer:
>Felis_catus
GCACTGGGTTCACTGGAGGCTGCAGTGCCGCGG
>Felis_silvestris
GCACTGGGTTCACTGGAGGCTGCAGTGCCGCGG

#################################################################################################
Bovine

Bison_bison             AATGGAAGATTGGACTT-CCCTGGGCCAAACACAAGAGG--CATC-CTGAATTCCCCGT- 55
Bos_taurus_taurus       AATGGAAGATTGGACTT-CC-TGGGC--AACACAAGAAGGCCATCACTGATTTCCCCGTT 56
Bos_grunniens           -ATGGAAGATTGGACTT-CCCTGGGCC-AACACAAGAGG--CATC-CTGAATTCCCCGT- 53
Bos_taurus_indicus      AATGGAAGATTGAACTTGCCCTGGGCC-AAC-CAAAAGG--CATC-CTGACTTCCCCGT- 54
                         *********** **** ** *****  *** *** * *  **** **** ********

Bison_bison             CGTAATTCGAGAATCC--TGCC-GACACTCGAGAAAATCCACGTG--TTCCCCCGTCATC 110
Bos_taurus_taurus       CGTAACTCGAGATTCCGCCGCC-AA-ACTCGAGACA-CCAACGTGGATTCCCCCGTCATC 113
Bos_grunniens           CGTAACTCGAGAATCC--CGCC-GCAACTCGAGAAAAACCACGTGG-TTCCCCCGTCATC 109
Bos_taurus_indicus      CGTAACTC-AGAATCC-CCGCCCGTAACTCGAGAAAAACCACGTGG-CTCCCCGTCCATC 111
                        ***** ** *** ***   ***    ******** *  * *****    *****   ****

Bison_bison             -GCA--AGAT-GAAGCCCTTTCCCGCTACAGT--GTCT--CAGGAGAAGTCCCACGTTAG 162
Bos_taurus_taurus       -GCACAAGATTGAAGCCCTTTCC-GCTACAGC--GTCT--CAGGAGAAGTCCCACGTTAG 167
Bos_grunniens           -GCA--AGAT-GAAGCCCTTTCCCGCTACAGC--GTCT--CAGGAGAAGTCCCACGTTAG 161
Bos_taurus_indicus      CGCA--AGAT-GAAGGCCCTTCCCGCAACAGCCGGCCCTCCAGGAGAAGTCCCACGT-AG 167
                         ***  **** **** ** **** ** ****    * *   **************** **

Bison_bison             GTATTGGAGGTCGAAACGGTACTTGGC-ACC-TTGATG-CGACC--ACAAAGTGCCCCGA 217
Bos_taurus_taurus       GAATTGAGCGGTGGAACGGTACTTGGC-ACCCTTGACGGCGGCCCCACAAAGTTCCCCGG 226
Bos_grunniens           GAATTGGAGGTCGAAACGGTACTTGGCCACCCTTGATG-CGACCC-ACAAAGTTCCC-GA 218
Bos_taurus_indicus      G-ATTGGAGGTCGAAAGGGCACTTGGC-GCCCTTGATG-CGACCC-ACAAAGTTCCCCGA 223
                        * ****   *  * ** ** *******  ** **** * ** **  ******* *** *

Bison_bison             CATCCCGGTCCTCCCTCGAGAGGAACACC-GAAGTTTTCCGGC-ACCACTTCCTCTGAGC 275
Bos_taurus_taurus       CATCCCG--TCTCCCTGGAGAGGAACACC-GAG--TTTCCGGC-ACCACTTCCTCTGAGC 280
Bos_grunniens           CATCCCGGTCTCCCTCGGAGAGGAACCCCTGAGGTTTTCCGGCCACCACTTCCTCTGAGC 278
Bos_taurus_indicus      AATCCCCGGTCTCCCTCGAGAGGAACACT-GAGGCTTTC--GC-ACCCCCTCCTCTGACC 279
                         *****      **   ********* *  **   ****  ** *** * ******** *

Bison_bison             CCCTTCTACCC-TCCTGA-TCTGGACAGG-AGGGTCGACTCCCCTGCTTTGTCTGGAAGG 332
Bos_taurus_taurus       CCCTTCTACCC-TCCTGA-TCTCGACAGGGAGGGTCGACTCCCCTGCTTTGTCTGGAAGG 338
Bos_grunniens           CCCTTCTACCC-TCCTGA-TCTGGACAGG-AGGGTCGACTCCCCTGCTTTTTCTGGAAGG 335
Bos_taurus_indicus      CTTTTCTCCCCCTCCTGAATCTGGACAGG--AGGTCGACTCCCCTGCTT-GTCTGGAAGG 336
                        *  **** *** ****** *** ******   ****************  *********

Bison_bison             GGTTCCCGACCTTCCGGTCGCACCTCAGGATGAGGCCGGGTCTCA-CGACGAC-ATTTCA 390
Bos_taurus_taurus       GGTTCCCGACCTTCCGGTCGCACCTCAGGATGAGGCCGGG-CTCA-CGACGAC-ATTCCA 395
Bos_grunniens           GGTTCCCGACCTTCCGGTCGCACCTCAGGATGAGGCCGGT-CTCA-CGACGAC-ATTCCA 392
Bos_taurus_indicus      GGTTCC-GACCTTCCGTCCACCTCCAGGAATGAGGCCGGT-CTCAACGAAGACCATTCCA 394
                        ****** *********  * *  *   * *********  **** *** *** *** **
```

```
Bison_bison             G-ACGTGGCCT--CGTGGGTGGTTCC-ACATTCCGAAGGAC-CCCGATTTCCCGGTCCCC 445
Bos_taurus_taurus       G-ACGTGGCCT--CGTGGGTGCTTCC-ACATTGCGAAGCAC-CCCGATTTCCCGGTCCCC 450
Bos_grunniens           GGACGTGGCCT--CGTGGGTGGTTCC-ACATTGCGTAAGACACCCGATTTCCCGGTCCCC 449
Bos_taurus_indicus      --ACGTGCCCCTCCGTGGGTGGTTCCCACATT-CGTAGGAC-CCCGATTCCC--GTCCCC 448
                          ***** **   ******** **** ***** ** *  ** ******* **  ******

Bison_bison             TCTTG-ATAAGAACCCGATGCCCGGACACCTCTT-CGAACCTCCACCCTGTGAATGAAGT 503
Bos_taurus_taurus       TCTTG-GTAAGAACCCGATGCCCGGACACCTCT--CGAAC-TCCACCCTGTGAATGAGGT 506
Bos_grunniens           TCTTG-ATAAGAACCCGATGCCCGGACACCTCTTCCGAAC-TCCACCCTGTGAATGAAGT 507
Bos_taurus_indicus      TCCTGGATAAGAACCCGATGCCGGACCACCTCTC-CGAACTCCAACCCTG-GAATGAAGT 506
                        ** **  *************** *  *******  *****  * ****** ****** **

Bison_bison             CAACACGAAAGGGCAGTGACTCGCCCGTGCATCGTCGGGAAAAAACCCCCAGGTTCCAAA 563
Bos_taurus_taurus       CAACACGAAGGGGCAGTG-CCCGCCCGTGCATCGTCCGGAAAGAACCCCCAGGTTCCAAA 565
Bos_grunniens           CAACACGAAGGGGCAGTGACTCGCCCGTGCATCGTTCGGGGAAAACCCCCAGGTTCCAAA 567
Bos_taurus_indicus      CAACACGAAGGGCCAAT---TTTTCCGTGCATCGTTCAGAAAAA---CCCAG-TTCCAAA 559
                        ********* ** ** *       ***********   *  * *   ***** *******

Bison_bison             TACCGCTCGACAAGT-GGCCTGTCT-CCCCGGGAAACACCTCGAGAGGCAAGCGGAGTTC 621
Bos_taurus_taurus       TACAGCTCGACAAGT-GGCCTCTCT-CCCCGGGGA-CACCTCGAGAGGCAAGCGGAGTTC 622
Bos_grunniens           TACAGCTCGACAATG-GGCCTCTCT-CCCCGGGGA-CACCTCGAGAGGCAAGCGGAGTTC 624
Bos_taurus_indicus      TACAGCTCGACAAGCCGGCTCTTCTTCCCCCGGGACCATCTCGAGATGCAAGCGGAGTTC 619
                        *** *********   ***   *** **** ** * ** ******* *************

Bison_bison             CATGCCTCA--CCCAAGACGAGG-CCTGACTCTCCCTGTCCCAAGTCTGCAGGGACCTTG 678
Bos_taurus_taurus       CATGCCTCAA-CCCAAGACGAGGACCTGACTCTCC-TGTCCCCAGTCTGCAGGGACCCTG 680
Bos_grunniens           CATGCCTCAA-CCCAAGACGAGG-CCTGACTCTCC-TGTCCCAAGTCTGCAGGGACCCTG 681
Bos_taurus_indicus      CATGCCTCAAACCAAAGACAG----CCGACTCTCC-TGTCCCAGTCCTGCAGA---CCTG 671
                        *********  ** *****      * ******** ******    ******    * **

Bison_bison             CGAT-CAGAGTCTGAAGTCAGAGGAACCCTGAGGTTCC-TGCCTCAACTGGAGATG-AGG 735
Bos_taurus_taurus       CGATTCGGAGTCTGAAATCAGAGGAACCCTGAGGTTCC-TGCCTCAACTGGAGATGGAGG 739
Bos_grunniens           CGAT-CGG--TCTGAAATCAGAGGAACCCTGAGGTTCC-TGCCTCAACTGGAGATG-AGG 736
Bos_taurus_indicus      CGATCGGAAGTCTGAA-TCA-AGGTACCCTGCGGTTCCGCCCCTCAACTGGAGATG-AGG 728
                        ****      ****** *** *** ****** ******   *************** ***

Bison_bison             CCCTCTTCCAATGCACCAAA-CCCAGTGGGAGTGCCGAGAAGCCC-TCCCA-CCTCCAGT 792
Bos_taurus_taurus       CCCTCTTCCAATGCACCAAAACCCCGTGG-AGTCCCGAGAGGCCCCTCCCA-CCTCCAGT 797
Bos_grunniens           CCCTCTTCCAATGCACCAA--CCCCATGG-AGTCCCGAGAGGCCCCTCCCAACCTCCAGT 793
Bos_taurus_indicus      CC-TCTTCCAATGCACCAAAACCCAGTGG-GGTCCCGAGAGGCCCTTCCCA-CCTCCAGG 785
                        ** ****************  ***  ***  ** ****** **** ***** ******* 

Bison_bison             TTTCCCTGACTTCTCAGAGCC-ACCATGAGAAGCCCCCTGAGGTCACCTGCA-CAAGTCA 850
Bos_taurus_taurus       TC-CGCTGA-TTCTCAGAGCCCACCATGAGAAGCCCCCTGAGGTCACCTGCA-CAAGTCG 854
Bos_grunniens           TT-CCCTGACTTCTCAGAGCC-ACCATGAGAAGCCCCCTGAGGTCACCTGCA-CAAGTCG 850
Bos_taurus_indicus      TT-CCCTG-CTTCTCAGAGCC-ACCATGGAGAAGCCCCTGAGGTCACCTGCAACAAGTCG 842
                        *  * ***  *********** ******   *  ****************** ****** 

Bison_bison             AGGGAAGCCA--GGTTTTCTGCCTCAACCCGAGAAAGACC---TCGAGAGACCTTCTTCA 905
Bos_taurus_taurus       AGGGAAGCCAAGGGTTCCCTGCCTCAACCCGAGAAAGACC---TCGAGAGACCTTCTTCA 911
Bos_grunniens           AGGGAA-CCA-GTGTTTCCTGCCTCAACC-GAGAAAGACCCCCTCGAGAGACCTTCTTCA 907
Bos_taurus_indicus      AGG-AACCCA-GGGTTTCCTGCCTCAACC--GAAAAGACC---TCGAGAGACCTTCT-CA 894
                        *** ** ***   ***  ***********    *******   ************** **

Bison_bison             ACA-CGTCTC-GAG-CCAGATTCCCCTAACAGTGACTC-GAGA-GCAATGACGCGCTCCC 960
Bos_taurus_taurus       ACATCGTCTC-GAGGCCAGGTTCCC-TACCAG-GACTC-GAGG-GCAATGACGCGCTCCC 966
Bos_grunniens           ACA-CGTCTC-GAGGCCAGATTCCCCTACCAT-GACTCCGAGGAGCAATGACGCGCTCCC 964
Bos_taurus_indicus      ACA-CGTCTCTGAGGCCAA--TCCC-TAACAT-GGCTCGGGAATCCAGTGACGCGCTCCC 949
                        *** ****** *** ***   **** ** **  * *** *     ** ************

Bison_bison             CCTCGCCATTGCGCA-TGGAGACCCCGACTTCCCTGGCGCCC-ACGAGAG--CTCACTGA 1016
Bos_taurus_taurus       CCTCGCCACT-CGCG-TGGAGACCC-GACTTCCCTGGCGCCCCACGAGAGG-CTCACTGA 1022
Bos_grunniens           CCTCGCCACT-TGCA-TGGAGACCC-GACTTCCCTGG-GCCCCACGAGAGG-CTCACTGA 1019
Bos_taurus_indicus      CCCCACCACTTCGCACTGGAGAACCCGACTTCCCTGGCACCCCACCAGAGGCCTCACTGA 1009
                        ** * *** *  **  ****** ** ***********  *** ** ****  ********
```

```
Bison_bison             CCTCGCCGTCGTACACT--AGGA-AAAACCGCACACTGTGGC-GCCAGCTCGA-GAACAA 1071
Bos_taurus_taurus       CCTCGCCGTCGCACCTC--GTGA-GAAACCGCACCCTGGGGCCGCCGGCTCGA-GAACCA 1078
Bos_grunniens           CCTCGCCGTCGTACCTC--GTGA-GAAACCGCACCCTGGGGCCGCC-GCTCGA-GAACAA 1074
Bos_taurus_indicus      CCTCGCCG-CGTACCTCCAGTGAAGAAAACAC-CACCGGGGCCGCC-GCTCGAAGAACAA 1066
                        ******** ** **       **  *** * * * * * *** *** ****** **** *

Bison_bison             --CCCTGAGCCCTCC-CCATCATCGCG-AGTTGAGGG-CCTTCGTCTCCTGTATGG-CCT 1125
Bos_taurus_taurus       --CCC-GAGACTCCC-CCGTCATCGCGGAGATGAGGG-CCTTCGTCTCCTGCATGGGCCT 1133
Bos_grunniens           --CCCCGAGACTCCC-CCGTCATCGCG-AGATGAGGGGCCTTCGTCTCCTGCATGG-CCT 1129
Bos_taurus_indicus      ACCCCCGAGAATTCCACCCTCATCGAG-AGATGAGGGCCCTCCGCCCCCT---CAGGCCT 1122
                          *** ***    ** ** ****** * ** ****** *** ** * ***      * ***

Bison_bison             AGA--GACCAA-TCT--CCTCGACTCTCTCTCAAACGCCTCAGGAGGCTTGACTCCCTTT 1180
Bos_taurus_taurus       AGA--G-CCAA-CCT--CGCGACCTCTCTC-CAAACGCCTCAGGAGGCT-GACTCCCATT 1185
Bos_grunniens           AGA--GACCAAATCT--CGCGACCTCTCTC-CAAGCGCCTCAGGAGGCTTGACTCCCTTT 1184
Bos_taurus_indicus      AGAAGGCCCAATCCTTCCGCGAACCTCTCCAAAAACGCCTCAGA--GCCTGACTCCCTTC 1180
                        ***  * ****  **  *      *      *  ** ********   **  ******* *

Bison_bison             ---AGTCC-ACCCAGTGAGCTCCAAGAGAT-ACCCGTCGC--GACTCGAGAGC-AGAGCG 1232
Bos_taurus_taurus       ---TGTCC-ACCCAGTGGAGCTCAAGAGAT-ACCCGTCGC--GACTCGGAGGC-AGAAAG 1237
Bos_grunniens           ---AGTCC-ACCCAGTGAGCTCCAAGAGAT-ACCCGTCGC--GATTCGAGAGC-AGGCGG 1236
Bos_taurus_indicus      GCGAGTCCCACCCAGTGGAGCCCAAGAGATGAACCGTCGCCTGATTCGAGAGCCAGAGCG 1240
                            **** ********     ******** * *******  ** ***   ** **   *

Bison_bison             --GGGTTCTTTGCTTCCACT--CGACA-TGAATGC--TGTCTCCCCGGGTGCGTCTGGAA 1285
Bos_taurus_taurus       TTGGGTTCGTTGCTTCCCCT--CGAGA-TGAATGCC-TGTCTCCCCGGGTGCGTCTTGGA 1293
Bos_grunniens           --GGGTTCTTTGCTTCCACT--CGAGAATGAATGCC-TGTCTCCCCGGGTGCGTCT-GGA 1290
Bos_taurus_indicus      --GCCCTCTTTGCTTCCACTTCCGAGGTGAAATGCCTTGTCTCCCCGG-TG-GTCT--GA 1294
                          *   ** ******** **  ***      *****  *********** ** ****   *

Bison_bison             TTGCAACCCTGAGATCCCTTTCGCCCCCTGGAGAGGAA-CACTGG-CTTCTGGACACGAA 1343
Bos_taurus_taurus       ATGCCACCCCGAG-TCCCGGTCGCCCC-TGGAGAGGAA-CCTCGGGCTTCTGGGCAC-AA 1349
Bos_grunniens           ATGCAACCC-GAGATCCCTGTCGGCCC-TGGAGAGGAA-CATTGG-CTTCTGGACAC-AA 1345
Bos_taurus_indicus      ATGC-ACCCCGAGATCC-TGTCGCCCT---GAGAGAAAACAATTG-CTTCTG--CACAAA 1346
                         *** **** *** ***   *** **    ***** ** *    * ******  *** **

Bison_bison             GCCTAGA-TGA-GGTCTATT-GGCCCTGCA-GT-CACTCGA-GAGCAATCCCC-AGCTTT 1396
Bos_taurus_taurus       GCCTAGA-TGAAGGTCTATCAGGGCCTGCAAGT-CACTCTG-GAGCAATCCCC-AGCTTT 1405
Bos_grunniens           GCCTAGA-TGA-TGTCTATT-GGCCCTGCA-GT-CACTCGA-GAGCAATCCCC-AGCTTT 1398
Bos_taurus_indicus      GCCTAGAATGA-GTCCTATT-GGCCCTCCA-GTACACTCGATGAGCAATCCCCCAGCTTT 1403
                        ******* ***    ****  ** *** ** ** *****   *********** ******

Bison_bison             CCTT-CGCAACTCG- 1409
Bos_taurus_taurus       TCTTTCGCAACTCG- 1419
Bos_grunniens           CCTT-CGCAACTCGA 1412
Bos_taurus_indicus      CCCTTCGCAACTCCA 1418
                         * * ********
```

Candidate centromere repeat monomer:

```
>Bison_bison
AATGGAAGATTGGACTTCCCTGGGCCAAACACAAGAGGCATCCTGAATTCCCCGTCGTAA
TTCGAGAATCCTGCCGACACTCGAGAAAATCCACGTGTTCCCCCGTCATCGCAAGATGAA
GCCCTTTCCCGCTACAGTGTCTCAGGAGAAGTCCCACGTTAGGTATTGGAGGTCGAAACG
GTACTTGGCACCTTGATGCGACCACAAAGTGCCCCGACATCCCGGTCCTCCCTCGAGAGG
AACACCGAAGTTTTCCGGCACCACTTCCTCTGAGCCCCTTCTACCCTCCTGATCTGGACA
GGAGGGTCGACTCCCCTGCTTTGTCTGGAAGGGGTTCCCGACCTTCCGGTCGCACCTCAG
GATGAGGCCGGGTCTCACGACGACATTTCAGACGTGGCCTCGTGGGTGGTTCCACATTCC
GAAGGACCCCGATTTCCCGGTCCCCTCTTGATAAGAACCCGATGCCCGGACACCTCTTCG
AACCTCCACCCTGTGAATGAAGTCAACACGAAAGGGCAGTGACTCGCCCGTGCATCGTCG
GGAAAAAACCCCCAGGTTCCAAATACCGCTCGACAAGTGGCCTGTCTCCCCGGGAAACAC
CTCGAGAGGCAAGCGGAGTTCCATGCCTCACCCAAGACGAGGCCTGACTCTCCCTGTCCC
AAGTCTGCAGGGACCTTGCGATCAGAGTCTGAAGTCAGAGGAACCCTGAGGTTCCTGCCT
CAACTGGAGATGAGGCCCTCTTCCAATGCACCAAACCCAGTGGGAGTGCCGAGAAGCCCT
CCCACCTCCAGTTTTCCCTGACTTCTCAGAGCCACCATGAGAAGCCCCCTGAGGTCACCT
GCACAAGTCAAGGGAAGCCAGGTTTTCTGCCTCAACCCGAGAAAGACCTCGAGAGACCTT
CTTCAACACGTCTCGAGCCAGATTCCCCTAACAGTGACTCGAGAGCAATGACGCGCTCCC
CCTCGCCATTGCGCATGGAGACCCCGACTTCCCTGGCGCCCCACGAGAGCTCACTGACCTC
GCCGTCGTACACTAGGAAAAACCGCACACTGTGGCGCCAGCTCGAGAACAACCCTGAGCC
CTCCCCATCATCGCGAGTTGAGGGCCTTCGTCTCCTGTATGGCCTAGAGACCAATCTCCT
CGACTCTCTCTCAAACGCCTCAGGAGGCTTGACTCCCTTTAGTCCACCCAGTGAGCTCCA
AGAGATACCCGTCGCGACTCGAGAGCAGAGCGGGGTTCTTTGCTTCCACTCGACATGAAT
```

```
GCTGTCTCCCCGGGTGCGTCTGGAATTGCAACCCTGAGATCCCTTTCGCCCCCTGGAGAG
GAACACTGGCTTCTGGACACGAAGCCTAGATGAGGTCTATTGGCCCTGCAGTCACTCGAG
AGCAATCCCCAGCTTTCCTTCGCAACTCG
>Bos_taurus_taurus
AATGGAAGATTGGACTTCCTGGGCAACACAAGAAGGCCATCACTGATTTCCCCGTTCGTA
ACTCGAGATTCCGCCGCCAAACTCGAGACACCAACGTGGATTCCCCCGTCATCGCACAAG
ATTGAAGCCCTTTCCGCTACAGCGTCTCAGGAGAAGTCCCACGTTAGGAATTGAGCGGTG
GAACGGTACTTGGCACCCTTGACGGCGGCCCCACAAAGTTCCCCGGCATCCCGTCTCCCT
GGAGAGGAACACCGAGTTTCCGGCACCACTTCCTCTGAGCCCCTTCTACCCTCCTGATCT
CGACAGGGAGGGTCGACTCCCCTGCTTTGTCTGGAAGGGGTTCCCGACCTTCCGGTCGCA
CCTCAGGATGAGGCCGGGCTCACGACGACATTCCAGACGTGGCCTCGTGGGTGCTTCCAC
ATTGCGAAGCACCCCGATTTCCCGGTCCCCTCTTGGTAAGAACCCGATGCCCGGACACCT
CTCGAACTCCACCCTGTGAATGAGGTCAACACGAAGGGGCAGTGCCCGCCCGTGCATCGT
CCGGAAAGAACCCCCAGGTTCCAAATACAGCTCGACAAGTGGCCTCTCTCCCCGGGGACA
CCTCGAGAGGCAAGCGGAGTTCCATGCCTCAACCCAAGACGAGGACCTGACTCTCCTGTC
CCCAGTCTGCAGGGACCCTGCGATTCGGAGTCTGAAATCAGAGGAACCCTGAGGTTCCTG
CCTCAACTGGAGATGGAGGCCCTCTTCCAATGCACCAAAACCCCGTGGAGTCCCGAGAGG
CCCCTCCCACCTCCAGTTCCGCTGATTCTCAGAGCCCACCATGAGAAGCCCCCTGAGGTC
ACCTGCACAAGTCGAGGGAAGCCAAGGGTTCCCTGCCTCAACCCGAGAAAGACCTCGAGA
GACCTTCTTCAACATCGTCTCGAGGCCAGGTTCCCTACCAGGACTCGAGGGCAATGACGC
GCTCCCCCTCGCCACTCGCGTGGAGACCCGACTTCCCTGGCGCCCCACGAGAGGCTCACT
GACCTCGCCGTCGCACCTCGTGAGAAACCGCACCCTGGGGCCGCCGGCTCGAGAACCACC
CGAGACTCCCCCGTCATCGCGGAGATGAGGGCCTTCGTCTCCTGCATGGGCCTAGAGCCA
ACCTCGCGACCTCTCTCCAAACGCCTCAGGAGGCTGACTCCCATTTGTCCACCCAGTGGA
GCTCAAGAGATACCCGTCGCGACTCGGAGGCAGAAAGTTGGGTTCGTTGCTTCCCCTCGA
GATGAATGCCTGTCTCCCCGGGTGCGTCTTGGAATGCCACCCCGAGTCCCGGTCGCCCCT
GGAGAGGAACCTCGGGCTTCTGGGCACAAGCCTAGATGAAGGTCTATCAGGGCCTGCAAG
TCACTCTGGAGCAATCCCCAGCTTTTCTTTCGCAACTCG
>Bos_taurus_indicus
AATGGAAGATTGAACTTGCCCTGGGCCAACCAAAAGGCATCCTGACTTCCCCGTCGTAAC
TCAGAATCCCCGCCCGTAACTCGAGAAAAACCACGTGGCTCCCCGTCCATCCGCAAGATG
AAGGCCCTTCCCGCAACAGCCGGCCCTCCAGGAGAAGTCCCACGTAGGATTGGAGGTCGA
AAGGGCACTTGGCGCCCTTGATGCGACCCACAAAGTTCCCCGAAATCCCCGGTCTCCCTC
GAGAGGAACACTGAGGCTTTCGCACCCCCTCCTCTGACCCTTTTCTCCCCCTCCTGAATC
TGGACAGGAGGTCGACTCCCCTGCTTGTCTGGAAGGGGTTCCGACCTTCCGTCCACCTCC
AGGAATGAGGCCGGTCTCAACGAAGACCATTCCAACGTGCCCCTCCGTGGGTGGTTCCCA
CATTCGTAGGACCCCGATTCCCGTCCCCTCCTGGATAAGAACCCGATGCCGGACCACCTC
TCCGAACTCCAACCCTGGAATGAAGTCAACACGAAGGGCCAATTTTTCCGTGCATCGTTC
AGAAAAACCCAGTTCCAAATACAGCTCGACAAGCCGGCTCTTCTTCCCCCGGGACCATCT
CGAGATGCAAGCGGAGTTCCATGCCTCAAACCAAAGACAGCCGACTCTCCTGTCCCAGTC
CTGCAGACCTGCGATCGGAAGTCTGAATCAAGGTACCCTGCGGTTCCGCCCCTCAACTGG
AGATGAGGCCTCTTCCAATGCACCAAAACCCAGTGGGGTCCCGAGAGGCCCTTCCCACCT
CCAGGTTCCCTGCTTCTCAGAGCCACCATGGAGAAGCCCCTGAGGTCACCTGCAACAAGT
CGAGGAACCCAGGGTTTCCTGCCTCAACCGAAAAGACCTCGAGAGACCTTCTCAACACGT
CTCTGAGGCCAATCCCTAACATGGCTCGGGAATCCAGTGACGCGCTCCCCCCCACCACTT
CGCACTGGAGAACCCGACTTCCCTGGCACCCCACCAGAGGCCTCACTGACCTCGCCGCGT
ACCTCCAGTGAAGAAAACACCACCGGGGCCGCCGCTCGAAGAACAAACCCCCGAGAATTC
CACCCTCATCGAGAGATGAGGGCCCTCCGCCCCCTCAGGCCTAGAAGGCCCAATCCTTCC
GCGAACCTCTCCAAAAACGCCTCAGAGCCTGACTCCCTTCGCGAGTCCCACCCAGTGGAG
CCCAAGAGATGAACCGTCGCCTGATTCGAGAGCCAGAGCGGCCCTCTTTGCTTCCACTTC
CGAGGTGAAATGCCTTGTCTCCCCGGTGGTCTGAATGCACCCCGAGATCCTGTCGCCCTG
AGAGAAAACAATTGCTTCTGCACAAAGCCTAGAATGAGTCCTATTGGCCCTCCAGTACAC
TCGATGAGCAATCCCCCAGCTTTCCCTTCGCAACTCCA
>Bos_grunniens
ATGGAAGATTGGACTTCCCTGGGCCAACACAAGAGGCATCCTGAATTCCCCGTCGTAACT
CGAGAATCCCGCCGCAACTCGAGAAAAACCACGTGGTTCCCCCGTCATCGCAAGATGAAG
CCCTTTCCCGCTACAGCGTCTCAGGAGAAGTCCCACGTTAGGAATTGGAGGTCGAAACGG
TACTTGGCCACCCTTGATGCGACCCACAAAGTTCCCGACATCCCGGTCTCCCTCGGAGAG
GAACCCCTGAGGTTTTCCGGCCACCACTTCCTCTGAGCCCCTTCTACCCTCCTGATCTGG
ACAGGAGGGTCGACTCCCCTGCTTTTTCTGGAAGGGGTTCCCGACCTTCCGGTCGCACCT
CAGGATGAGGCCGGTCTCACGACGACATTCCAGGACGTGGCCTCGTGGGTGGTTCCACAT
TGCGTAAGACACCCGATTTCCCGGTCCCCTCTTGATAAGAACCCGATGCCCGGACACCTC
TTCCGAACTCCACCCTGTGAATGAAGTCAACACGAAGGGGCAGTGACTCGCCCGTGCATC
GTTCGGGGAAAACCCCCAGGTTCCAAATACAGCTCGACAATGGGCCTCTCTCCCCGGGGA
CACCTCGAGAGGCAAGCGGAGTTCCATGCCTCAACCCAAGACGAGGCCTGACTCTCCTGT
CCCAAGTCTGCAGGGACCCTGCGATCGGTCTGAAATCAGAGGAACCCTGAGGTTCCTGCC
TCAACTGGAGATGAGGCCCTCTTCCAATGCACCAACCCCATGGAGTCCCGAGAGGCCCCT
CCCAACCTCCAGTTTCCCTGACTTCTCAGAGCCACCATGAGAAGCCCCCTGAGGTCACCT
GCACAAGTCGAGGGAACCAGTGTTTCCTGCCTCAACCGAGAAAGACCCCCTCGAGAGACC
TTCTTCAACACGTCTCGAGGCCAGATTCCCCTACCATGACTCCGAGGAGCAATGACGCGC
TCCCCCTCGCCACTTGCATGGAGACCCGACTTCCCTGGGCCCCACGAGAGGCTCACTGAC
```

```
CTCGCCGTCGTACCTCGTGAGAAACCGCACCCTGGGGCCGCCGCTCGAGAACAACCCCGA
GACTCCCCCGTCATCGCGAGATGAGGGGCCTTCGTCTCCTGCATGGCCTAGAGACCAAAT
CTCGCGACCTCTCTCCAAGCGCCTCAGGAGGCTTGACTCCCTTTAGTCCACCCAGTGAGC
TCCAAGAGATACCCGTCGCGATTCGAGAGCAGGCGGGGGTTCTTTGCTTCCACTCGAGAA
TGAATGCCTGTCTCCCCGGGTGCGTCTGGAATGCAACCCGAGATCCCTGTCGGCCCTGGA
GAGGAACATTGGCTTCTGGACACAAGCCTAGATGATGTCTATTGGCCCTGCAGTCACTCG
AGAGCAATCCCCAGCTTTCCTTCGCAACTCGA

#################################################################################################
Caprid

Capra_hircus      --TCCCATCGAGGCTTGCCCACGGGGCCTCTC----GGGATTCCTCTCCCGTCGATGCCG 54
Ovis_aries        TTTTCCCTCGTGTCTTTCCCACGAGGCTTTCCCACGAGGCTTTCCCACAGGGCTGTCCCA 60
                    * ** *** * *** ****** *** *  *     ** ** * * *  * *  * **

Capra_hircus      GGGCCTAAGACCTTGTGTGGAGTCGGTGCCGGAACCTGAGGATTCCTCTCCAGTGCAG-A 113
Ovis_aries        CGTGC--ACACGTGGTG-GGAGTCGAT-CCTCGGCTTGAA-------CGTCAAGGCAGTG 109
                   *  *  * ** * *** ******* * **    * ***        *  **  ****

Capra_hircus      CAGGGATAATGGGGTTCCCTGGAGTCGCCTCAGGGGATCAGGCCTCCTCTCGAATGGTGG 173
Ovis_aries        CAGGGAAAACAGGTTCCTCTGGAATGGACTGACACATCTGGGGGACTCTTGGAATGGTGG 169
                  ****** **  ** * * ***** * * ** *        **   *   * *********

Capra_hircus      CACGAACCTGGAGTTCCTCTCGCCTTTCCTGTGGAGAGCGCCTCCTCTTGCGATGCGACG 233
Ovis_aries        CACGACCCTGGAGTTCCTCTCGCCTTTCCTGTGGAGAGCGCCTCCTCTTGAGATGCGACG 229
                  ***** ******************************************** *********

Capra_hircus      GGAATCCCGGGAATTCTTTCCCTCC-ACGCAGGCACACGAGCCCTCCCCACGAGCTTACA 292
Ovis_aries        GGAACGCCGGGAATTCTTTCCCTACGAAACAGGGAAAGGATCCCTCATCTCGAGCTCG-G 288
                  ****  ***************** * *  **** * * ** *****  * ******

Capra_hircus      AGGCGCAAACGGGCCACCTCTGGATGTGAGCGCGACCCTCGTGCTTAAACTCGAGTGCAG 352
Ovis_aries        AGGCGGAAACGGGGCTCCCCTGGATGTGTTCGGGACCCTCGTGCTTCCTCTCGAGTGGAG 348
                  ***** ******* * ** *********  ** *************   ******** **

Capra_hircus      ACAGGGAAAACAGGGAACTCCTGGAATGGCAGCAAACATCTGAAGGACCCTTGGGAAGGT 412
Ovis_aries        AC-GGGTGTGTCGGGAACTTCTTGAGTTGCAGCAAGGGTGTGAAGGACCCTTTGGAAGTT 407
                  ** ***      ******* ** ** * *******   * ************ ***** *

Capra_hircus      CCACACGTAACCTGTGAGTAGCCTCGAGACGCCCCAGCGGAAATGCGCCTCATCTCGACA 472
Ovis_aries        CCAGGGGTTAGATGTGATTAGCCTCGAGAAGCCTCAGCGGAAATGGGCCTCATCTCGCCT 467
                  ***   ** *  ***** *********** *** *********** *********** *

Capra_hircus      GGAGACGAGAACCGCCGGGATTTTCCCGACACGCGGAAGGTCCTCTCGACCTACGACGCG 532
Ovis_aries        GGAGGGCAAAACCTCCTGGATTTTCTCGAGTTGCGGCAGGTGCTCTCGACTTACGACGGG 527
                  ****   * **** ** ******** ***   **** **** ******** ******* *

Capra_hircus      GACAACAGGGACCCGCTCTGGTGGCCGCAGGAAACGCCAGTCCCCATGCGAGTTGCACAC 592
Ovis_aries        GCCCTCAGGGACCCGCTCTGGTGGCCTCAGGAAAGGCCAGTCCCCATGCGAGTTCCTCGG 587
                  * *  ********************* ******* ******************* * *

Capra_hircus      GGGCATCTCGGGAAACCTCTCCAGTCGAAGCAAGGGCCTAAGACCCTGTGTGAAGTCCCA 652
Ovis_aries        GGGCCTTTCGGAATTCCTCTCCCGCTGATGCCGGGGCCTAAGACCTTGTGTGAACTCAGG 647
                  **** * **** *  ******* *  ** **  ************ ******** **

Capra_hircus      GACGGAACCTGTG-ATTACCCTCCAGCGCTCACACGGATAATGGGCCACATCTG-AGTCT 710
Ovis_aries        GCCGGAACCTGAGGATTCCTCTCCAGTGCTGACATGGATCTTGGGGTACTTCTGGAGTCT 707
                  * ********* * *** * ****** *** *** ****  ****  ** **** *****

Capra_hircus      CCCCAGGGGAGAAACTCCTCGTCTCGAGTGGCGGCATGCG--TGCGCTCGACTCACGAGC 768
Ovis_aries        CCCCAGGGGAGTCAGTCCTCGTCTCGAATGCGGGCATGCACTTGCGCTTTCCTCCAGAGC 767
                  ***********  * ************ **  *******   ******   ***  ****

Capra_hircus      GGGAACAGCAGGGACACGCTTCCCGTCGCCTGCAGCAAAGGACCAGTGC 817
Ovis_aries        GGTAGCAGCAGTGTCACGCAGTCCGCCCCGTGGATCAAAGCATCTATGG 816
                  ** * ****** * *****   *** * * ** * ***** * *  **

Centromere repeat monomer:
>Capra_hircus
TCCCATCGAGGCTTGCCCACGGGGCCTCTCGGGATTCCTCTCCCGTCGATGCCGGGGCCT
AAGACCTTGTGTGGAGTCGGTGCCGGAACCTGAGGATTCCTCTCCAGTGCAGACAGGGAT
```

```
AATGGGGTTCCCTGGAGTCGCCTCAGGGGATCAGGCCTCCTCTCGAATGGTGGCACGAAC
CTGGAGTTCCTCTCGCCTTTCCTGTGGAGAGCGCCTCCTCTTGCGATGCGACGGGAATCC
CGGGAATTCTTTCCCTCCACGCAGGCACACGAGCCCTCCCCACGAGCTTACAAGGCGCAA
ACGGGCCACCTCTGGATGTGAGCGCGACCCTCGTGCTTAAACTCGAGTGCAGACAGGGAA
AACAGGGAACTCCTGGAATGGCAGCAAACATCTGAAGGACCCTTGGGAAGGTCCACACGT
AACCTGTGAGTAGCCTCGAGACGCCCCAGCGGAAATGCGCCTCATCTCGACAGGAGACGA
GAACCGCCGGGATTTTCCCGACACGCGGAAGGTCCTCTCGACCTACGACGCGGACAACAG
GGACCCGCTCTGGTGGCCGCAGGAAACGCCAGTCCCCATGCGAGTTGCACACGGGCATCT
CGGGAAACCTCTCCAGTCGAAGCAAGGGCCTAAGACCCTGTGTGAAGTCCCAGACGGAAC
CTGTGATTACCCTCCAGCGCTCACACGGATAATGGGCCACATCTGAGTCTCCCCAGGGGA
GAAACTCCTCGTCTCGAGTGGCGGCATGCGTGCGCTCGACTCACGAGCGGGAACAGCAGG
GACACGCTTCCCGTCGCCTGCAGCAAAGGACCAGTGC
>Ovis_aries
TTTTCCCTCGTGTCTTTCCCACGAGGCTTTCCCACGAGGCTTTCCCACAGGGCTGTCCCA
CGTGCACACGTGGTGGGAGTCGATCCTCGGCTTGAACGTCAAGGCAGTGCAGGGAAAACA
GGTTCCTCTGGAATGGACTGACACATCTGGGGGACTCTTGGAATGGTGGCACGACCCTGG
AGTTCCTCTCGCCTTTCCTGTGGAGAGCGCCTCCTCTTGAGATGCGACGGGAACGCCGGG
AATTCTTTCCCTACGAAACAGGGAAAGGATCCCTCATCTCGAGCTCGGAGGCGGAAACGG
GGCTCCCCTGGATGTGTTCGGGACCCTCGTGCTTCCTCTCGAGTGGAGACGGGTGTGTCG
GGAACTTCTTGAGTTGCAGCAAGGGTGTGAAGGACCCTTTGGAAGTTCCAGGGGTTAGAT
GTGATTAGCCTCGAGAAGCCTCAGCGGAAATGGGCCTCATCTCGCCTGGAGGGCAAAACC
TCCTGGATTTTCTCGAGTTGCGGCAGGTGCTCTCGACTTACGACGGGGCCCTCAGGGACC
CGCTCTGGTGGCCTCAGGAAAGGCCAGTCCCCATGCGAGTTCCTCGGGGGCCTTTCGGAA
TTCCTCTCCCGCTGATGCCGGGGCCTAAGACCTTGTGTGAACTCAGGGCCGGAACCTGAG
GATTCCTCTCCAGTGCTGACATGGATCTTGGGGTACTTCTGGAGTCTCCCCAGGGGAGTC
AGTCCTCGTCTCGAATGCGGGCATGCACTTGCGCTTTCCTCCAGAGCGGTAGCAGCAGTG
TCACGCAGTCCGCCCCGTGGATCAAAGCATCTATGG

#################################################################################################
Primates

Gorilla_gorilla_graueri            CAAAAAGAGTGTTTCAAA-CTGCTGTATCAAAAGAAAGGTTCAACTCTGT 49
Gorilla_gorilla_gorilla            CAAAAAGAGGGTTTCAAAACTGCTCTGTCAAAAGAAAGGTTAAACTCTGT 50
homo_sapiens                       CAAAAAGAGTGTTTCAAAACTGCTCAATCAAAAGAAAGGTTCAACTCTGT 50
pongo_abelii                       CAAAAAGAGTGTTTCAAAACTGCTCAATCAAAAGAAAGGTTCAACTCTGT 50
nomascus_leucogenys                CAAAAAGAGTGTTTCAAAACTGCTCAGTCAAAAGAAAGGTTCAACTCTGT 50
pongo_pygmaeus                     CAAAAAGACTGTTTCCAAACTGCTCAATCAAAAGAAAGGTTCAACTCTGT 50
hylobates_concolor                 CAAAAAGAGTGTTTATAAACTGCTCAATCAAAAGAAAGGTTCAACTCTGT 50
pan_paniscus                       CAAAAAGAGTGTTTCAAAACTGCTCTATGAAAAGAAAGGTTCAACTCTGT 50
pan_troglodytes_schweinfurtii      CAAAAAGAGTGTTTCAAAACTGCTCTATCAAAAGAAAGGTTCAACTCTGT 50
pan_troglodytes_troglodytes        CAAAAAGAGTGTTTCAAAACTGCTCTATCAAAAGAAAGGTTCAACTCTGT 50
macaca_mulatta                     CAAGAACTGGCT-AGCGAAAGGCTCCTTGAAAG-AAAGATGTAACTCTGT 48
Macaca_fascicularis                CAAGAACTGTCTTAGC-AAAGGCTTCTTGAGGGGAAAGCTGTAACTCTGT 49
papio_hamadryas                    CAAGAAATAGGCTAGCGAAAGGATCCATGAGAAGAAAGATGTAACTCTGT 50
                                   *** **           *   * *   * *    **** *  ********

Gorilla_gorilla_graueri            TAGTTGAGGACACACATCACAAAGAAGTTTCTGAGAATGCTTCTGTCTAG 99
Gorilla_gorilla_gorilla            GAGTGGAACACACACAACACAAAGAAGTTACTGAGAATGATTCTCTCTAG 100
homo_sapiens                       GAGTTGAATGCACACATCACAAAGAAGTTTCTCAGAATGCTTCTGTCTAG 100
pongo_abelii                       GAGATGAATGCACACATCACAAAGAAGTTTCTCAGAATGCTTCTGTCTAA 100
nomascus_leucogenys                TAGATGAATGCACAGATTAGAAAGAAGTTTCACAGAATGCTTCTGTGTAG 100
pongo_pygmaeus                     GAGATGAATGCACACATCACAAAGAAGTTTCTCAGAATGCTTCTGTCTAG 100
hylobates_concolor                 GAGATGAATGCACACATCACAAAGAAGTTTCTCAGAATGCTTCTGTCTAG 100
pan_paniscus                       GAGTTGAAAGCACACATCACAAAGAAGTTTCTGAGAATGCTTCTGTGTAG 100
pan_troglodytes_schweinfurtii      GAGTTGAATGCACACATCACAAAGAAGTTTCTGAGAATGCTTCTGTCTAG 100
pan_troglodytes_troglodytes        GAGTTGAATGCACACATCACAAAGAAGTTTCTGAGAATGCTTCTGTCTAG 100
macaca_mulatta                     GAGATGAATTCACAGAACACAAAGAAGTTTCTCAGAAAGCTTCTTTCTCT 98
Macaca_fascicularis                GAGATGATTTCACAGAACACAAAGCAGTTTCTCAGAAAGCTTCTTTCTCT 99
papio_hamadryas                    GAGATGAATTCACAGAACACAAAGCAGTTTCTCAGAAAGCTTCTTTCCAG 100
                                    **  **   **** *  * **** **** *  **** * **** *

Gorilla_gorilla_graueri            ATTTTATATGAAGATATTCCCGTTTCCAACGAAATCTTCAGAGTA-TCC- 147
Gorilla_gorilla_gorilla            TCATTAGACGAAGATAATCCCGTTTCCAACGAAAGCCCCAAAGAGCTCC- 149
homo_sapiens                       TTTTTATGTGAAGATATTTCCTTTTCCACCATAGGCCTCAAAGCGCTCC- 149
pongo_abelii                       TTTTTATGTGAAGATATTTCCTTTTCCACCATAGGCCTCAAAGCGCTCC- 149
nomascus_leucogenys                TTTTTATTTGAAGATATTTCCTTTTTCACCATAGGCCTCAAAGCGCTCC- 149
pongo_pygmaeus                     TTTTTATGTGAAGATATTTCCTTTTTCACCATAGGCCTCAAAGCGCTCC- 149
hylobates_concolor                 TTTTTATGTGAAGATATTTCCTTTTCCACTATAGGCCCTAAAGTGCTCC- 149
pan_paniscus                       TTTTTATGTGAAGATATTTCCTTTTTCAAAGTAGGCCTCAAAGCGCTCC- 149
pan_troglodytes_schweinfurtii      TTTTTATGTGAAGATATTTCCTTTTCCAACATAGGCCTCAAAGCGCTCC- 149
pan_troglodytes_troglodytes        TTTTTATGTGAAGATATTTCCTTTTCCACCATAGGCCTCAAAGGGCTCC- 149
```

```
macaca_mulatta                   TTTTTATCGGAGGATATTTCCTTTGGCACCATAGCCCTCAAAGGGATCCC 148
Macaca_fascicularis              TTTTTATCTGAGGATATTTCCTTTTTCCCTATAGTCTTCTATGGGCTTCG 149
papio_hamadryas                  TTTTCATCTGAGGATATTTCCTTTTTCACCATAGCCCTCAATGGGCTTCC 150
                                    * *   ** **** * ** **  *      *  *       *   * *

Gorilla_gorilla_graueri          AAATATCCACTTGCAGATTCTA 169
Gorilla_gorilla_gorilla          AAATATCCACTTGCAAACTCCA 171
homo_sapiens                     AAATATCCACATGCAGATTCTA 171
pongo_abelii                     AAATATCCACTTGCAGATTCTA 171
nomascus_leucogenys              AAATGTCCACTTGCAGATTCTA 171
pongo_pygmaeus                   AAATATCCCTTTGCAGATTCTA 171
hylobates_concolor               AATTATCCACTTGCAGATCCTA 171
pan_paniscus                     AAATGTCCACTTGCAGATTCTA 171
pan_troglodytes_schweinfurtii    AAATATCCACTTGCAGATTCTA 171
pan_troglodytes_troglodytes      AAATATCCACTTGCAGATTCTA 171
macaca_mulatta                   AAATATCACTTCGCCGATTCCA 170
Macaca_fascicularis              AAATATCTCTTTTCCAATTCCA 171
papio_hamadryas                  AAATATCACTTTTCAAATTCCA 172
                                 ** * **      *  *  * *
```

```
callithrix_jacchus       GAAGGAACGTGCTCC--ACTCTCTGAATCTAAACGCGGATTCAAC--TCCCTTAGTTAAG 56
Saimiri_boliviensis      GAAGAAACGTGTTTCTACCTCGCTGAGACT-------GTTTCTACATTCTCCTAAGAAAG 53
ateles_geoffroyi         CACAGAAAGGGTTTCTAAACTGCTGCAAACACACACGAGTTTAAC--TCTGTAAGTTGAA 58
                          *   ** * * * *       ***                **  **  **    *    *

callithrix_jacchus       CGAACGCACAGGAGAGCAGTTTGAAGGATACGTTCGTTTTGTGTTTTAGACGGAGATTTA 116
Saimiri_boliviensis      -GGTTTCTTAACTGCGTGCTT--ACACACATCTTTTTTCCGTGTTTGAAACGCATTCTTA 110
ateles_geoffroyi         TGCACACATTGACAAGCATTCAGTTAGATTGCTCCGTTTTG-GTTTTAGAGCGAGATATT 117
                          *    *        *   *       *    *   **  * **** * *   *    *

callithrix_jacchus       CAGCTATTTCTGAAATGCTTGGATTTGGCTGGAAAAATGCGTTTTCCGC-TGCTTCACAG 175
Saimiri_boliviensis      CTGCTAACTCTTATATGCGTCTCTAGGGCTGGAAATGACTGTTT-CTGCATTCTCCAAAG 169
ateles_geoffroyi         CCTCTGTACAGCAAATGCTCCCTTTCGCTCTGGAAAATCCAAACGCAG--TTCCTAGAAG 175
                         *  **       * ****     *  *    * **          * *  * *     **

callithrix_jacchus       AAAGGGTTTACAAACTC-CTGCAAACGCATATCAGTTTAACTCTGTAAGTTGAA-TGCAC 233
Saimiri_boliviensis      AAAGGGTTTCTGAACTG-CTGCAAACACACATCAGTTTAACTCTGGAAGTTGAA-TACAC 227
ateles_geoffroyi         AAACGTGTTGCTAACTCGCTG-AATGAAACTCCGCTTCAGCTCCCTTAGTTGAAACGAAC 234
                         *** *  **    ****  *** **    *   *  ** * ***    *******    **

callithrix_jacchus       ACATTGAAGACCGTTCTGTTAGGTAGCTGCGTTTTG-GTTCTAGAAGCA--GATATTTCT 290
Saimiri_boliviensis      ACATTGACAACCATTCTGCTAGATAGCTCCGTTTTG-GTTTTAGAGCAA--GGTATTTCT 284
ateles_geoffroyi         ACACAGCAGAGCAGTTTGAATGATGCGTTTCTTCTGTGTTTTAGACGCACTATTACTGCT 294
                         ***  *   * *  * **   * *   *   ** ** *** ****   *    ** * **

callithrix_jacchus       CAATACAGCGTTTACTCCCATTCACGCTGGAA-TAGCAAAAC-GCAGCTCCTA 341
Saimiri_boliviensis      CTTTACAGCCTGTTCTCCCATCCACTCTGGAA-TGTCCAAAC-GCAGCTTTTA 335
ateles_geoffroyi         TTTTCTGACGTG--CCTGGATTTG-GCTGGAAATGACAGTTTTGCATCTCTG- 343
                            *    * *   *    **     ***** *  *      *** **
```

>homo_sapiens
CAAAAAGAGTGTTTCAAAACTGCTCAATCAAAAGAAAGGTTCAACTCTGTGAGTTGAATGCACACATCACAAAGAAGTTTCTCAGAATGCTTCTGTC
TAGTTTTTATGTGAAGATATTTCCTTTTCCACCATAGGCCTCAAAGCGCTCCAAATATCCACATGCAGATTCTA
>Gorilla_gorilla_graueri
CAAAAAGAGTGTTTCAAACTGCTGTATCAAAAGAAAGGTTCAACTCTGTTAGTTGAGGACACACATCACAAAGAAGTTTCTGAGAATGCTTCTGTCT
AGATTTTATATGAAGATATTCCCGTTTCCAACGAAATCTTCAGAGTATCCAAATATCCACTTGCAGATTCTA
>Gorilla_gorilla_gorilla
CAAAAAGAGGGTTTCAAAACTGCTCTGTCAAAAGAAAGGTTAAACTCTGTGAGTGGAACACACACAACACAAAGAAGTTACTGAGAATGATTCTCTC
TAGTCATTAGACGAAGATAATCCCGTTTCCAACGAAAGCCCCAAAGAGCTCCAAATATCCACTTGCAAACTCCA
>hylobates_concolor
CAAAAAGAGTGTTTATAAACTGCTCAATCAAAAGAAAGGTTCAACTCTGTGAGATGAATGCACACATCACAAAGAAGTTTCTCAGAATGCTTCTGTC
TAGTTTTTATGTGAAGATATTTCCTTTTCCACTATAGGCCCTAAAGTGCTCCAATTATCCACTTGCAGATCCTA
>macaca_mulatta
CAAGAACTGGCTAGCGAAAGGCTCCTTGAAAGAAAGATGTAACTCTGTGAGATGAATTCACAGAACACAAAGAAGTTTCTCAGAAAGCTTCTTTCTC
TTTTTTATCGGAGGATATTTCCTTTGGCACCATAGCCCTCAAAGGGATCCCAAATATCACTTCGCCGATTCCA
>nomascus_leucogenys
CAAAAAGAGTGTTTCAAAACTGCTCAGTCAAAAGAAAGGTTCAACTCTGTTAGATGAATGCACAGATTAGAAAGAAGTTTCACAGAATGCTTCTGTG
TAGTTTTTATTTGAAGATATTTCCTTTTTCACCATAGGCCTCAAAGCGCTCCAAATGTCCACTTGCAGATTCTA
>pan_paniscus

CAAAAAGAGTGTTTCAAAACTGCTCTATGAAAAGAAAGGTTCAACTCTGTGAGTTGAAAGCACACATCACAAAGAAGTTTCTGAGAATGCTTCTGTG
TAGTTTTTATGTGAAGATATTTCCTTTTTCAAAGTAGGCCTCAAAGCGCTCCAAATGTCCACTTGCAGATTCTA
>pan_troglodytes_troglodytes
CAAAAAGAGTGTTTCAAAACTGCTCTATCAAAAGAAAGGTTCAACTCTGTGAGTTGAATGCACACATCACAAAGAAGTTTCTGAGAATGCTTCTGTC
TAGTTTTTATGTGAAGATATTTCCTTTTCCACCATAGGCCTCAAAGGGCTCCAAATATCCACTTGCAGATTCTA
>pan_troglodytes_schweinfurtii
CAAAAAGAGTGTTTCAAAACTGCTCTATCAAAAGAAAGGTTCAACTCTGTGAGTTGAATGCACACATCACAAAGAAGTTTCTGAGAATGCTTCTGTC
TAGTTTTTATGTGAAGATATTTCCTTTTCCAACATAGGCCTCAAAGCGCTCCAAATATCCACTTGCAGATTCTA
>papio_hamadryas
CAAGAAATAGGCTAGCGAAAGGATCCATGAGAAGAAAGATGTAACTCTGTGAGATGAATTCACAGAACACAAAGCAGTTTCTCAGAAAGCTTCTTTC
CAGTTTTCATCTGAGGATATTTCCTTTTTCACCATAGCCCTCAATGGGCTTCCAAATATCACTTTTCAAATTCCA
>pongo_abelii
CAAAAAGAGTGTTTCAAAACTGCTCAATCAAAAGAAAGGTTCAACTCTGTGAGATGAATGCACACATCACAAAGAAGTTTCTCAGAATGCTTCTGTC
TAATTTTTATGTGAAGATATTTCCTTTTCCACCATAGGCCTCAAAGCGCTCCAAATATCCACTTGCAGATTCTA
>pongo_pygmaeus
CAAAAAGACTGTTTCCAAACTGCTCAATCAAAAGAAAGGTTCAACTCTGTGAGATGAATGCACACATCACAAAGAAGTTTCTCAGAATGCTTCTGTC
TAGTTTTTATGTGAAGATATTTCCTTTTTCACCATAGGCCTCAAAGCGCTCCAAATATCCCTTTGCAGATTCTA
>Macaca_fascicularis
CAAGAACTGTCTTAGCAAAGGCTTCTTGAGGGGAAAGCTGTAACTCTGTGAGATGATTTCACAGAACACAAAGCAGTTTCTCAGAAAGCTTCTTTCT
CTTTTTTATCTGAGGATATTTCCTTTTTCCCTATAGTCTTCTATGGGCTTCGAAATATCTCTTTTCCAATTCCA
>ateles_geoffroyi
CACAGAAAGGGTTTCTAAACTGCTGCAAACACACACGAGTTTAACTCTGTAAGTTGAATGCACACATTGACAAGCATTCAGTTAGATTGCTCCGTTT
TGGTTTTAGAGCGAGATATTCCTCTGTACAGCAAATGCTCCCTTTCGCTCTGGAAAATCCAAACGCAGTTCCTAGAAGAAACGTGTTGCTAACTCGC
TGAATGAAACTCCGCTTCAGCTCCCTTAGTTGAAACGAACACACAGCAGAGCAGTTTGAATGATGCGTTTCTTCTGTGTTTTAGACGCACTATTACT
GCTTTTTCTGACGTGCCTGGATTTGGCTGGAAATGACAGTTTTGCATCTCTG
>callithrix_jacchus
GAAGGAACGTGCTCCACTCTCTGAATCTAAACGCGGATTCAACTCCCTTAGTTAAGCGAACGCACAGGAGAGCAGTTTGAAGGATACGTTCGTTTTG
TGTTTTAGACGGAGATTTACAGCTATTTCTGAAATGCTTGGATTTGGCTGGAAAAATGCGTTTTCCGCTGCTTCACAGAAAGGGTTTACAAACTCCT
GCAAACGCATATCAGTTTAACTCTGTAAGTTGAATGCACACATTGAAGACCGTTCTGTTAGGTAGCTGCGTTTTGGTTCTAGAAGCAGATATTTCTC
AATACAGCGTTTACTCCCATTCACGCTGGAATAGCAAAACGCAGCTCCTA
>Saimiri_boliviensis
GAAGAAACGTGTTTCTACCTCGCTGAGACTGTTTCTACATTCTCCTAAGAAAGGGTTTCTTAACTGCGTGCTTACACACATCTTTTTTCCGTGTTTG
AAACGCATTCTTACTGCTAACTCTTATATGCGTCTCTAGGGCTGGAAATGACTGTTTCTGCATTCTCCAAAGAAAGGGTTTCTGAACTGCTGCAAAC
ACACATCAGTTTAACTCTGGAAGTTGAATACACACATTGACAACCATTCTGCTAGATAGCTCCGTTTTGGTTTTAGAGCAAGGTATTTCTCTTTACA
GCCTGTTCTCCCATCCACTCTGGAATGTCCAAACGCAGCTTTTA

####################################################################################################
Mice

```
Mus_musculus              TAGGACGTGAAATATGGCGAGAAAACTGAAAATCATGGAAAATGAGAAAT 50
Mus_famulus_domesticus    TAGGACCTGGAATATGGCAAAAAACCTGAAAAACGTGGAAAATGAGAAAA 50
                          ****** ** ******** * *** ******* * ** ***********

Mus_musculus              GCCCACTT 58
Mus_famulus_domesticus    TCACACTG 58
                           * ****
```

Candidate centromere repeat monomer:
>Mus_musculus
TAGGACGTGAAATATGGCGAGAAAACTGAAAATCATGGAAAATGAGAAATGCCCACTT
>Mus_famulus_domesticus
TAGGACCTGGAATATGGCAAAAAACCTGAAAAACGTGGAAAATGAGAAAATCACACTG

####################################################################################################
Flying lemurs

```
cynocephalus_variegatus   CTCCCTGGAGGTTCAGCTTCTGCAAGCCACTTTAACTTTAAGAAATGAGA 50
cynocephalus_volans       CTCCCCGGAGGTTGAGCTTTCAGAAGCTAGTTTAACTCTAAGGGAGGAAG 50
                          ***** ******* *****    **** * ******* ****  * **

cynocephalus_variegatus   GAGCAGAATATGCTTCTCACATCTAGCATGCAGGAATTCAGGTTTGCAAG 100
cynocephalus_volans       CAGCAGAATGTGCCTCTCACATCTAGCCTGTAGGAATTCAGGTTGGCAAG 100
                           ******** *** ************* ** ************* *****

cynocephalus_variegatus   TTCTATGTGAAAGCCCATTGAAGCCTATGGGGGAAACGGCTTCTGAGACT 150
cynocephalus_volans       TTCTATGGGGAAGCCCATTGCTGCCTATGGTGGAAACGGCATGGGAGATT 150
                          ******* * **********  ******** ********* *  **** *

cynocephalus_variegatus   TCTGTTTCAAAATCTCTGTCAAAAAGCATCAGAGAAGCTTCCCCGCTTAG 200
cynocephalus_volans       TCAGTTTCAAAATCTCTGTGAAAAAGCATCAGAGAAGCTTCTCCGTTTAG 200
                          ** **************** ********************* *** ****
```

```
cynocephalus_variegatus      AATCCAGCATAACTCACTGAACCGAAGATGTGGCAGTTACAAGAGAAGTT 250
cynocephalus_volans          AATCCAGCATATCTCACAGAACCGAAGAAGTGGTAGTTAGAAGGGAGTTA 250
                             *********** ***** ********** **** ***** *** **  *

cynocephalus_variegatus      GAGCTAAGAGAAGTTTGTTTTCTGAAATTAGAAGCAAAATCACACTGCAA 300
cynocephalus_volans          GAGCTGCAAGCAGTTTGCTTTGGGAAATTAGACGCGAAATGAAACTGCAA 300
                             *****   ** ****** ***  ********* ** **** * *******

cynocephalus_variegatus      ACGGCATTCAAGCAGGCTTG-AAAAGTGTGGTGCATATCTCTGCCAAAAA 349
cynocephalus_volans          ATG-CCTTCCAACAGGCTTGGAAAAGTGTGTTCTATATCTCTGCCAAACA 349
                             * * * *** * ******** ********* *  ************** *

cynocephalus_variegatus      CA 351
cynocephalus_volans          AA 351
                              *
```

Candidate centromere repeat monomer:

```
>cynocephalus_variegatus
CTCCCTGGAGGTTCAGCTTCTGCAAGCCACTTTAACTTTAAGAAATGAGAGAGCAGAATA
TGCTTCTCACATCTAGCATGCAGGAATTCAGGTTTGCAAGTTCTATGTGAAAGCCCATTG
AAGCCTATGGGGGAAACGGCTTCTGAGACTTCTGTTTCAAAATCTCTGTCAAAAAGCATC
AGAGAAGCTTCCCCGCTTAGAATCCAGCATAACTCACTGAACCGAAGATGTGGCAGTTAC
AAGAGAAGTTGAGCTAAGAGAAGTTTGTTTTCTGAAATTAGAAGCAAAATCACACTGCAA
ACGGCATTCAAGCAGGCTTGAAAAGTGTGGTGCATATCTCTGCCAAAAACA
>cynocephalus_volans
CTCCCCGGAGGTTGAGCTTTCAGAAGCTAGTTTAACTCTAAGGGAGGAAGCAGCAGAATG
TGCCTCTCACATCTAGCCTGTAGGAATTCAGGTTGGCAAGTTCTATGGGGAAGCCCATTG
CTGCCTATGGTGGAAACGGCATGGGAGATTTCAGTTTCAAAATCTCTGTGAAAAAGCATC
AGAGAAGCTTCTCCGTTTAGAATCCAGCATATCTCACAGAACCGAAGAAGTGGTAGTTAG
AAGGGAGTTAGAGCTGCAAGCAGTTTGCTTTGGGAAATTAGACGCGAAATGAAACTGCAA
ATGCCTTCCAACAGGCTTGGAAAAGTGTGTTCTATATCTCTGCCAAACAAA
```

Supplementary Figure S2: Centromere repeats in primates only conserved among apes and monkeys
The centromere repeat sequences of the more basal primates (tarsiers and prosimians) do not show sequence similarity. Closer inspection of basal primate genomes did not reveal any sequences with sequence similarity to the ape andmonkey centromere repeat.

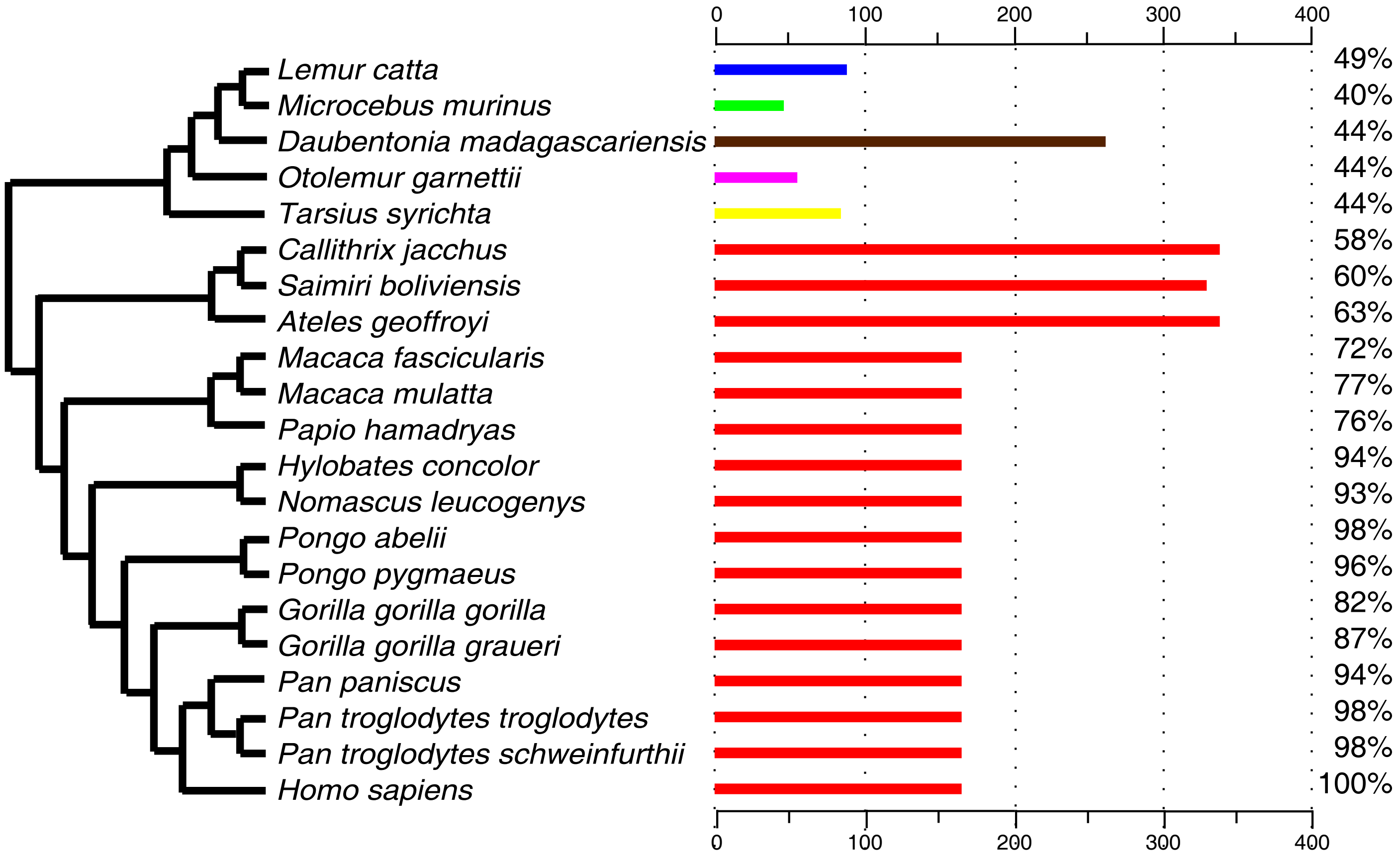

Supplementary Figure 3A: No correlation was observed between repeat length, GC content, kingdom or genomic fraction.

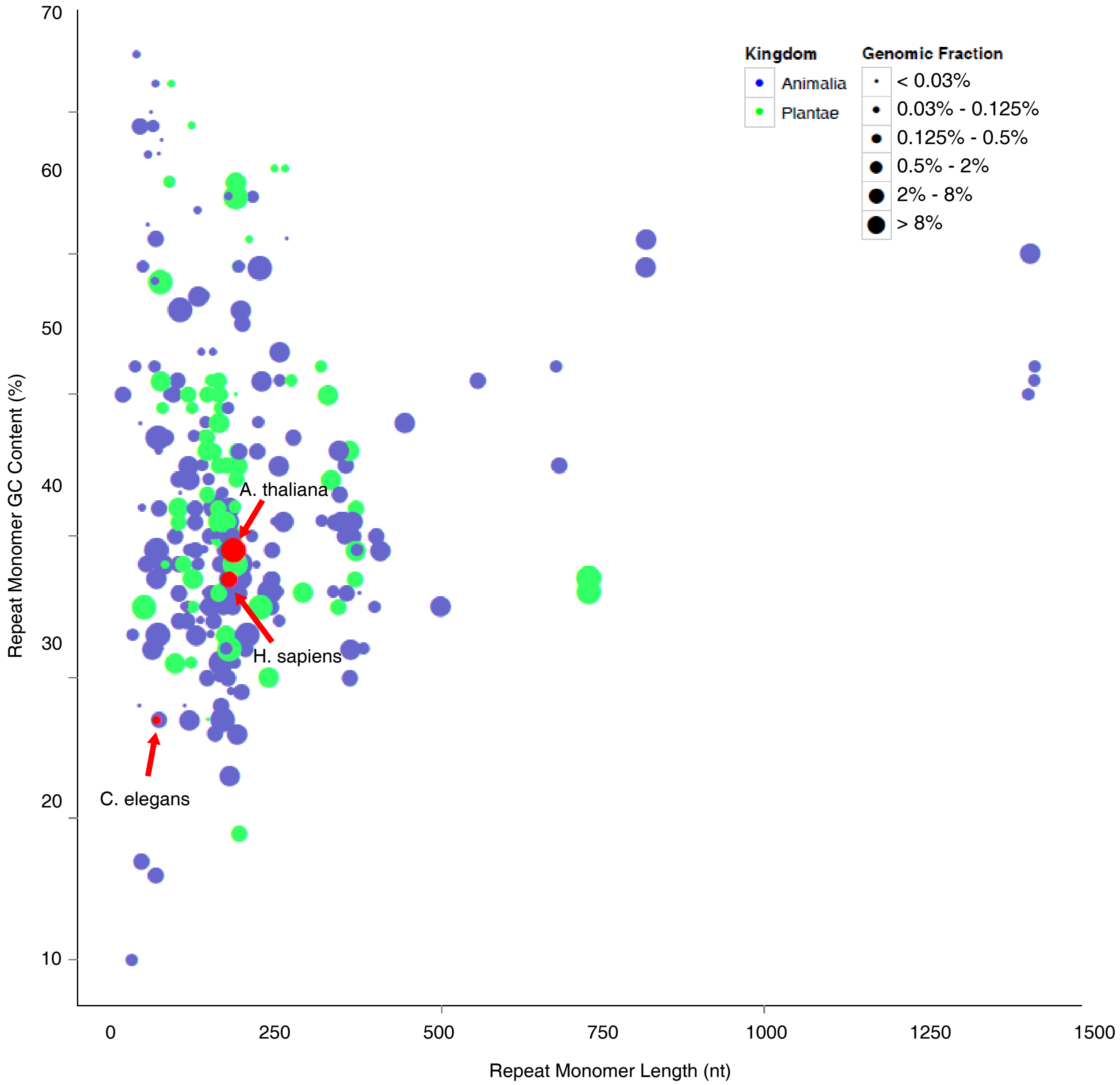

Supplementary Figure S3B: We did not observe a correlation between genome size and chromosome number versus either repeat length, GC content or genomic fraction.

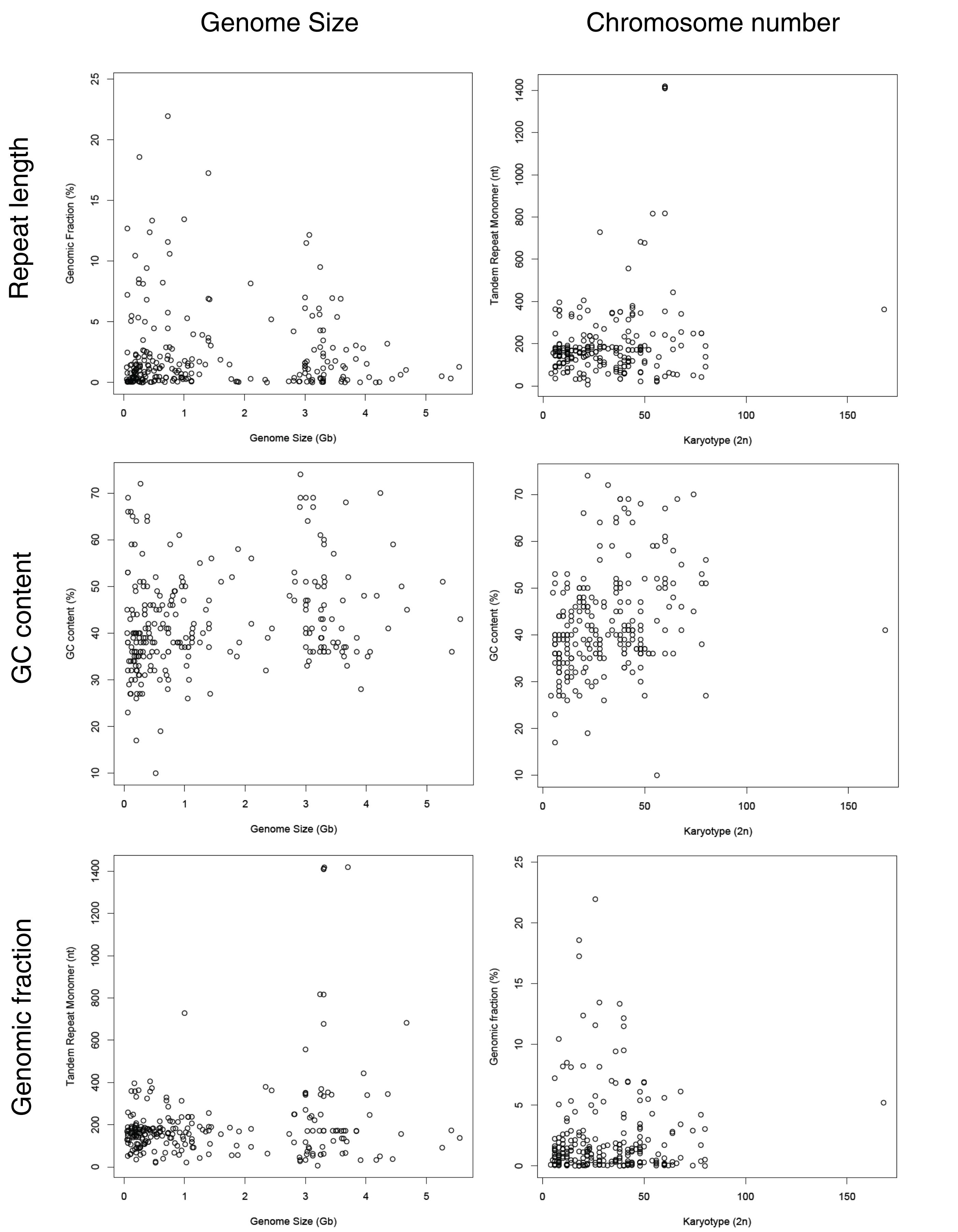

**Supplementary Figure S4** – Nile tilapia contains related tandem repeats that differ by a 29 bp indel.

```
tandem-71       TGTGTTTCAGGCGAGAACTCCATTCAAATGCATGTAATAGCGAGAAACGCACTGTCTCGC 60
tandem-519      TGTGTTTCAGGCGAGAA-----------------------------ACGCACTGTCTCGC 31
                *****************                             **************

tandem-71       CGAAATAAAGCGATTTTCACTAAGTTCATAAAGACAGCTGTAACTTTTGATAGAAGACTC 120
tandem-519      CGAAATAAGGCGATTTTCACCAAGTCCATAAAGACAGCTGTAACTTTTGATAGGAGACTC 91
                ******** *********** **** *************************** ******

tandem-71       GGACACACATATTTCAGGCTTGGCCTTATAGAATTCAGCATTTCTATGCTGGGGAAATAG 180
tandem-519      GGACACACATATTTCAGGCTTGGCCTTATAGAATTCAGCATTTCCATGCTGGGGAAATAG 151
                ******************************************** ***************

tandem-71       GTTTTGCAGCTGTTTGAGCTAAGATTTTCAAGTTATTCACAAAATGAAAACCCATAA 237
tandem-519      GTTTTGCCACTGTTTGAGCTAAGATTTTCA-GTTATTGACTAAATGAAAACC-ATAA 206
                *******  ********************* ****** ** *********** ****
```

```
>tandem-71
TGTGTTTCAGGCGAGAACTCCATTCAAATGCATGTAATAGCGAGAAACGCACTGTCTCGC
CGAAATAAAGCGATTTTCACTAAGTTCATAAAGACAGCTGTAACTTTTGATAGAAGACTC
GGACACACATATTTCAGGCTTGGCCTTATAGAATTCAGCATTTCTATGCTGGGGAAATAG
GTTTTGCAGCTGTTTGAGCTAAGATTTTCAAGTTATTCACAAAATGAAAACCCATAA
>tandem-519
TGTGTTTCAGGCGAGAAACGCACTGTCTCGCCGAAATAAGGCGATTTTCACCAAGTCCAT
AAAGACAGCTGTAACTTTTGATAGGAGACTCGGACACACATATTTCAGGCTTGGCCTTAT
AGAATTCAGCATTTCCATGCTGGGGAAATAGGTTTTGCCACTGTTTGAGCTAAGATTTTC
AGTTATTGACTAAATGAAAACCATAA
```

**Figure S5 - Relative ratios of short repeat (~680 bp) and long repeat (~1410 bp) in the five *Bovideae* species.**

| Species | Short repeat | Long repeat |
|---|---|---|
| ***Bos taurus taurus*** | **29%** | **71%** |
| ***Bos taurus indicus*** | **32%** | **68%** |
| ***Bos grunniens*** | **40%** | **60%** |
| ***Bison bison*** | **41%** | **59%** |
| ***Bubalus bubalis*** | **57%** | **43%** |

**Figure S6A**. Tabular comparison to figure 6A of sequence comparison. Empty cells show sequence comparison with less than 40% sequence similarity. The three letter abbreviation corresponds to the species in top-to-bottom order as shown in Figure 6A.

| | Gac | Oni | Mze | Mau | Lfu | Res | Pny | Nbr |
|---|---|---|---|---|---|---|---|---|
| **Gac** | 100 | | | | | | | |
| **Oni** | | 100 | | | | | | |
| **Mze** | | 66 | 100 | | | | | |
| **Mau** | | 62 | 94 | 100 | | | | |
| **Lfu** | | 67 | 95 | 94 | 100 | | | |
| **Res** | | 60 | 93 | 92 | 93 | 100 | | |
| **Pny** | | 74 | 94 | 95 | 97 | 93 | 100 | |
| **Nbr** | | | | | | | | 100 |

**Figure S6B**. Tabular comparison to figure 6C of sequence comparison. Empty cells represent sequence comparison with less than 40% sequence similarity. The three letter abbreviation corresponds to the species in top-to-bottom order as shown in Figure 6C.

| | Hvu | Ata | Bdi | Obr | Osa | Svi | Sit | Pha | Pca | Pvi | Zma | Zlu | Sbi | Spr | Mgi |
|---|---|---|---|---|---|---|---|---|---|---|---|---|---|---|---|
| **Hvu** | 100 | | | | | | | | | | | | | | |
| **Ata** | | 100 | | | | | | | | | | | | | |
| **Bdi** | | | 100 | | | | | | | | | | | | |
| **Obr** | | | | 100 | | | | | | | | | | | |
| **Osa** | | | | | 100 | | | | | | | | | | |
| **Svi** | | | | | 57 | 100 | | | | | | | | | |
| **Sit** | | | | | 59 | 98 | 100 | | | | | | | | |
| **Pha** | | | | | 62 | 68 | 67 | 100 | | | | | | | |
| **Pca** | | | | | 65 | 63 | 63 | 94 | 100 | | | | | | |
| **Pvi** | | | | | 41 | 44 | 45 | 56 | 55 | 100 | | | | | |
| **Zma** | | | | | 50 | 56 | 55 | 67 | 67 | 42 | 100 | | | | |
| **Zlu** | | | | | 49 | 54 | 53 | 65 | 66 | 40 | 98 | 100 | | | |
| **Sbi** | | | | | | | | | | | | | 100 | | |
| **Spr** | | | | | | | | | | | | | 91 | 100 | |
| **Mgi** | | | | | | | | | | | | | 87 | 84 | 100 |